\documentclass[%
 aip,
 amsmath,amssymb,
 reprint,%
]{revtex4-1}

\usepackage{graphicx}
\usepackage{dcolumn}
\usepackage{bm}

\usepackage[utf8]{inputenc}
\usepackage[T1]{fontenc}
\usepackage{mathptmx}
\usepackage[binary-units=true]{siunitx}
\usepackage{amsmath}
\usepackage{mathtools}
\usepackage{amssymb}
\usepackage{booktabs}
\usepackage{subcaption}
\usepackage{graphicx}
\usepackage{hyperref}
\usepackage{color}

\def \rb {\mathbf{r}}
\def \dr {\, \mathrm{d}\mathbf{r}}

\makeatother
\newcommand*\diff{\mathop{}\!\mathrm{d}}

\begin{document}

\preprint{AIP/123-QED}

\title[Predicting Solvation Free Energies in Non-Polar Solvents using cDFT based on PC-SAFT]{Predicting Solvation Free Energies in Non-Polar Solvents using Classical Density Functional Theory based on the PC-SAFT equation of state}

\author{Johannes Eller}
\affiliation{Institute of Thermodynamics and Thermal Process Engineering, University of Stuttgart, Pfaffenwaldring 9,70569 Stuttgart, Germany
}%

\author{Tanja Matzerath}%
\affiliation{Institute of Thermodynamics and Thermal Process Engineering, University of Stuttgart, Pfaffenwaldring 9,70569 Stuttgart, Germany
}%

\author{Thijs van Westen}%
\affiliation{Institute of Thermodynamics and Thermal Process Engineering, University of Stuttgart, Pfaffenwaldring 9,70569 Stuttgart, Germany
}%

\author{Joachim Gross}
\homepage{http://www.itt.uni-stuttgart.de}
\email{joachim.gross@itt.uni-stuttgart.de}
\affiliation{Institute of Thermodynamics and Thermal Process Engineering, University of Stuttgart, Pfaffenwaldring 9,70569 Stuttgart, Germany
}%

\date{\today}

\begin{abstract}
We propose a predictive Density Functional Theory (DFT) for the calculation of solvation free energies.
Our approach is based on a Helmholtz free-energy functional that is consistent with the perturbed-chain SAFT (PC-SAFT) equation of state.
This allows a coarse-grained description of the solvent, based on an inhomogeneous density of PC-SAFT segments.
The solute, on the other  hand, is described in full detail by atomistic Lennard-Jones interaction sites. The approach is entirely predictive, as it only takes the PC-SAFT parameters of the solvent and the force-field parameters of the solute as input.
No adjustable parameters or empirical corrections are involved.
The framework is applied to study self-solvation of \textit{n}-alkanes and to the calculation of residual chemical potentials in binary solvent mixtures.
Our DFT approach accurately predicts solvation free energies of small molecular solutes in three different solvents.
Additionally, we show the calculated solvation free energies agree well with those obtained by molecular dynamics simulations and with the residual chemical potential calculated by the bulk PC-SAFT equation of state.
We observe higher deviations for the solvation free energy of systems with significant solute-solvent Coulomb interactions.
\end{abstract}

\maketitle

\section{Introduction}
Predicting solvation free energies (SFE) is of central interest in physical chemistry and biology. 
Solvation plays an important role in biological and chemical processes, such as protein-ligand binding, solvent-mediated chemical reactions, molecular recognition, and binding affinity\cite{hirata2003molecular,feig2004performance}. 
Solvation free energies grant access to physical properties like relative solubilities, binding free energies, partition coefficients \cite{garrido2012prediction}, and activity coefficients. Especially the calculation of solvation free energies in water, i.e.\ the hydration free energy, remains a challenging task and an active field of research \cite{jorgensen1985monte,riquelme2018hydration}. 

Several theoretical approaches for estimating solvation free energies have been developed. 
Rather accurate estimates, depending on the considered force field, are obtained by molecular dynamics or Monte Carlo simulations. Such simulations allow an explicit treatment of solute and solvent molecules, in full atomistic detail.
Although this is appealing for many applications, for screening solute-solvent combinations, or for optimizing solute-solvent interaction parameters, such calculations can become too computationally expensive.
Simpler methods, such as mean-field theory or classical density functional theory, can in that context be very useful.

Phenomenological mean-field models describe the solvent only implicitly, as a polarizable medium of dielectric constant $\varepsilon$. 
The solvation free energy is then calculated from a cavitation free energy, with surface and Coulomb contributions added. 
Typically, the Coulomb contributions are modelled by Poisson-Boltzmann theory \cite{roux1999implicit} or the empirical generalized Born model \cite{bashford2000generalized,mongan2007generalized,marenich2007self}. 
Phenomenological methods neglect the effect of solutes on the microscopic structure of the solvent. 
Though this is computationally efficient, it often leads to poor agreement between calculated SFEs and those obtained by explicit simulation methods or experiments.

Classical Density Functional Theory (DFT) is a powerful approach for calculating the equilibrium density profile of molecules or molecular fragments (also called 'interaction sites') in inhomogeneous systems. 
Compared to mean-field models, DFT more faithfully captures the effect of the solute-solvent interactions on the local (microscopic) density of the solvent around a solute.
Molecular density functional theory  \cite{gendre2009classical,zhao2011molecular,borgis2012molecular,jeanmairet2013molecular} (MDFT) is a variant of classical DFT, capable of resolving much of the molecular detail of the solvent. 
This includes the orientational distribution function of molecules, which enters the theory through an angle-dependent single-particle density. 
MDFT is based on the hypernetted-chain approximation, and a second-order Taylor expansion of the excess contribution to the Helmholtz energy functional around a homogeneous reference system. 
The theory requires additional higher-order bridge functionals\cite{jeanmairet2015molecular,zhao2011new} and empirical pressure corrections \cite{luukkonen2020hydration} for improving its accuracy with respect to experiments and molecular simulations. 
MDFT has been applied to the estimation of hydration free energies of monovalent ions and small organic molecules from the FreeSolv database\cite{mobley2014freesolv}, and has been extended to include mutual polarization of the molecular solvent and solute by a coupling to electronic DFT \cite{jeanmairet2020tackling}. 
A drawback of MDFT is that it is formulated in terms of the solvent's direct correlation function (DCF), which is rather difficult to calculate with sufficient accuracy. Typically, the calculation procedure for the DCF comprises a numerical solution of the molecular Ornstein-Zernike\cite{blum1972invariant1,blum1972invariant2} (OZ) equation, after which extensive additional input from molecular simulations is needed to further increase the accuracy of the results. This 'refinement' of the OZ results needs to be performed for each each state point individually, rendering a rather computationally expensive method.
Furthermore, due to the nature of the utilized functional, MDFT does not include intramolecular flexibility of solvent molecules into its model, and thus only applies to rigid molecular models.
This deficiency makes MDFT unsuitable for the application to chain-like solvents.

Another route to a DFT of molecules is via an interaction-site model. Here each molecule is described by a set of interaction sites, where each site is ascribed its own inhomogeneous number density. The approach, championed by Chandler, McCoy and Singer \cite{chandler1986density1,chandler1986density2}, was extended by Liu and Wu for the calculation of hydration free energies of amino-acid side chains\cite{liu2013site} and small molecules\cite{liu2013high}. 
Though the approach is applicable to both, rigid and flexible molecules, the underlying functional requires empirical modifications to the hard-sphere contribution and depends on intramolecular distributions functions that have to be calculated using molecular simulations of bulk fluids.\cite{yu2002structures}
This, combined with the fact that the number of intramolecular correlation functions drastically increases with the number of solvent sites, makes the approach not trivially applicable to longer, chain-like solvent molecules.

Density functional theory based on the PC-SAFT equation of state describes molecules as chains of tangentially bonded, coarse-grained segments of equal size. Each coarse-grained segment is ascribed its own inhomogeneous number density, similar as in site-based DFT. As such, it is suitable for a treatment of chain like molecules. 
The Statistical Associating Fluid Theory (SAFT)\cite{jackson1988phase1,chapman1988phase2,chapman1989saft,chapman1990new} relies on Wertheim's first-order thermodynamic perturbation theory \cite{wertheim1984fluids1,wertheim1984fluids2,wertheim1986fluids1,wertheim1986fluids2} for describing the effects of chain formation, while the different kinds of interactions between the segments are usually handled based either on perturbation theory, or empirical expressions correlated to simulation data. Gross and Sadowski formulated the PC-SAFT \cite{gross2001perturbed} equation of state, by applying a Barker-Henderson perturbation theory to a hard chain reference fluid.
Sauer and Gross extended PC-SAFT to inhomogeneous system using a weighted density approach \cite{sauer2017classical}. The Helmholtz energy functional was successfully applied to the prediction of contact angles\cite{sauer2018prediction}, adsorption isotherms\cite{sauer2019prediction} and the calculation of Tolman lengths\cite{rehner2018surface}.

Here we extend PC-SAFT DFT to the calculation of solvation free energies. Only the solvent is described using the coarse-grained molecular model of PC-SAFT, whereas the solute is described in full atomistic detail, based on a force-field for molecular simulations. The approach requires no empirical adjustments based on input from molecular simulations and only takes the molecular parameters of PC-SAFT (solvent) and the force-field (solute) as input. It is thus computationally much cheaper than available methods, while~\textemdash most importantly~\textemdash being fully predictive.
This work should be considered as an initial study, in which we asses the predictive capability of the method for the calculation of solvation free energies in non-polar solvents. We study self-solvation of \textit{n}-alkanes in pure solvents and binary solvent mixtures. We select molecules from different chemical groups and perform SFE calculations of these molecules in three different solvents. Calculated SFEs are compared to the results of molecular dynamics simulations performed in this work and experiments from the literature.

\section{\label{sec:SolvationTD}Solvation Thermodynamics}

The solvation free energy (SFE) is defined as the reversible work needed to transfer a solute molecule of species $s$ from a \textit{fixed position} in an ideal gas phase to a \textit{fixed position} in a fluid phase.\cite{ben2013solvation}
Experiments and MD simulations are typically conducted in the isothermal-isobaric ensemble, with a specified number of solvent molecules $N=\left\lbrace N_i,i=1,\dots,\nu\right\rbrace$, pressure $p$ and temperature $T$, where $\nu$ is the number of solvent species. The appropriate ensemble for the study of solvation using classical DFT is the semi-grand canonical ensemble.
Here, the volume $V$, the temperature $T$, the chemical potentials of all solvent species $\mu=\left\lbrace\mu_i,i=1,\dots,\nu\right\rbrace$, and additionally, the number of solute molecules $N_s$ are specified.
We show how the free energy of solvation can be calculated in various ensembles. The systems are considered sufficiently large, in order for the results obtained in one ensemble to be consistent with the results from other ensembles.

The semi-grand canonical potential $\Tilde{\Omega}$ can be expressed in terms of the grand-canonical potential $\Omega$ or the Helmholtz energy $F$, as

\begin{equation}
\begin{aligned}
    \Tilde{\Omega}(N_s,\mu,V,T)&\equiv \Omega(\mu,V,T)+\mu_s~N_s\\
    &\equiv F(N_s,N,V,T)-\sum_{i=1}^{\nu} \mu_i~N_i
    \label{eq:T Omega}
\end{aligned}
\end{equation}

\noindent
where the subscript 's' denotes the solute species $s$. The SFE in the semi-grand canonical ensemble $\Delta \Tilde{\Omega}_\mathrm{solv}$ is defined as \cite{ben2013solvation}

\begin{equation}
    \Delta \Tilde{\Omega}_\mathrm{solv}\left(N_s,\mu,V,T\right)=\mu_s^\ast\left(N_s,\mu,V,T\right)-\mu_s^{\ast,\mathrm{ig}}
    \label{eq:Delta Omega_solv}
\end{equation}

\noindent
where $\mu_s^\ast$ and $\mu_s^{\ast,\mathrm{ig}}$ are the pseudo-chemical potentials of the fluid phase and the ideal gas, respectively. 
The pseudo-chemical potential corresponds to the change in the semi-grand canonical potential by adding an additional solute molecule at a fixed location. In our case, where $N_s$ solute molecules are present in the system, we add the $(N_s+1)$-th solute molecule at fixed position $\mathbf{r}_{N_s+1}=\mathbf{r}_0$ to the system, that is

\begin{align}
    \mu_s^\ast\left(N_s,\mu,V,T\right)&=\Tilde{\Omega}\left(N_s+1,\mu,V,T;\rb_0\right)-\Tilde{\Omega}\left(N_s,\mu,V,T\right) \nonumber\\
    &=F(N_s+1,N,V,T;\rb_0)-F(N_s,N,V,T)
    \label{eq:mu*(N,mu,V,T)_1}
\end{align}

\noindent
where we invoked eq. \eqref{eq:T Omega}. We limit consideration to homogeneus systems, where the value of $\mu_s^\ast\left(N_s,\mu,V,T\right)$ is invariant with the choice of $\mathbf{r}_0$, which is why we do not make the location $\mathbf{r}_0$ explicit in the variable list of $\mu_s^\ast$.
The pseudo-chemical potential in the semi-grand canonical ensemble thus equals the pseudo-chemical potential in the canonical ensemble $\mu_s^\ast\left(N_s,N,V,T\right)$, provided the specified number of solvent molecules $N$ of the canonical ensemble agree with the ensemble average of $N$ from the grand-canonical ensemble. The first line of eq.~\eqref{eq:mu*(N,mu,V,T)_1} is used for calculating the SFE from classical DFT, as shown in the next section.

We continue with the analysis of the canonical ensemble, with
\begin{equation}
    \mu_s^\ast\left(N_s,\mu,V,T\right)=-k_B T\ln{\frac{Q(N_s+1,N,V,T;\rb_0)}{Q(N_s,N,V,T)}}
\end{equation}
where $k_B$ is Boltzmann's constant and $Q$ is the canonical partition function.
Following the detailed derivation given in the supporting information, the above equation can be rewritten to obtain the expression for the solute's pseudo-chemical potential 
\begin{multline}
\mu_s^\ast(N_s,\mu,V,T)=-k_B T \ln{\left(q_s~q_s^\mathrm{intra}\right)}\\
    -k_B T\ln{\left\langle \left\langle \exp{\left(-\beta U_B\left(\mathbf{X}^N,\mathbf{X}^{N_s};\mathbf{R}_{N_s+1}\right)\right)}\right\rangle_{\mathbf{R}_{N_s+1}}\right\rangle_{N_s NVT}}
    \label{eq:mu*(N,mu,V,T)_2}   
\end{multline}
where $\beta=\left(k_B T\right)^{-1}$ is the inverse temperature, $q_s$ is the solute's intramolecular partition function due to nuclear and electronic degrees of freedom and $q_s^\mathrm{intra}$ is the solute's intramolecular partition function due to bonded and non-bonded intramolecular interactions.
The binding energy $U_B$ is the intermolecular potential energy of the newly added solute molecule at center of mass position $\mathbf{r}_0$ and of intramolecular configuration $\mathbf{R}_{N_s+1}$ with respect to the remaining solvent molecules. 
We introduce the short-hand notation $\mathbf{X}=\left\lbrace\rb,\mathbf{R}\right\rbrace$ with the center of mass position $\rb$ and the intramolecular configuration $\mathbf{R}$ relative to the center of mass position and we further use the notation $\mathbf{X}^N$ for $\mathbf{X}_1,..\mathbf{X}_N$. The angle brackets $\left\langle~\right\rangle_{N_sNVT}$ indicate a canonical ensemble average over the configurations of all solvent and solute molecules other than the newly introduced $(N_s+1)$-th solute molecule, whereas $\left\langle~\right\rangle_{\mathbf{R}_{N_s+1}}$ denotes an average over the intramolecular configurations $\mathbf{R}_{N_s+1}$ of the new solute molecule.
Due to the absence of intermolecular interactions in the ideal-gas phase, the ideal-gas pseudo-chemical potential $\mu_s^{\ast,ig}$, follows from eq.~\eqref{eq:mu*(N,mu,V,T)_2}, as
\begin{equation}
    \mu_s^{\ast,ig}=-k_B T\ln{\left(q_s~q_s^\mathrm{intra}\right)}
    \label{eq:mu_s^ig}
\end{equation}
The ideal gas pseudo-chemical potential $\mu_s^{\ast,ig}$ is for a solute molecule at fixed center of mass position and without intermolecular interactions to other molecules.
Substitution of eqs.~\eqref{eq:mu*(N,mu,V,T)_2} and \eqref{eq:mu_s^ig} in eq.~\eqref{eq:Delta Omega_solv} gives the solvation free energy in the canonical ensemble,
\begin{multline}
    \Delta F_\mathrm{solv}\left(N_s,N,V,T\right)=\\-k_B T\ln{\left\langle \left(\langle\exp{\left(-\beta U_B\left(\mathbf{X}^N,\mathbf{X}^{N_s};\mathbf{R}_{N_s+1}\right)\right)}\right\rangle_{\mathbf{R}_{N_s+1}}\right\rangle_{N_s NVT}}
    \label{eq:Delta T Omega_solv}
\end{multline}
which is equal to the solvation free energy in the semi-grand canonical ensemble, $\Delta \Tilde{\Omega}_\mathrm{solv}=\Delta F_\mathrm{solv}$, according to eq.~\eqref{eq:Delta Omega_solv} and \eqref{eq:mu*(N,mu,V,T)_1}.
As shown in the supporting information, namely eq. (S17), the SFE is also equal to the residual chemical potential within the canonical ensemble, leading to
\begin{equation}
    \Delta F_\mathrm{solv}\left(N_s,N,\mu,V,T\right)=\mu_s^\mathrm{res}\left(Ns,N,V,T\right)
    \label{eq:residual_chemical_potential}
\end{equation}

The SFE in the \textit{NpT} ensemble, i.e.\ the Gibbs energy of solvation $\Delta G_\mathrm{solv}(N_s,N,p,T)$, is defined similarly, as
\begin{equation}
    \Delta G_\mathrm{solv}(N_s,N,p,T)=\mu_s^\ast(N_s,N,p,T)-\mu_s^{\ast,\mathrm{ig}}
    \label{eq:Delta G_solv}
\end{equation}
The pseudo-chemical potential $\mu_s^\ast(N_s,N,p,T)$ corresponds to the change in the Gibbs energy due to adding one additional solute molecule at a fixed location $\mathbf{r}_0$ to the system,
\begin{align}
    \mu_s^\ast(N_s,N,p,T)&=\mathrm{G}(N_s+1,N,p,T;\mathbf{r}_0)-\mathrm{G}(N_s,N,p,T)\nonumber\\
    &=-k_B T\ln{\frac{\Delta(N_s+1,N,p,T;\rb_0)}{\Delta(N_s,N,p,T)}}
\end{align}
with the isobaric-isothermal partition function $\Delta(N_s,N,p,T)$. As shown in the supporting material of this work, the above expression for the pseudo-chemical potential can be expressed in terms of the following \textit{NpT} ensemble average,
\begin{multline}
    \mu_s^\ast(N_s,N,p,T)=-k_B T\ln{\left(q_s~q_s^\mathrm{intra}\right)}\\
    -k_B T\ln{\left\langle\left\langle(\exp{\left(-\beta U_B\left(\mathbf{X}^{N_s},\mathbf{X}^N;\mathbf{R}_{N_s+1}\right)\right)}\right\rangle_{\mathbf{R}_{N_s+1}}\right\rangle_{N_sNpT}}
    \label{eq:mu*(Ns,N,p,T)}
\end{multline}
The pseudo-chemical potential of the ideal gas phase in the \textit{NpT} ensemble is thus the same as in the semi-grand canonical ensemble, see eq. \eqref{eq:mu_s^ig}.
Combining equations \eqref{eq:mu_s^ig}, \eqref{eq:Delta G_solv} and \eqref{eq:mu*(Ns,N,p,T)}, we obtain the following expression for the Gibbs energy of solvation
\begin{multline}
    \Delta G_\mathrm{solv}(N_s,N,p,T)=\\-k_B T\ln{\left\langle\left\langle\exp{\left(-\beta U_B\left((\mathbf{X}^{N_s},\mathbf{X}^N;\mathbf{R}_{N_s+1}\right)\right)}\right\rangle_{\mathbf{R}_{N_s+1}}\right\rangle_{N_sNpT}}
    \label{eq:Delta G_solv2}
\end{multline}
In the thermodynamic limit, i.e.\ for sufficiently large systems, the ensemble averages $\left\langle~\right\rangle_{N_sNVT}$ and $\left\langle~\right\rangle_{N_sNpT}$ become equivalent for appropriately chosen variables; therefore, eqs. \eqref{eq:Delta Omega_solv}, \eqref{eq:mu*(N,mu,V,T)_1}, \eqref{eq:Delta T Omega_solv} and \eqref{eq:Delta G_solv2} imply that the solvation free energy does not depend on the ensemble in which it is calculated, that is
\begin{equation}
    \Delta \Tilde{\Omega}_\mathrm{solv}(N_s,\mu,V,T)=\Delta F_\mathrm{solv}(N_s,N,V,T)=\Delta G_\mathrm{solv}(N_s,N,p,T)
\end{equation}
We can thus directly compare solvation free energies in the semi-grand canonical $(N_s,\mu,V,T)$, canonical $(N_s,N,V,T)$ and the isobaric-isothermal $(N_s,N,p,T)$ ensemble. In the remainder, we use the term "solvation free energy" as a general term, denoting either the semi-grand potential, Helmholtz energy or Gibbs energy of solvation.

\section{\label{sec:DFT_solvation}Density Functional Theory of Solvation}
In this section, we summarize some of the basics of classical density functional theory (DFT) and its application to solvation. 
The solvent (consisting of $\nu$ species) is represented by the inhomogeneous density field $\rho(\rb)=\left\{\rho_i(\rb),i=1,\dots,\nu\right\}$. The solute molecule newly inserted into the system is described through an external potential $V_i^\mathrm{ext}(\rb)$ that acts on this density field.
The grand-canonical potential $\Omega$ can be expressed as a unique functional of the solvent density $\rho(\rb)$, as
\begin{equation}
    \Omega[\rho(\rb)]=F[\rho(\rb)]-\sum_{i=1}^{\nu} \int \rho_i(\rb) \left(\mu_i-V_i^\mathrm{ext}(\rb)\right) \mathrm{d}\mathbf{r}
    \label{eq:Omega_DFT}
\end{equation}
where $\mu_i$ is the chemical potential of solvent species $i$ and $F[\rho(\rb)]$ is the solvent's Helmholtz energy functional. The equilibrium density profile minimizes the functional and can be calculated by solving the Euler-Lagrange equation
\begin{equation}
    \frac{\delta F[\rho(\rb)]}{\delta \rho_i(\rb)}=\mu_i-V_i^\mathrm{ext}(\rb)
    \label{eq:ELE}
\end{equation}
The solvent-solvent interactions are described by the Helmholtz energy functional $F[\rho(\rb)]$ based on the PC-SAFT equation of state. The underlying molecular model of the PC-SAFT equation of state coarse grains molecules as chains of tangentially bound spherical segments. 
The theory contains no intramolecular potential energy contributions; the chains are thus fully flexible. 
For the non-polar, non-associating solvent molecules studied in this work, only the Helmholtz-energy contributions due to hard-sphere interactions \cite{roth2002fundamental,yu2002structures}, hard-sphere-chain formation \cite{tripathi2005microstructure1,tripathi2005microstructure2}, and dispersion interactions \cite{sauer2017classical} have to be considered, leading to
\begin{align}
    F[\rho(\rb)]&=
    F^\mathrm{ig}[\rho(\rb)]+F^\mathrm{res}[\rho(\rb)]\\
    F^\mathrm{res}[\rho(\rb)]&=F^\mathrm{hs}[\rho(\rb)]+F^\mathrm{hc}[\rho(\rb)]+F^\mathrm{disp}[\rho(\rb)]
    \label{eq:F_terms}
\end{align}
The ideal gas contribution $F^\mathrm{ig}[\rho(\rb)]$ is known exactly from statistical mechanics, with
\begin{equation}
    F^\mathrm{ig}[\rho(\rb)]=k_B T\sum_{i=1}^\nu \int \rho_i(\rb)\left(\ln{\left(\rho_i(\rb)\frac{\Lambda_i^3}{q_i~q_i^\mathrm{intra}}\right)}-1\right) \dr
\end{equation}
with the de Broglie wavelength $\Lambda_i$ of molecule $i$.
The pure-component parameters required for describing the other Helmholtz-energy contributions are the number of segments per molecule $m_i$, the segment size parameter  $\sigma_i$ and the dispersive energy parameter $\varepsilon_i$.
The Helmholtz-energy functional applied in this work does not distinguish individual segments of a chain.
The density profile of component $i$ is determined as an average over the density profiles of individual segments $\alpha_i$,
\begin{equation}
    \rho_i(\rb)=\frac{1}{m_i}\sum_{\alpha_i}^{m_i} \rho_{\alpha_i}(\rb)
\end{equation}
leading to $\rho_i(\rb)=\rho_{\alpha_i}(\rb)$ for homosegmented chains.
A description beyond this approximation is possible. The connectivity of chains can be accounted for using the Thermodynamic Perturbation Theory of Wertheim\cite{wertheim1984fluids1,wertheim1984fluids2, wertheim1986fluids1,wertheim1986fluids2,zmpitas2016detailed} as proposed by Jain et al.\cite{jain2007modified} and applied with a similar functional by Mairhofer et al.\cite{mairhofer2018classical}.
A detailed guide for the implementation of the utilized functionals and for solving the occurring convolution integrals in 3-dimensions using Fast Fourier Transforms can be found in our previous work \cite{stierle2020guide}. Further details of the DFT calculations performed in this work are given in Appendix \ref{app:implementation}. 

\begin{figure}
    \centering
    \begin{subfigure}[b]{0.27\textwidth}
        \centering
        \includegraphics[clip, trim=0 50 0 50,width=0.8\textwidth]{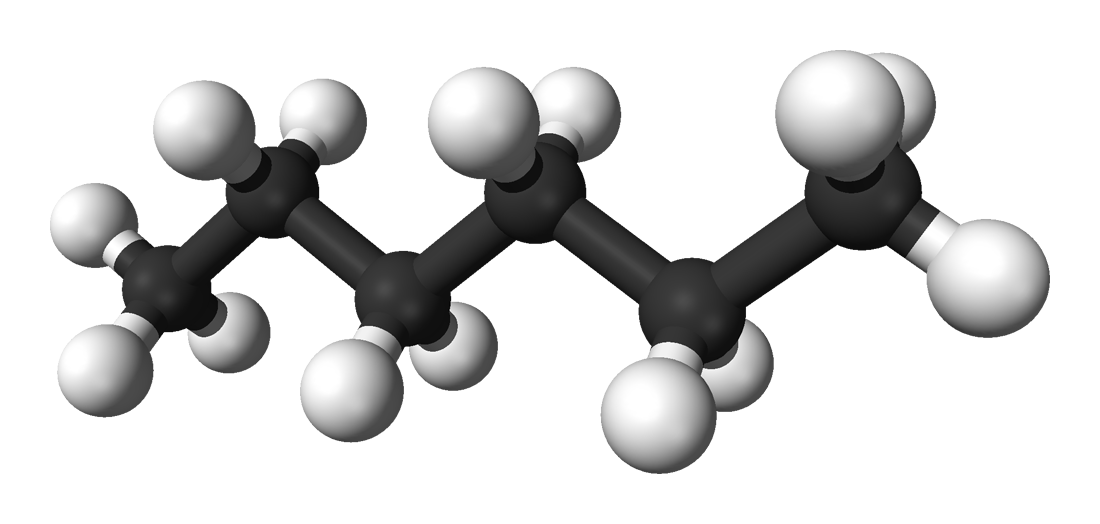}
    \end{subfigure}
    \begin{subfigure}[b]{0.2\textwidth}
        \centering
        \includegraphics[clip, trim=50 10 80 50,width=0.5\textwidth]{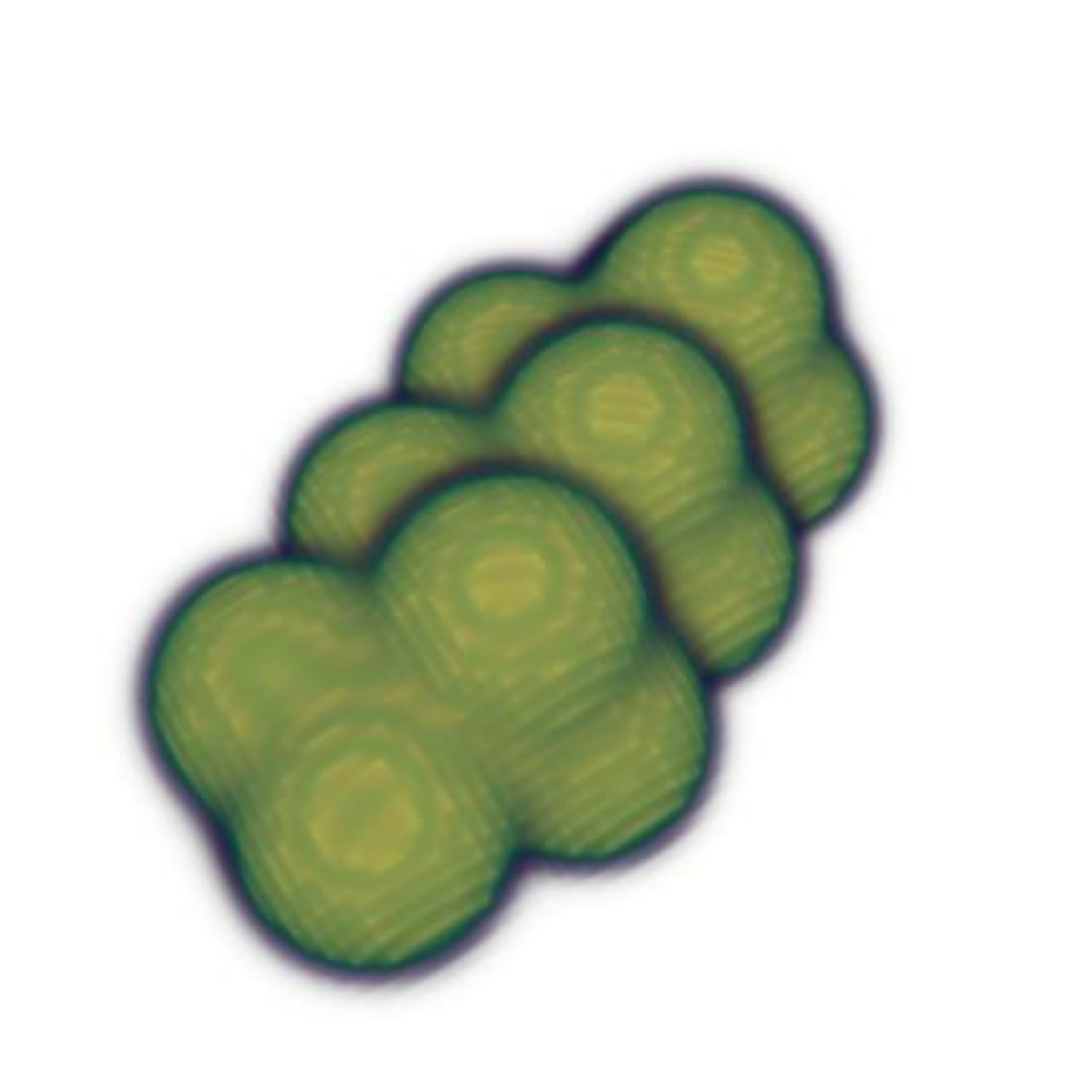}
    \end{subfigure}
    \caption{Atomistic \textit{n}-hexane structure with individual Lennard-Jones interaction sites (left) and external potential surface (right) resulting from eq. \eqref{eq:ext_pot} using the GAFF force field.}
    \label{fig:hexaneDFT_MD}
\end{figure}

We account for the (only) van-der-Waals solute-solvent interactions (corresponding to hard-sphere and dispersive contributions in eq.~\eqref{eq:F_terms}) through the external potential $V_i^\mathrm{ext}(\rb)$. This allows us to describe the solute in full microscopic detail, based on atomistic Lennard-Jones interaction sites, according to
\begin{equation}
    V_i^\mathrm{ext}(\rb)=m_i\sum_{\alpha=1}^M 4 \varepsilon_{\alpha i} \left( \left(\frac{\sigma_{\alpha i}}{|\mathbf{r}_{\alpha}-\mathbf{r}|}\right)^{12}-\left(\frac{\sigma_{\alpha i}}{|\mathbf{r}_{\alpha}-\mathbf{r}|}\right)^{6}\right)
    \label{eq:ext_pot}
\end{equation}
Here, $M$ is the total number of solute interaction sites, $\sigma_{\alpha i}$ and $\varepsilon_{\alpha i}$ are the LJ parameters for interactions between an atomistic interaction site $\alpha$ of the solute and a PC-SAFT segment of solvent molecule $i$, and $\mathbf{r}_{\alpha}$ are the coordinates of the solute interaction sites. The interaction parameters $\sigma_{\alpha i}$ and $\varepsilon_{\alpha i}$ are calculated using the Berthelot-Lorentz combining rules
\begin{equation}
\begin{aligned}
    \sigma_{\alpha i}&=(\sigma_\alpha+\sigma_i)/2\\
    \varepsilon_{\alpha i}&=\sqrt{\varepsilon_\alpha \varepsilon_i}  
\end{aligned}
\end{equation}
using the PC-SAFT parameters $\sigma_i$ and $\varepsilon_i$, and the Lennard-Jones parameters $\sigma_\alpha$ and $\varepsilon_\alpha$ of solute site $\alpha$ taken from the molecular force-field. A visual representation of the individual Lennard-Jones interaction sites of \textit{n}-hexane and the resulting external potential surface calculated with eq. \eqref{eq:ext_pot} using the General Amber (GAFF) force field is given in figure \ref{fig:hexaneDFT_MD}.
We only consider one representative (intramolecular) configuration of the solute for the DFT calculations. 
For the simple weakly-polar solute molecules considered in this work, using more than just a single configuration did not lead to significant changes in the calculated solvation free energies.
We do not include solute-solvent Coulomb interactions, as individual PC-SAFT segments do not carry partial charges. Solvents with dipole and quadrupole moments or hydrogen bond forming solvents such as water and alcohols are not considered in this study.

The solvation free energy is calculated using eqs. \eqref{eq:Delta Omega_solv} and \eqref{eq:mu*(N,mu,V,T)_1}, but with the solute molecule added to the system represented by the external potential $V^\mathrm{ext}(\rb)=\left\{V^\mathrm{ext}_i(\rb),i=1,\dots,\nu\right\}$, leading to
\begin{multline}
    \Delta \tilde{\Omega}_\mathrm{solv}(N_s,\mu,V,T)=\Tilde{\Omega}\left(N_s,\mu,V,T;V^\mathrm{ext}(\rb)\right)\\
    -\Tilde{\Omega}\left(N_s,\mu,V,T;V^\mathrm{ext}(\rb)=0\right)
\end{multline}
where $\Tilde{\Omega}\left(N_s,\mu,V,T;V^\mathrm{ext}(\rb)\right)$ is the semi-grand canonical potential of the inhomogeneous system that contains the additional solute molecule and $\Tilde{\Omega}\left(N_s,\mu,V,T;V^\mathrm{ext}(\rb)=0\right)$ is the semi-grand canonical potential of the homogeneous bulk system in the absence of the solute molecule.
In this equation we have not subtracted the pseudo-chemical potential of the ideal gas $\mu_s^{\ast,\mathrm{ig}}$ as eq.~\eqref{eq:Delta Omega_solv} might initially suggest. That is because the solute molecule introduced through the external potential $V^\mathrm{ext}(\rb)$ does not carry a free energy contribution due to intramolecular energies and consequently we do not have to subtract this contribution (namely $\mu_s^{\ast,\mathrm{ig}}$).
By using the first line of eq. \eqref{eq:T Omega}, the solvation free energy $Delta \tilde{\Omega}_\mathrm{solv}(N_s,\mu,V,T)$ can be expressed as the difference in the grand-canonical potential, according to
\begin{align}
   \Delta \tilde{\Omega}_\mathrm{solv}(N_s,\mu,V,T)=&\left(\Omega\left(\mu,V,T;V^\mathrm{ext}(\rb)\right)+\mu_s N_s\right)\nonumber\\
    &\qquad-\left(\Omega\left(\mu,V,T;V^\mathrm{ext}(\rb)=0\right)+\mu_s N_s\right)\nonumber\\
    =&~~\Omega\left[\rho(\rb);V^\mathrm{ext}(\rb)\right]-\Omega\left(\mu,V,T\right)\nonumber\\
    =& \Delta \Omega_\mathrm{solv}(\mu,V,T)
    \label{eq:SFE_DFT}
\end{align}
where $\Omega\left[\rho(\rb);V^\mathrm{ext}(\rb)\right]$ is the grand-canonical functional of the inhomogeneous system, given by eq.\eqref{eq:Omega_DFT}, and $\Omega\left(\mu,V,T\right)$ is the grand-canonical potential of the homogeneous bulk system.

\section{\label{sec:MD}Molecular Dynamics simulation: Alchemical Free Energy Calculations}

To assess the accuracy of the solvation free energies predicted by PC-SAFT DFT, we calculated benchmark solvation free energies using molecular dynamics simulations in combination with alchemical free energy methods \cite{rodinger2005enhancing,shirts2007alchemical}.

To explain the principle of such methods, consider a system characterized by the generalized positions and momenta of molecules $\mathbf{q}$ and $\mathbf{p}$, at two different states $A$ and $B$. The difference between the two states is governed by a coupling parameter $\lambda$; the respective Hamiltonians are written as $\mathcal{H}_A\left(\mathbf{q},\mathbf{p};\lambda\right)$ and $\mathcal{H}_B\left(\mathbf{q},\mathbf{p};\lambda\right)$. Alchemical methods provide the means to calculate the free energy difference between states $A$ and $B$ utilizing a combined Hamiltonian given by 
\begin{equation}
    \mathcal{H}\left(\mathbf{q},\mathbf{p};\lambda\right)=f(\lambda)\mathcal{H}_A\left(\mathbf{q},\mathbf{p};\lambda\right)+g(\lambda)\mathcal{H}_B\left(\mathbf{q},\mathbf{p};\lambda\right)
\end{equation}
where $f(\lambda)$ and $g(\lambda)$ are the mixing functions of the Hamiltonians, with $\mathcal{H}=\mathcal{H}_A$ for $\lambda=0$ and $\mathcal{H}=\mathcal{H}_B$ for $\lambda=1$. 
For the calculation of solvation free energies, $\mathcal{H}_A$ corresponds to the system with full solute-solvent interactions and $\mathcal{H}_B$ to the system without solute-solvent interactions. 
The coupling parameter $\lambda$ thus only acts on the solute-solvent interactions. Any small contribution to the solvation free energy due to changes in the solute's average intramolecular potential energy upon turning on the solute-solvent interactions are thereby neglected.
The choice of intermediate $\lambda$ states and the resulting phase-space overlap between neighboring states significantly increases the efficiency and accuracy of the solvation free energy calculations.
To prevent overlapping of opposite charges during the annihilation of the solute-solvent interactions, the mixing functions $f(\lambda)$ and $g(\lambda)$ first turn off the Coulomb interactions and then decrease the Lennard-Jones interactions using a soft-core potential\cite{beutler1994avoiding,zacharias1994separation}.
We calculate SFEs using a thermodynamic integration path of 20 intermediate states, where the first five states were used to turn off the Coulomb potential, whereas the last 15 states are used to decrease the Lennard-Jones interactions.

The solvation free energy $\Delta G_\mathrm{solv}$ is then calculated in a post-processing step, after performing molecular simulations at discrete $\lambda $ values using alchemical free energy methods like thermodynamic integration\cite{kirkwood1935statistical}, exponential averaging \cite{zwanzig1954high}, Bennet's acceptance ratio (BAR) \cite{bennett1976efficient} and multistate Bennet's acceptance ratio (MBAR) \cite{shirts2008statistically}. An extensive comparison between all alchemical free energy methods can be found in the literature \cite{shirts2005comparison,ytreberg2006comparison,paliwal2011benchmark}.
MBAR showed a consistent good performance, and is thus our method of choice. The aforementioned alchemical free energy methods are implemented in the Alchemical Analysis Python package\cite{Klimovich:2015er} which is used in this work to estimate the SFEs.

For each $\lambda$ state, we performed an energy minimization, $NVT$ equilibration and $NpT$ equilibration step before the production runs. The MD simulations of this work were performed using the GROMACS 2019.4 \cite{berendsen1995gromacs,lindahl2001gromacs,van2005gromacs} simulation package. Bonded and non-bonded interactions are described with the AMBER GAFF force field with AM1-BCC charges \cite{wang2004development}. Atom types, partial charges and the input files for GROMACS for the aforementioned workflow are taken directly from the FreeSolv database \cite{mobley2014freesolv}, version 0.52.
MD simulations are performed in a periodic box with 156 solvent molecules. In the case of self-solvation, the total number of molecules is increased to 500. All bond lengths are kept rigid using the SHAKE algorithm \cite{ryckaert1977numerical} with a relative tolerance of $\SI{1.0e-4}{}$. The equations of motion are integrated by the leap frog algorithm \cite{hockney1970potential} using a $\SI{2}{\femto\second}$ time step with a total simulation time of $\SI{5}{\nano\second}$ for the production runs. Non-bonded interactions were neglected beyond a cutoff of $\SI{1.2}{\nano\meter}$. The temperature is maintained by an Andersen thermostat and pressure is kept constant by a Parrinello-Rahman barostat \cite{parrinello1980crystal}.

\section{\label{sec:results}Results}
\subsection{\label{subsec:selfsolvation}Self-Solvation of \textit{n}-alkanes}

\begin{figure}[h]
    \centering
    \includegraphics[width=.48\textwidth]{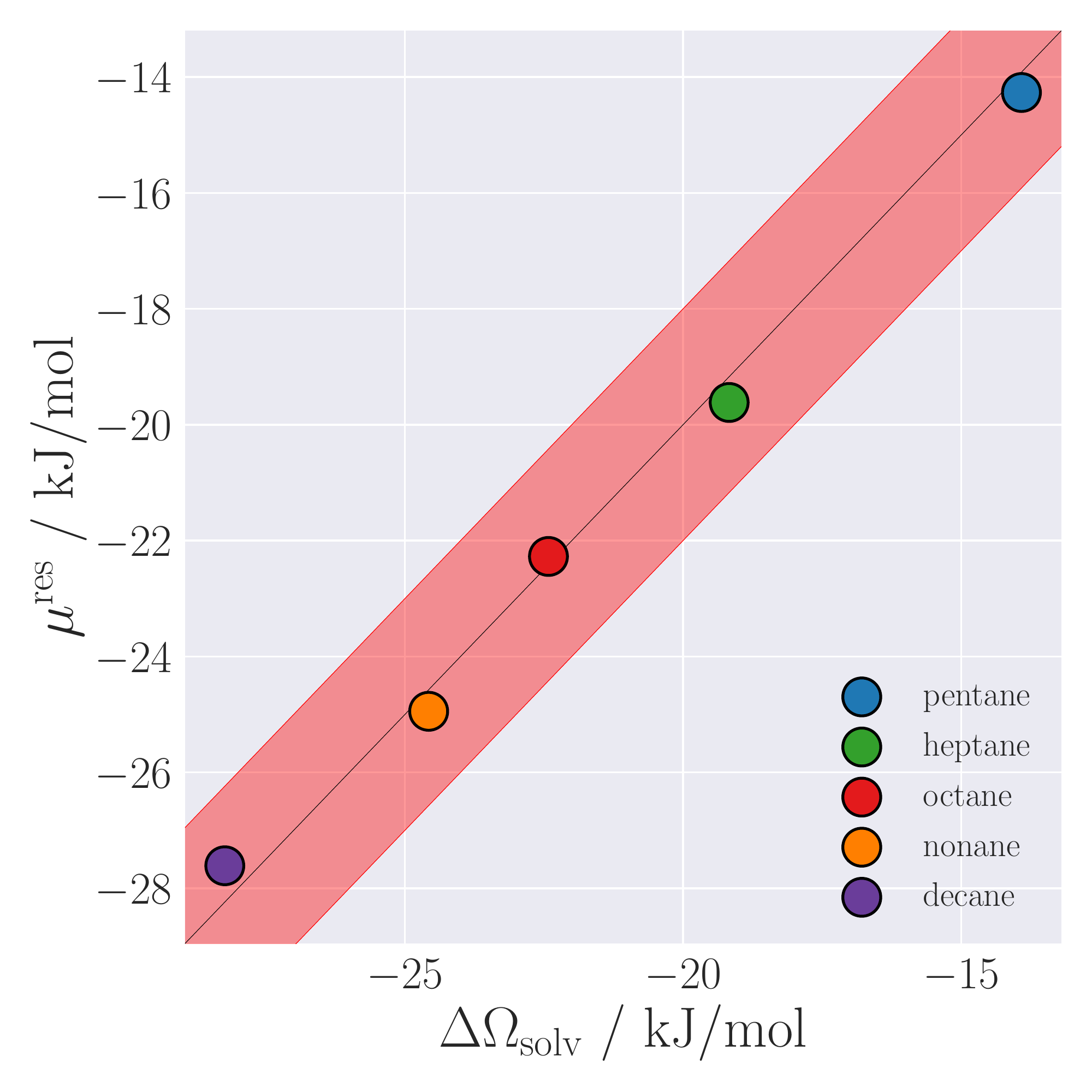}
    \caption{Correlation between self-solvation free energies by PC-SAFT DFT and residual chemical potentials $\mu^\mathrm{res}\left(T,\mathbf{\rho}_s^\mathrm{bulk}\right)$ calculated based on PC-SAFT for several \textit{n}-alkanes at $\SI{298}{\kelvin}$ and $\SI{1}{\bar}$. The red lines indicate a difference of ${\displaystyle \pm }\SI{2}{\kilo\joule\per\mol}$.}
    \label{fig:SS_alkanes298K1bar_SAFT}
\end{figure}

Firstly, we consider self-solvation where the solute is of the same species as the solvent. Conceptually this is similar to Percus' test particle theory\cite{percus1962approximation,percus1964equilibrium}, which involves introducing a test particle represented by an external potential equal to the pair potential of the solvent. In our PC-SAFT DFT approach the solute is not described using the (coarse-grained) pair potential of the solvent, instead, the solute is described in full atomistic detail, using coordinates and interaction parameters of Lennard-Jones interaction sites of the Amber GAFF force field.
The PC-SAFT parameters necessary to describe the solvents are taken from the literature\cite{gross2001perturbed} and can be found in the supporting material.
The self-solvation free energy $\Delta \Omega_\mathrm{solv}(\mu,V,T)$, following eqs. \eqref{eq:Delta Omega_solv}-\eqref{eq:mu*(N,mu,V,T)_1}, \eqref{eq:residual_chemical_potential}, and \eqref{eq:SFE_DFT}, then equals the residual chemical potential of a bulk fluid,
\begin{equation}
    \Delta \Omega_\mathrm{solv}(\mu,V,T)=\mu^\mathrm{res}\left(T,\mathbf{\rho}^\mathrm{bulk}\right)
    \label{eq:consistency}
\end{equation}
where the residual chemical potential is evaluated at given temperature $T$ and bulk density $\mathbf{\rho}^\mathrm{bulk}=\left\lbrace\rho_i^\mathrm{bulk},i=1,\dots,\nu\right\rbrace$.
Eq. \eqref{eq:consistency} provides a consistency test between the solvation free energy calculated by the inhomogeneous PC-SAFT DFT framework, molecular dynamics simulations and the bulk residual chemical potential of the PC-SAFT equation of state.
Here, the self-solvation free energy is calculated using three different solute-solvent representations; $\Delta \Omega_\mathrm{solv}$ is for an atomistic GAFF solute in a PC-SAFT DFT solvent, $\Delta G_\mathrm{solv}$ is for a GAFF solute in GAFF solvent and $\mu^\mathrm{res}$ is the self-solvation free energy according to the PC-SAFT equation of state (avoiding the DFT route, eq. \eqref{eq:SFE_DFT}).
In Table \ref{tab:FSE_alkanes}, we compare self-solvation free energies calculated with PC-SAFT DFT and molecular simulations to the respective residual chemical potential obtained by the bulk PC-SAFT equation of state for the homologous series of \textit{n}-alkanes from methane to decane at $\SI{298}{\kelvin}$ and $\SI{1}{\bar}$. We report excellent agreement between the self-solvation free energies from PC-SAFT DFT and molecular simulations, and the residual chemical potentials calculated with PC-SAFT, with a maximum deviation of less than $\SI{1}{\kilo\joule\per\mol}$. Figure \ref{fig:SS_alkanes298K1bar_SAFT} shows results from pentane to decane, the data points from C1 to C4 are omitted for clarity. The \textit{y}-axis shows the residual chemical potential calculated based on PC-SAFT and the \textit{x}-axis the self-solvation free energy by DFT. The red lines indicate a difference of ${\displaystyle \pm }\SI{2}{\kilo\joule\per\mol}$. All data points for the considered \textit{n}-alkanes are located close the main diagonal, implying excellent correlation between both approaches. Our findings confirm the adequacy of the Berthelot-Lorentz combining rules for calculating cross-interaction parameters between atomistic interaction sites of a force field (Amber GAFF in this case) and coarse-grained molecular segments of the PC-SAFT equation of state.

\begin{figure}
    \centering
    \includegraphics[width=0.45\textwidth]{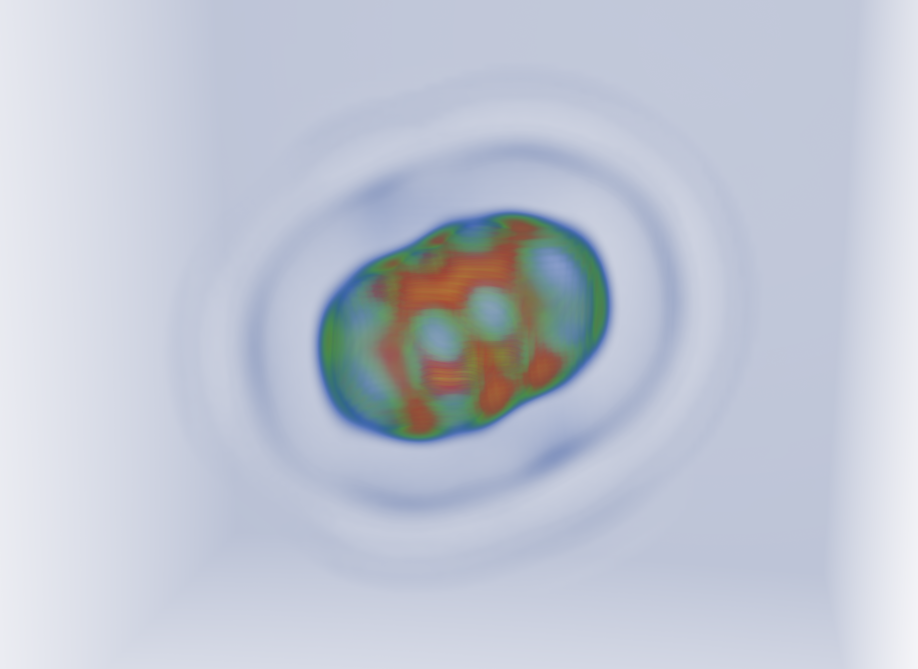}
    \caption{PC-SAFT DFT results for the density profile of a \textit{n}-hexane solvent surrounding an \textit{n}-hexane solute molecule, at $\SI{298}{\kelvin}$ and $\SI{1}{\bar}$. Yellow, red and green indicate areas of higher density, where as blue corresponds to densities close to the bulk density.}
    \label{fig:density_hexane}
\end{figure}

\begin{table}[]
    \centering
    \begin{tabular}{lccc}
    \toprule
    Solute & $\Delta \Omega_\mathrm{solv}/\SI{}{\kilo\joule\per\mol}$ & $\Delta G_\mathrm{solv}/\SI{}{\kilo\joule\per\mol}$ & $\mu^\mathrm{res}/\SI{}{\kilo\joule\per\mol}$ \\
    \midrule
        methane        &  -0.0055   & \hphantom{2}-0.021$\pm$0.001\hphantom{6} &-0.0093 \\
        ethane         &  -0.0318   & --- &-0.0376 \\
        propane        &  -0.0729   & --- &-0.0759 \\
        butane         &  -0.1391   & --- &-0.1287 \\
        pentane        & -13.9175   & -13,113$\pm$0.0036 &-14.267 \\
        hexane         & -16.2542   & -16,784$\pm$0,082\hphantom{6}  &-16.956 \\
        heptane        & -19.1715   & -19.269$\pm$0.051\hphantom{6} &-19.616 \\
        octane         & -22.4200   & \hphantom{2}-22.23$\pm$0.061\hphantom{6} &-22.273 \\
        nonane         & -24.5735   & -25.638$\pm$0.074\hphantom{6} &-24.945 \\
        decane         & -28.2381   & -28.524$\pm$0.086\hphantom{6} &-27.611 \\
    \bottomrule
    \end{tabular}
    \caption{Self-solvation free energies from PC-SAFT DFT, MD simulations and residual chemical potentials $\mu^\mathrm{res}\left(T,\mathbf{\rho}^\mathrm{bulk}\right)$ calculated with PC-SAFT of \textit{n}-alkanes at $\SI{298}{\kelvin}$ and $\SI{1}{\bar}$.}
    \label{tab:FSE_alkanes}
\end{table}
Figure \ref{fig:density_hexane} shows the 3-dimensional density profile of hexane surrounding a hexane solute molecule, as calculated by PC-SAFT DFT. The void space in the center of the box is a measure for the excluded volume of the solute molecule. Areas of high densities in the first solvation shell are marked in yellow and red. The figure also shows the inhomogeneous solvent structure forming multiple solvent layers in the vicinity of the solute.
\begin{figure}[h]
    \centering
    \includegraphics[width=.48\textwidth]{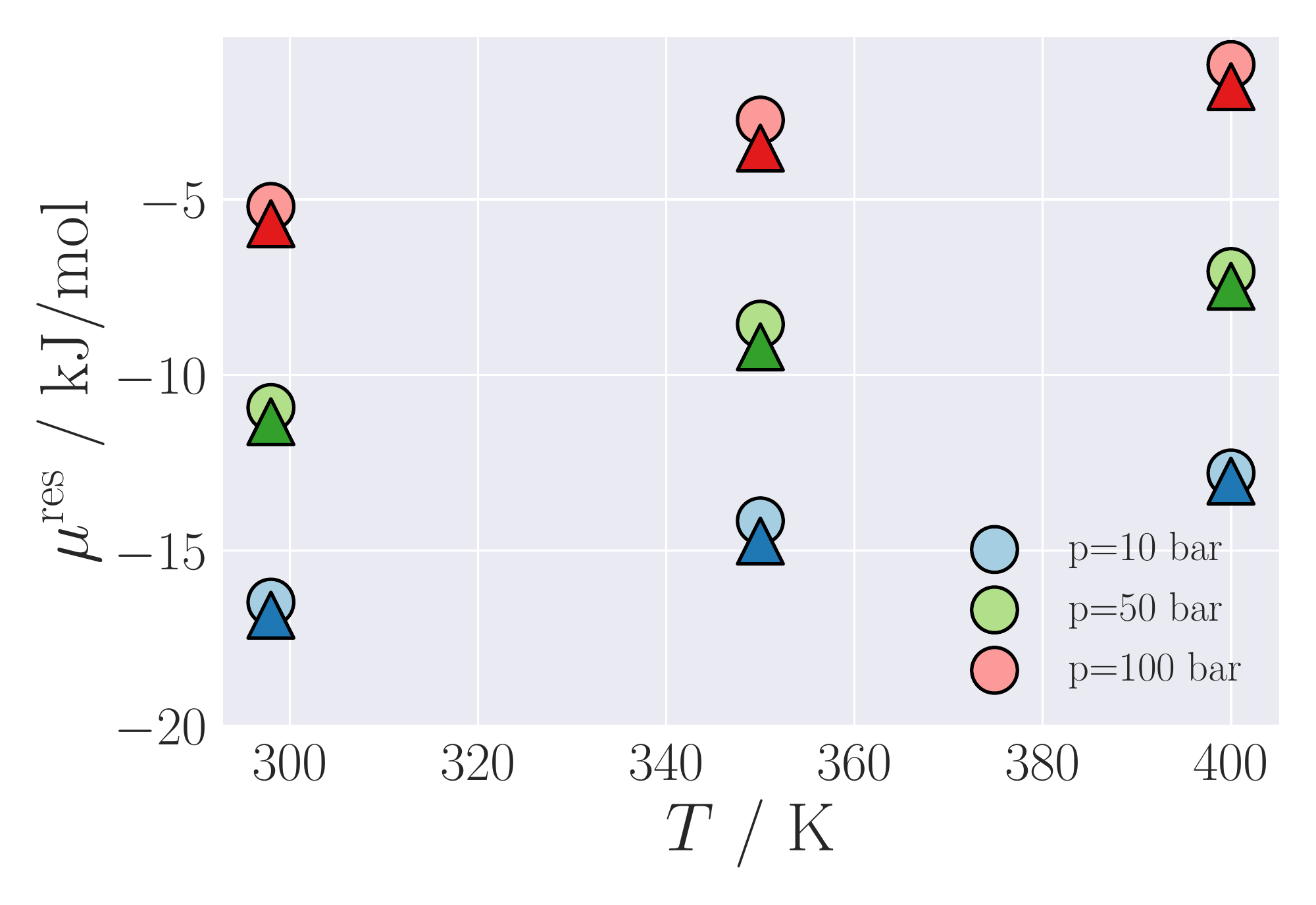}
    \caption{Comparison between self-solvation free energies of \textit{n}-hexane from PC-SAFT DFT ($\circ$) and residual chemical potentials $\mu^\mathrm{res}\left(T,\mathbf{\rho}^\mathrm{bulk}\right)$ calculated with PC-SAFT ($\triangle$) at varying temperatures and pressures. The values for $p=\SI{50}{\bar}$ are shifted by $+\SI{5}{\kilo\joule\per\mol}$ and the values for $p=\SI{100}{\bar}$ are shifted by $+\SI{10}{\kilo\joule\per\mol}$ for clarity.}
    \label{fig:SS_hexane}
\end{figure}
Further results for the self-solvation free energies of hexane at different temperatures and pressures ranging from $\SI{298}{\kelvin}$ to $\SI{400}{\kelvin}$ and $\SI{10}{\bar}$ to $\SI{100}{\bar}$ are shown in fig. \ref{fig:SS_hexane}. Here, we only consider a single (intramolecular) configuration of \textit{n}-hexane generated at \SI{298}{\kelvin} and \SI{1}{\bar} which is used for all temperatures and pressures. The triangles represent residual chemical potentials based on the PC-SAFT equation of state and circles are self-solvation free energies by DFT. We achieve excellent agreement between both methods over the whole pressure and temperature range, with a maximum deviation of only $\SI{1}{\kilo\joule\per\mol}$. The observation that our DFT formalism retains the correct temperature and pressure dependence of the SFE of hexane, when compared to the PC-SAFT equation of state, shows promise for calculating solvation enthalpies and entropies, which we will consider in future work.
Additionally, the intramolecular configuration of the \textit{n}-hexane solute molecule can be transferred from one temperature and pressure to another without significant errors in the calculated self-SFEs.
This allows the calculation of self-solvation free energies for different temperatures and pressures without the need for repeated molecular simulations.

\subsection{\label{subsec:selfsolvation_mixture}Self-Solvation of \textit{n}-alkanes in solvent mixtures}
We consider a binary solvent mixture of propane and hexane and introduce an additional propane/hexane solute molecule to the system. We continue to use the simple Berthelot-Lorentz combining rules, without employing $k_{ij}$ corrections for the solvent cross interactions. In Fig 5, we compare PC-SAFT DFT calculations for the self-solvation free energies of propane and hexane to residual chemical potentials calculated using PC-SAFT for six different mixtures at $\SI{298}{\kelvin}$ and $\SI{1}{\bar}$. The mixture exhibits a vapor-liquid phase transition where low hexane molar fractions correspond to the gas phase. The residual chemical potential in the gas phase is close to zero, as expected. We attain the correct size dependence of the residual chemical potential in the liquid phase, with hexane exhibiting a lower residual chemical potential. All calculated data points of hexane and propane are within $\SI{1}{\kilo\joule\per\mol}$. This shows the strength of our DFT formalism, as it is readily applicable to the prediction of residual chemical potentials in solvent mixtures without any adjustable parameters. 
\begin{figure}[h]
    \centering
    \includegraphics[width=.5\textwidth]{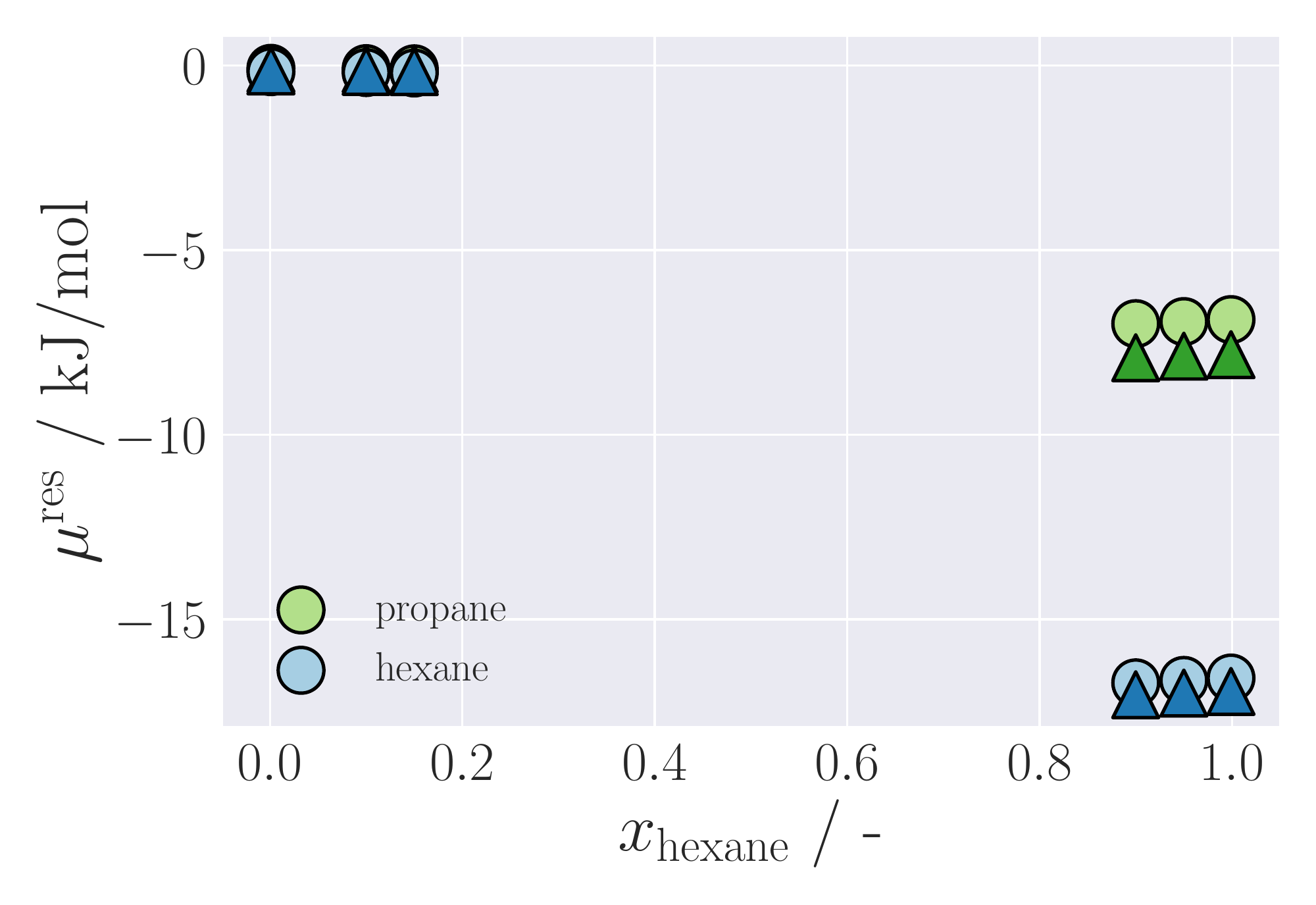}
    \caption{Comparison between self-solvation free energies of \textit{n}-propane and \textit{n}-hexane from PC-SAFT DFT ($\circ$) and residual chemical potentials $\mu^\mathrm{res}\left(T,\mathbf{\rho}^\mathrm{bulk}\right)$ calculated with PC-SAFT ($\triangle$) in a binary \textit{n}-propane/\textit{n}-hexane solvent mixture at $\SI{298}{\kelvin}$ and $\SI{1}{\bar}$.}
    \label{fig:SFEmixture}
\end{figure}
\subsection{\label{subsec:FSE}Solvation Free Energy}
\begin{figure}[h]
    \centering
    \includegraphics[width=.5\textwidth]{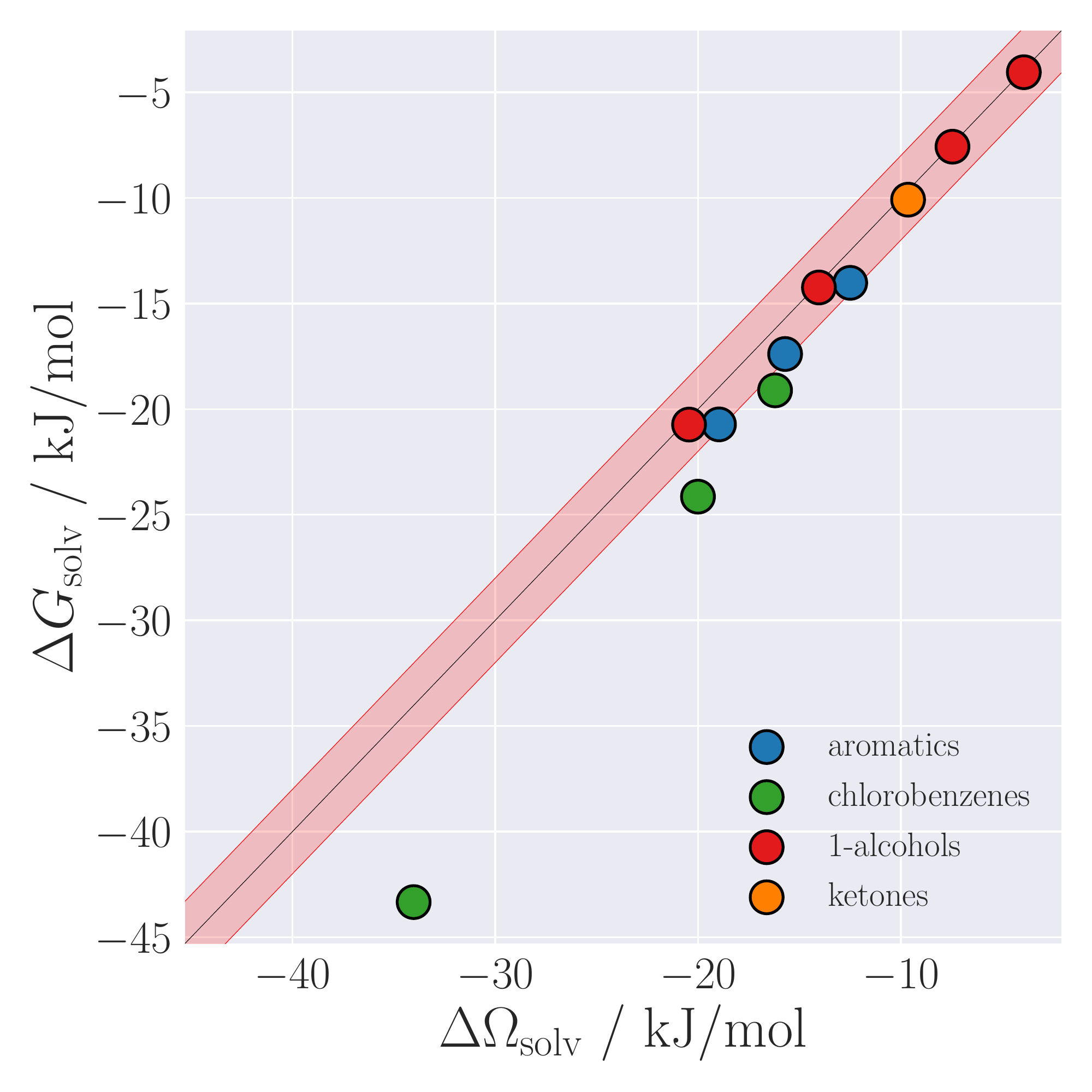}
    \caption{Correlation between solvation free energies from PC-SAFT DFT and MD simulations of the listed solutes from table \ref{tab:FSE_hexane} in \textit{n}-hexane at $\SI{298}{\kelvin}$ and $\SI{1}{\bar}$. The red lines indicate a difference of $\pm\SI{2}{\kilo\joule\per\mol}$.}
    \label{fig:SFEhexane_MD}
\end{figure}

In this section, we present results for solvation free energies of various molecules from different chemical groups in three solvents, namely \textit{n}-hexane, benzene and cyclohexane. The considered solutes and their SFEs in \textit{n}-hexane are given in table \ref{tab:FSE_hexane}. In figure \ref{fig:SFEhexane_MD} we show the correlation between the solvation free energy by DFT and molecular dynamics simulations of the listed solutes. Excellent agreement between DFT and molecular dynamics results is observed, with all data points (except chlorobenzenes) falling into the $\pm\SI{2}{\kilo\joule\per\mol}$ band. Coulomb interactions between chlorobenzenes and hexane add no significant contribution to the solvation free energy, see table \ref{tab:contributions_hexane}, and thus cannot account for the observed deviations. The GAFF force field uses a significantly higher Lennard-Jones energy parameter for chlorine atoms compared to carbon or oxygen instead. The observed deviations in the solvation free energy can be explained by the Lorentz-Berthelot combining rules for atomistic chlorine and PC-SAFT hexane segments which are not sufficient to account for the strong Lennard-Jones attraction. Halogenated bonds are highly directional non-covalent bonds between a halogen atom and another electronegative atom. They are formed due to a small region of positive electrostatic potential called $\sigma$-holes along the extensions of the covalent bonds \cite{politzer2013halogen}. Molecular force fields with atom-centered point charges cannot capture this effect as halogen atoms usually carry a negative charge and therefore interact repulsively with other electronegative atoms. This problem can be addressed by introducing a positive virtual charge to the halogen atom\cite{gutierrez2016parametrization} or including a $k_{\alpha,i}$ correction to the dispersive cross-energy between chlorine LJ sites and individual PC-SAFT segments of \textit{n}-hexane.
\begin{table}[h]
    \centering
    \begin{tabular}{lccc}
        \toprule
              Solute & $\Delta \Omega_\mathrm{solv}$ & $\Delta G_\mathrm{solv}$ & $\Delta G_\mathrm{solv}^\mathrm{exp}$ \\
        \midrule 
             \textbf{aromatics} &  &  &             \\
        \midrule 
             benzene &-12.497 &-14.024 $\pm$ 0.069 &-16.6 \\
             toluene &-15.707 &-17.392 $\pm$ 0.077 &-20.3 \\
            o-xylene &-19.381 &     ------         &-21.8 \\
            m-xylene &-18.968 &-20.724 $\pm$ 0.084 &-20.9 \\
        \midrule 
            \textbf{chlorobenzenes}&  &  &             \\
        \midrule 
            chlorobenzene &     -16.202 &    -19.113 $\pm$ 0.077 &       -12.5 \\
            1,4-dichlorobenzene &     -20.004 &    -24.145 $\pm$ 0.083 &       -23.8 \\
            hexachlorobenzene &     -34.024 &    -43.336 $\pm$ 0.110 &       -42.6 \\
        \midrule 
            \textbf{1-alcohols} &  &  &              \\
        \midrule 
            methanol &     -3.936 &     \phantom{1}-4.053 $\pm$ 0.043 &        -6.2 \\
            ethanol &      -7.454 &     \phantom{1}-7.579 $\pm$ 0.054 &       -11.0 \\
            butanol &     -14.040 &    -14.243 $\pm$ 0.072 &       -15.8 \\
            hexanol &     -20.443 &    -20.733 $\pm$ 0.086 &       -21.5 \\
        \midrule 
             \textbf{ketones} &  &  &               \\
        \midrule 
            acetone &      -9.643 &    -10.087 $\pm$ 0.061 &       -10.9 \\
    \bottomrule
    \end{tabular}
    \caption{Solvation free energies in $\SI{}{\kilo\joule\per\mol}$ in \textit{n}-hexane at $\SI{298}{\kelvin}$ and $\SI{1}{\bar}$ calculated with PC-SAFT DFT, MD simulations and from experiments \cite{garrido2012prediction}.}
    \label{tab:FSE_hexane}
\end{table}

\begin{table}[h]
    \centering
    \begin{tabular}{lcc}
    \toprule
    Solute &  $\Delta G_\mathrm{solv}^\mathrm{LJ}/\SI{}{\kilo\joule\per\mol}$ & $\Delta G_\mathrm{solv}^\mathrm{Coulomb}/\SI{}{\kilo\joule\per\mol}$ \\
    \midrule
    chlorobenzene & -19.055$\pm$0.077 & -0.058$\pm$0.001  \\
    \midrule
    1,4-dichlorobenzene & -24.103$\pm$0.083 & -0.042$\pm$0.001\\
    \midrule
    hexachlorobenzene & - 43.319$\pm$0.110 & -0.017$\pm$0.000\\
    \bottomrule
    \end{tabular}
    \caption{Lennard-Jones and Coulomb contributions to the solvation free energy from MD simulations in \textit{n}-hexane at $\SI{298}{\kelvin}$ and $\SI{1}{\bar}$.}
    \label{tab:contributions_hexane}
\end{table}

\begin{figure}[h]
    \centering
    \includegraphics[width=.5\textwidth]{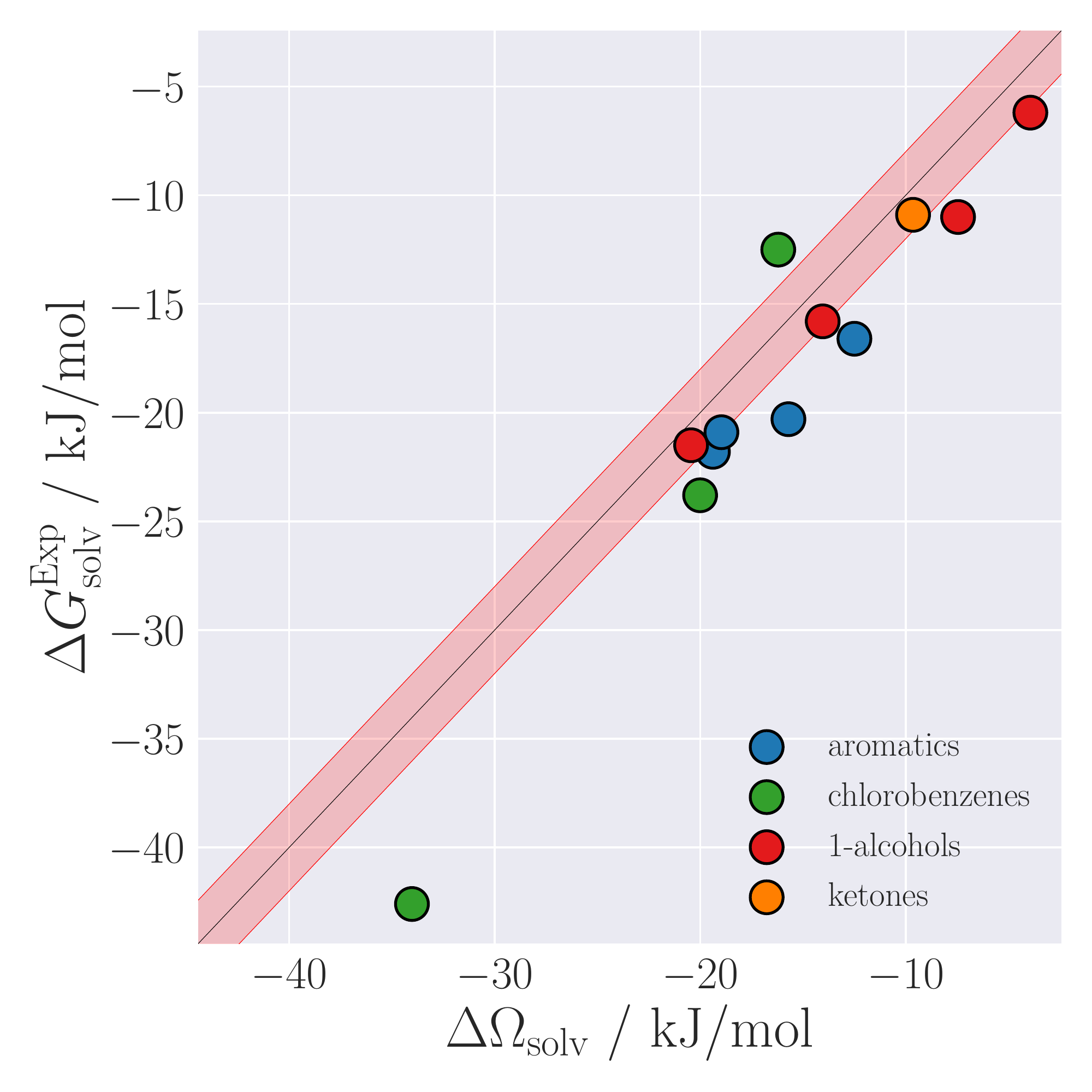}
    \caption{Correlation between solvation free energies from PC-SAFT DFT and experiments \cite{garrido2012prediction} of the listed solutes from table \ref{tab:FSE_hexane} in \textit{n}-hexane at $\SI{298}{\kelvin}$ and $\SI{1}{\bar}$. The red lines indicate a difference of ${\displaystyle \pm }\SI{2}{\kilo\joule\per\mol}$.}
    \label{fig:SFEhexane_exp}
\end{figure}

\begin{figure*}
    \centering
    \begin{subfigure}[b]{0.4\textwidth}
        \centering
        \includegraphics[clip, trim=450 0 450 0,width=\textwidth]{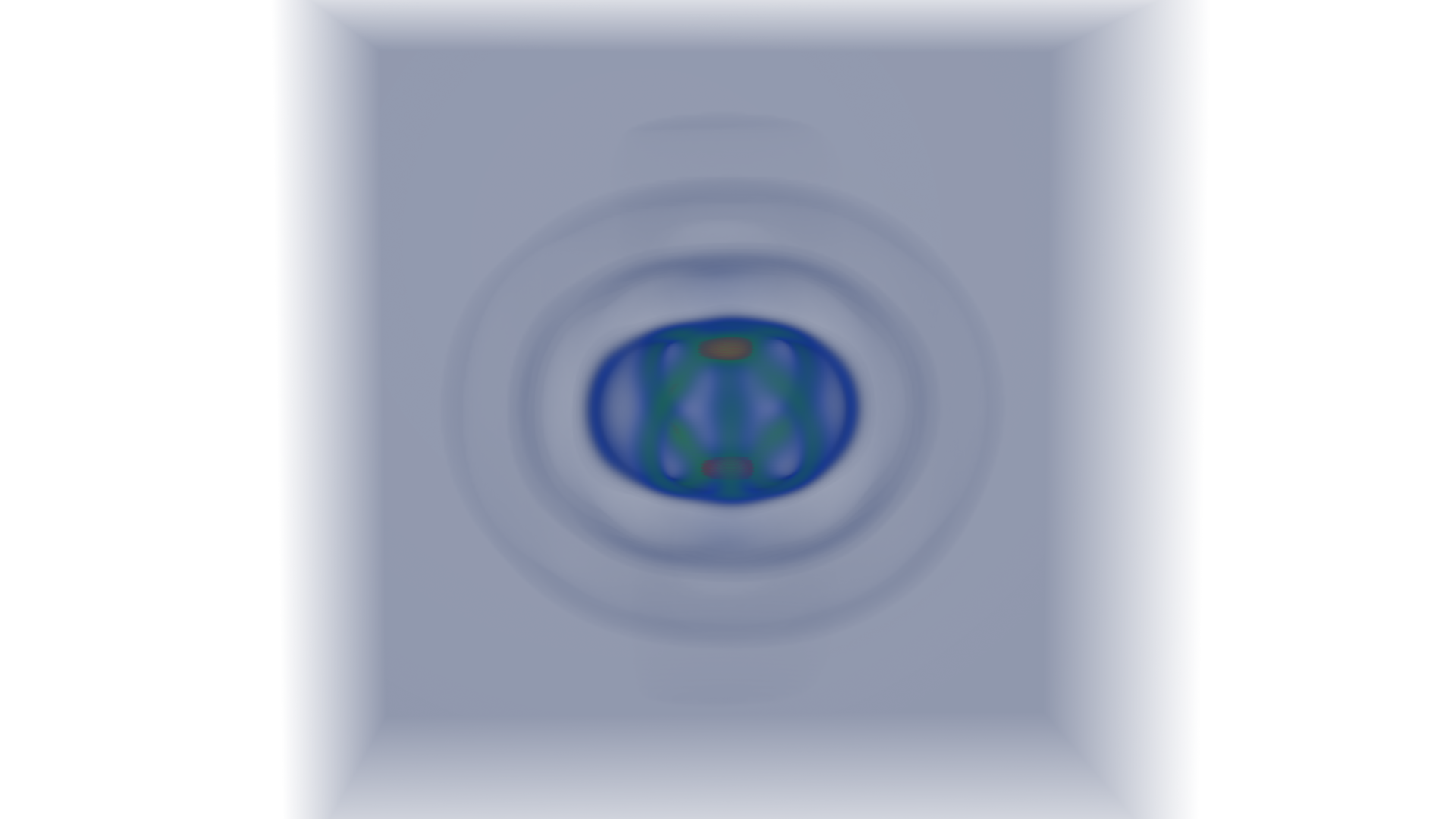}
        \caption{1-4 Dichlorobenzene}
        \label{fig:density_dcb}
    \end{subfigure}
    \hfill
    \begin{subfigure}[b]{0.4\textwidth}
        \centering
        \includegraphics[clip, trim=450 0 450 0,width=\textwidth]{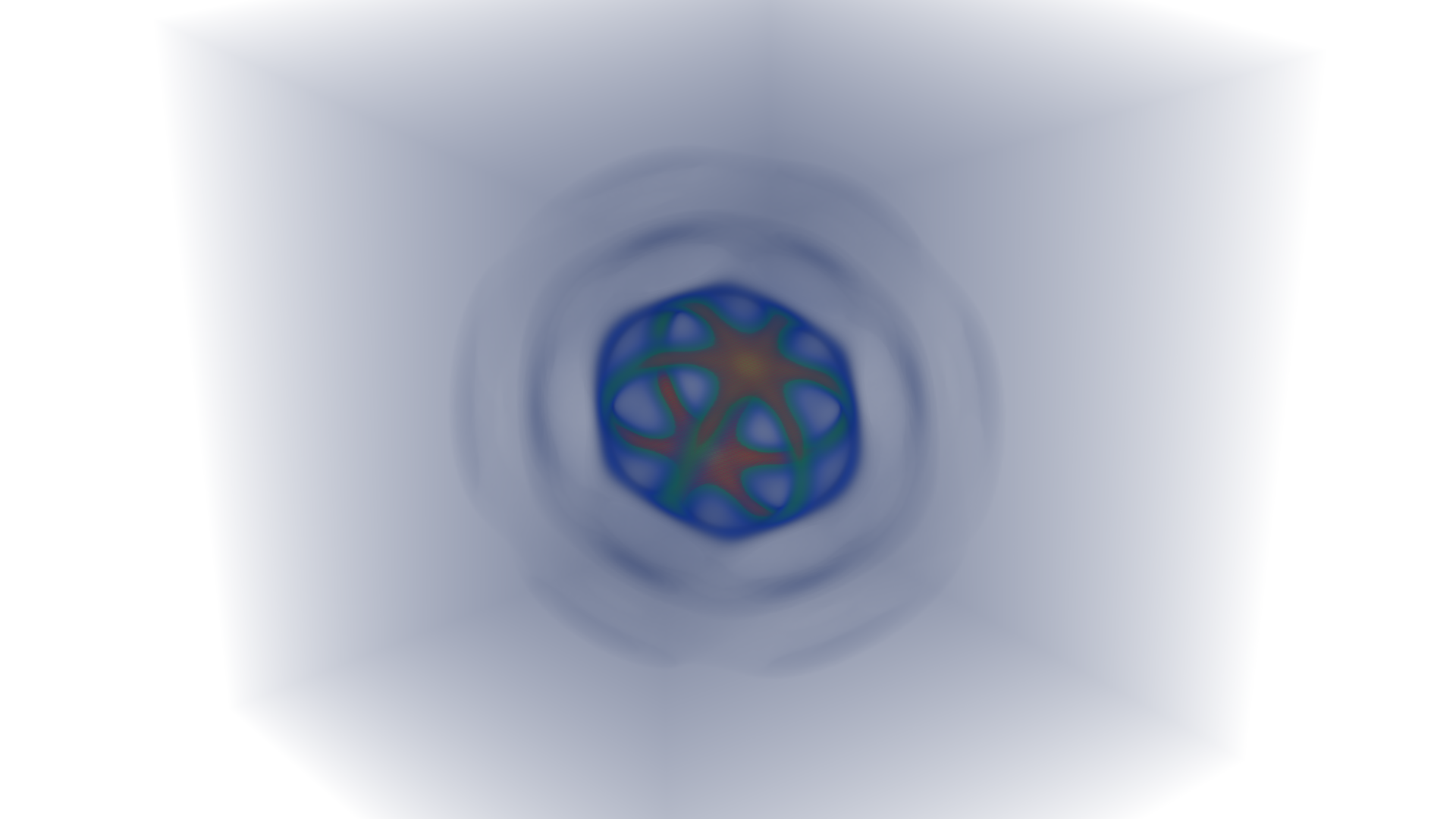}
        \caption{Hexachlorobenzene}
        \label{fig:density_hcb}
    \end{subfigure}
    \begin{subfigure}[b]{0.4\textwidth}
        \centering
        \includegraphics[clip, trim=450 0 450 0,width=\textwidth]{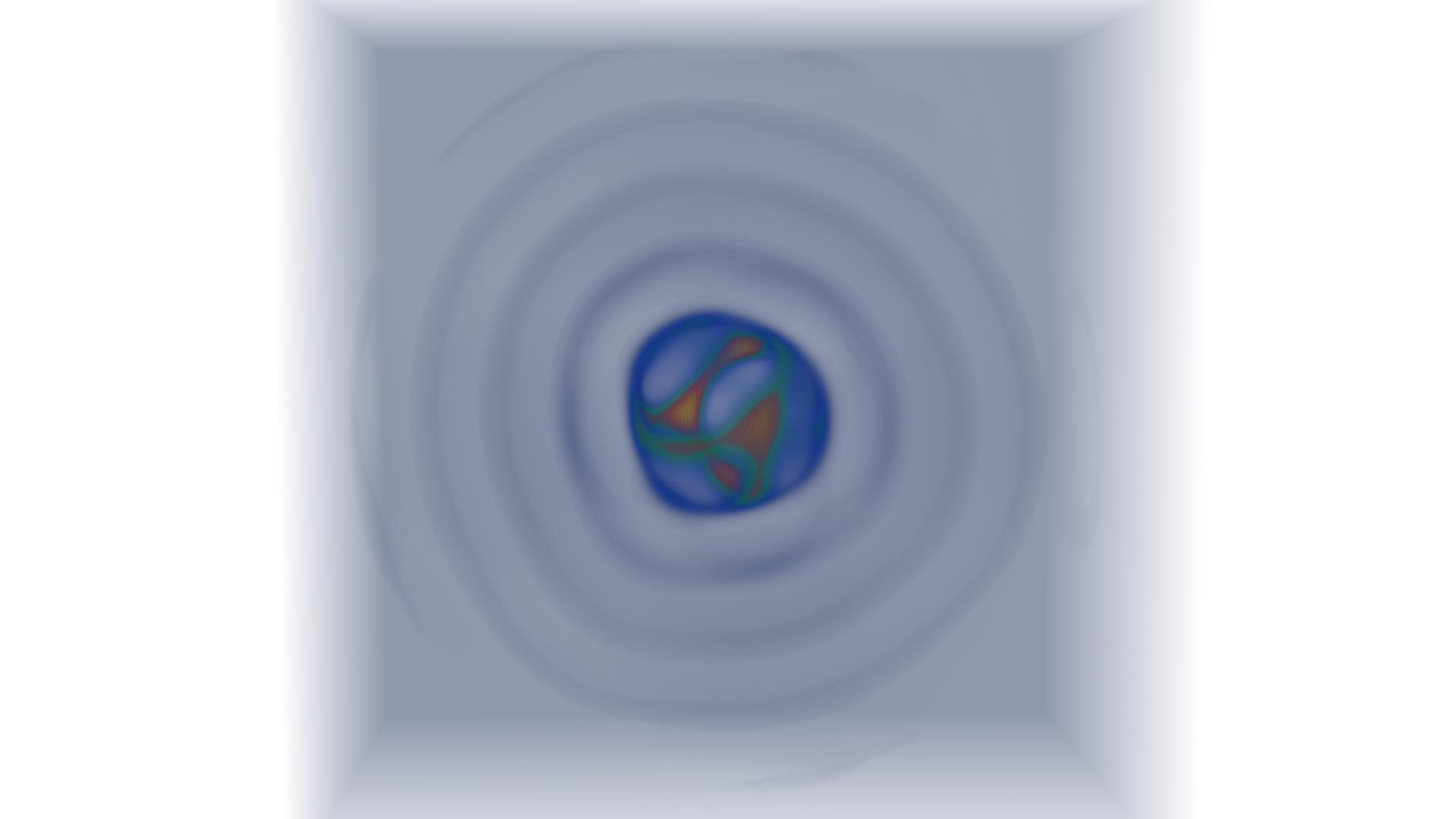}
        \caption{Acetone}
        \label{fig:density_acetone}
    \end{subfigure}
    \hfill
    \begin{subfigure}[b]{0.4\textwidth}
        \centering
        \includegraphics[clip, trim=450 0 450 0,width=\textwidth]{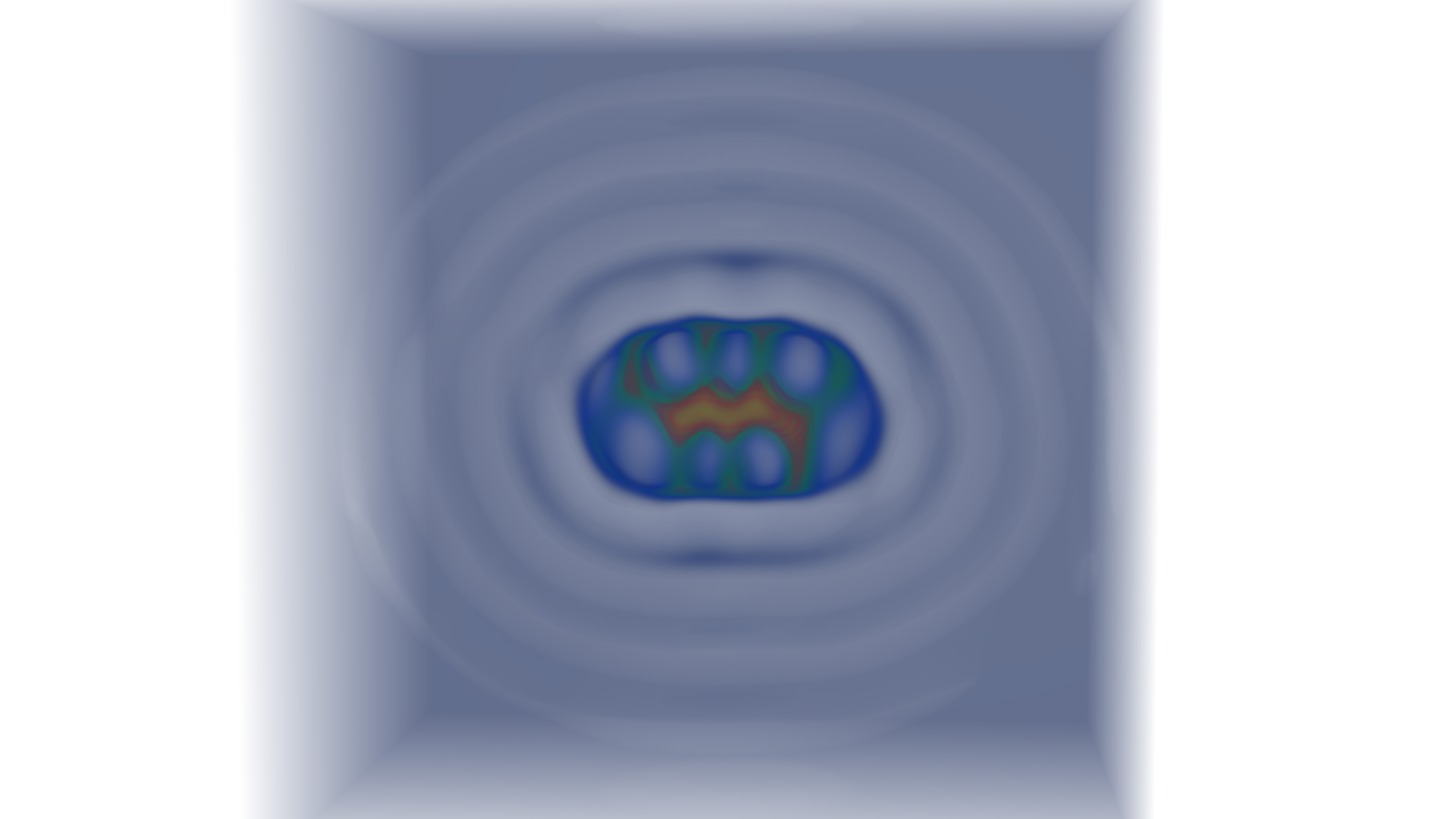}
        \caption{Hexanol}
        \label{fig:density_hexanol}
    \end{subfigure}
    \caption{PC-SAFT DFT results for the density profile of a \textit{n}-hexane solvent surrounding different solute molecules, at $\SI{298}{\kelvin}$ and $\SI{1}{\bar}$. Yellow, red and green indicate areas of higher density, where as blue corresponds to densities close to the bulk density.}
    \label{fig:density_profiles}
\end{figure*}

We compare solvation free energies by PC-SAFT DFT to experimental Gibbs energies of solvation \cite{garrido2012prediction} in figure \ref{fig:SFEhexane_exp}. The data points are scattered more widely with multiple points lying outside the $\pm\SI{2}{\kilo\joule\per\mol}$ range. The observed increased scattering is caused by deficiencies of the GAFF force field, which is not able to accurately reproduce the solvation free energies of the considered solutes.\\ 
In figure \ref{fig:density_profiles}, we show 3-dimensional representations of the solvent density profile of hexane calculated by Density Functional Theory at $\SI{298}{\kelvin}$ and $\SI{1}{\bar}$ for different solutes. The void in the center of the boxes corresponds to the excluded volume due to the presence of the solute. The density profile of hexane around dichlorobenzene \eqref{fig:density_dcb} shows distinct density maxima colored in yellow and red located at the central symmetry axis of the solute molecule. The blue connecting lines between the maxima trace the contours of the solute with the chlorine atoms sitting in the para positions on the right and left hand side. The density profile shows multiple solvation layers around the solute. A similar behaviour is observed for hexachlorobenzene in fig. \ref{fig:density_hcb} where all hydrogen atoms have been exchanged for chlorine. The solvation structure is more pronounced due to the increased size and strength of the Lennard-Jones interactions between chlorine and individual PC-SAFT \textit{n}-hexane segments. The density profile around the acetone solute molecule in fig. \ref{fig:density_acetone} is heavily influenced by its triangular shape. The favorable adsorption sites are located near the oxygen and carbon atom sites. The density profile subject to the external potential of the hexanol solute molecule, as shown in fig. \ref{fig:density_hexanol}, is predominantly affected by the elongated shape of the hexanol molecule. The highest density is observed along the carbon backbone of hexanol where the attractive Lennard-Jones potentials of the carbon interaction sites superimpose. 
\begin{figure}[ht]
    \centering
    \includegraphics[width=.5\textwidth]{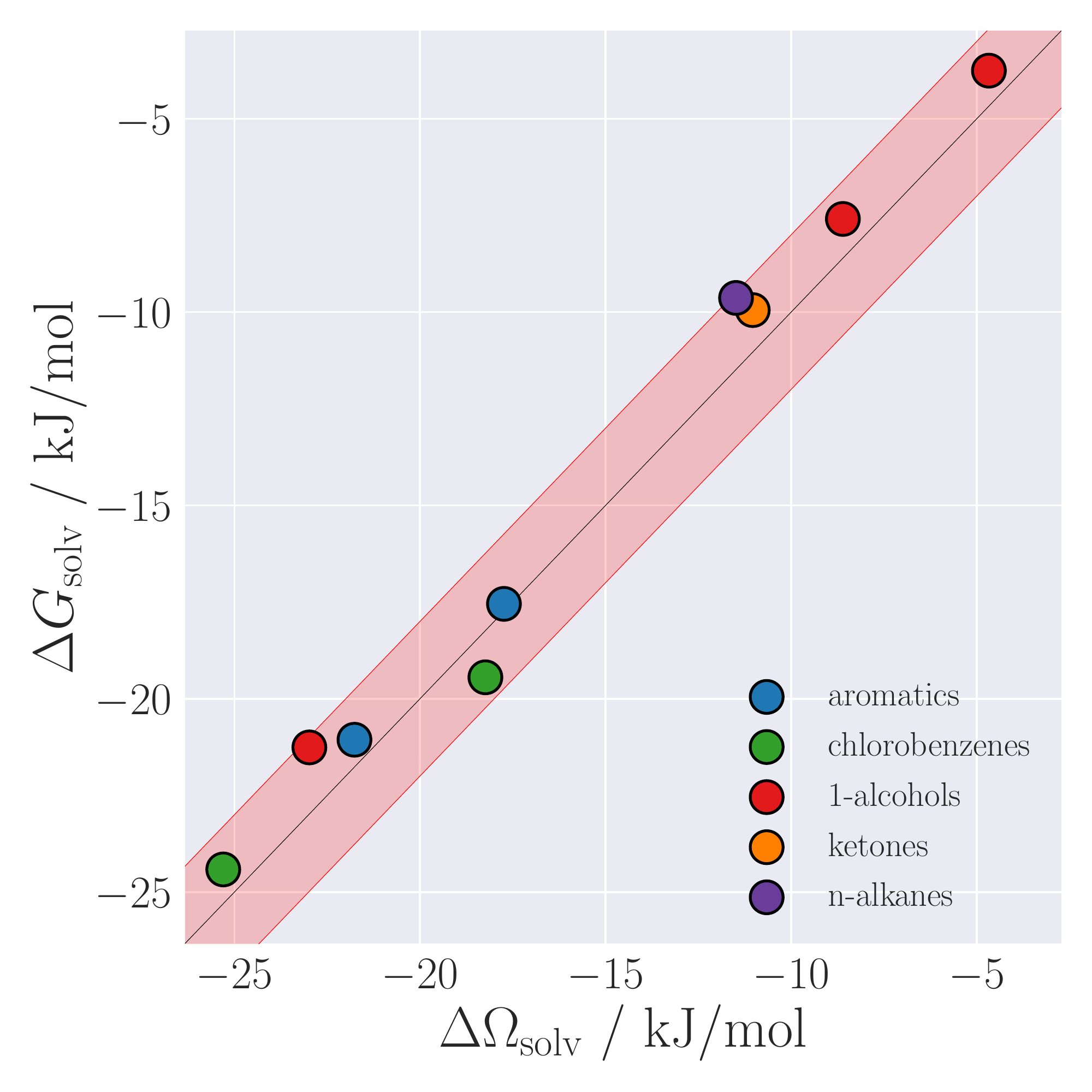}
    \caption{Correlation between solvation free energies from PC-SAFT DFT and MD simulations of the listed solutes from table \ref{tab:FSE_cyclohexane} in cyclohexane at $\SI{298}{\kelvin}$ and $\SI{1}{\bar}$. The red lines indicate a difference of ${\displaystyle \pm }\SI{2}{\kilo\joule\per\mol}$.}
    \label{fig:SFEcyclohexane}
\end{figure}
\begin{table}[h]
    \centering
    \begin{tabular}{lcc}
    \toprule
    Solute &          $\Delta \Omega_\mathrm{solv}/\SI{}{\kilo\joule\per\mol}$ & $\Delta G_\mathrm{solv}/\SI{}{\kilo\joule\per\mol}$ \\
    \midrule
    \textbf{aromatics} &            &            \\
    \midrule 
        toluene & -17.738 &     -17.546$\pm$0.108 \\
        o-xylene &   -21.765 &  -21.060$\pm$0.120 \\
    \midrule
    \textbf{chlorobenzenes} &            &            \\
    \midrule 
        chlorobenzene & -18.238 &     -19.445$\pm$0.107 \\
        1,4-dichlorobenzene & -25.300 &     -24.414$\pm$0.118 \\
    \midrule
    \textbf{1-alcohols} &            &            \\
    \midrule
        methanol &  -4.673 & \hphantom{2}-3.756$\pm$0.057 \\
        ethanol &  -8.607 &  \hphantom{2}-7.591$\pm$0.074 \\
        hexanol & -22.980 &     -21.258$\pm$0.122 \\
    \midrule
    \textbf{ketones} &            &            \\
    \midrule
        acetone &  -11.035 & \hphantom{2}-9.948$\pm$0.085 \\
    \midrule
    \textbf{\textit{n}-alkanes} &            &            \\
    \midrule
    butane & -11.485 & \hphantom{2}-9.634$\pm$0.099 \\
    \bottomrule
    \end{tabular}
    \caption{Solvation free energies from PC-SAFT DFT and MD simulations in cyclohexane at $\SI{298}{\kelvin}$ and $\SI{1}{\bar}$.}
    \label{tab:FSE_cyclohexane}
\end{table}

The next solvent investigated is cyclohexane. 
Calculated solvation free energies are listed in table \ref{tab:FSE_cyclohexane}. The correlations between solvation free energies by DFT and MD simulations are shown in figure \ref{fig:SFEcyclohexane}.
All data points are located within the $\pm\SI{2}{\kilo\joule\per\mole}$ band and we achieve excellent agreement between DFT and molecular dynamics results.
This further illustrates the strength of the density functional framework as we can correctly calculate the solvents solvation structure around complex solute molecules and additionally make physically meaningful predictions for solvation free energies for a wide range of solute-solvent combinations.

\begin{figure}[ht]
    \centering
    \includegraphics[width=.5\textwidth]{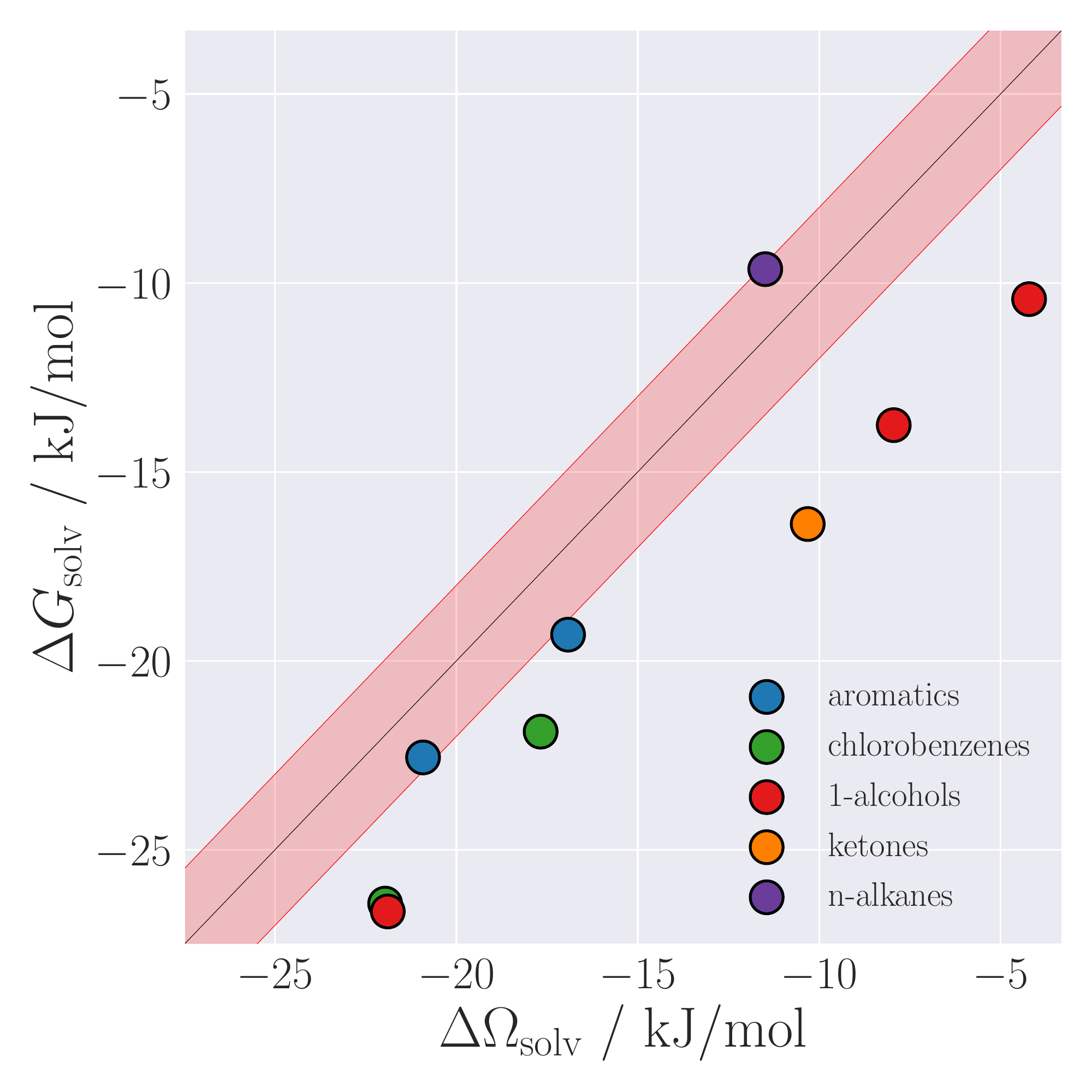}
    \caption{Correlation between solvation free energies from PC-SAFT DFT and MD simulations of the listed solutes from table \ref{tab:FSE_benzene} in benzene at $\SI{298}{\kelvin}$ and $\SI{1}{\bar}$. The red lines indicate a difference of ${\displaystyle \pm }\SI{2}{\kilo\joule\per\mol}$.}
    \label{fig:SFEbenzene}
\end{figure}

Next, we consider solvation free energies in benzene with results given in table \ref{tab:FSE_benzene}.
Fig. \ref{fig:SFEbenzene} summarizes solvation free energies by DFT compared to molecular dynamics simulations.
We observe small deviations of less than $\pm\SI{2}{\kilo\joule\per\kilo\gram}$ for butane as an \textit{n}-alkane and aromatics where the solute carries only small partial charges and the Coulomb contribution to the solvation free energy is comparatively small.
This is shown exemplary in table \ref{tab:contributions_benzene} for butane. Ketones, 1-alcohols and chlorobenzenes show higher deviations of $\sim\SI{5}{\kilo\joule\per\kilo\gram}$.
The molecular GAFF force field attributes partial charges to oxygen, chlorine and carbon atoms of benzene.
These partial charges yield significant Coulomb contributions to the solvation free energy in the same order of magnitude as the Lennard-Jones contributions, see table \ref{tab:contributions_benzene} for methanol.
The higher deviations for solute molecules with strong electrostatic interactions arise from limitations inherent to PC-SAFT, which (at this point of development) restricts segments to be charge neutral.
For further studies, complex solvents with multipole moments and hydrogen bond formation such as water and 1-alcohols are under consideration. Beside requiring additional contributions to the Helmholtz energy functional, these kind of solvents demand a proper treatment of the angular-dependent charge-multipole interactions.
We are currently investigating different approaches, such as an extension to heterosegmented chains using the iSAFT \cite{jain2007modified} theory, or the explicit resolution of the angular distribution function of dipolar segments \cite{frodl1992bulk,reindl2017electrolyte} similar to molecular density functional theory.
\begin{table}[h]
    \centering
    \begin{tabular}{lcc}
    \toprule
    Solute & $\Delta \Omega_\mathrm{solv}/\SI{}{\kilo\joule\per\mol}$ & $\Delta G_\mathrm{solv}/\SI{}{\kilo\joule\per\mol}$ \\
    \midrule
    \textbf{aromatics} &            &            \\
    \midrule 
        toluene &    -16.920&    -19.306$\pm$0.087  \\
        o-xylene &    -20.917&    -22.554$\pm$0.097  \\
    \midrule
    \textbf{chlorobenzenes} &            &            \\
    \midrule 
        chlorobenzene &    \-17.676 & -21.872$\pm$0.087 \\
        1,4-dichlorobenzene &    -21.962&-26.422$\pm$ 0.097   \\
    \midrule
    \textbf{1-alcohols} &            &            \\
    \midrule
        methanol &     -4.213 &    -10.428$\pm$0.056 \\
        ethanol &     -7.944 &    -13.762$\pm$0.065 \\
        hexanol &    -21.890 &    -26.630$\pm$0.101      \\
    \midrule
    \textbf{ketones} &            &            \\
    \midrule
        acteone &    -10.315  &    -16.379$\pm$0.075 \\
    \midrule
    \textbf{\textit{n}-alkanes} &            &            \\
    \midrule
        butane &    -10.366 & \hphantom{2}-9.121$\pm$0.077 \\
    \bottomrule
    \end{tabular}
    \caption{Solvation free energies from PC-SAFT DFT and MD simulations 
    in benzene at $\SI{298}{\kelvin}$ and $\SI{1}{\bar}$.}
    \label{tab:FSE_benzene}
\end{table}
\begin{table}[h]
    \centering
    \begin{tabular}{lcc}
    \toprule
    Solute &  $\Delta G_\mathrm{solv}^\mathrm{LJ}/\SI{}{\kilo\joule\per\mol}$ & $\Delta G_\mathrm{solv}^\mathrm{Coulomb}/\SI{}{\kilo\joule\per\mol}$\\
    \midrule
    butane & -9.062$\pm$0.058 & -0.076$\pm$0.00\hphantom{9} \\
    \midrule
    methanol & -3.304$\pm$0.037 & -7.137$\pm$0.029 \\
    \bottomrule
    \end{tabular}
    \caption{Lennard-Jones and Coulomb contributions to the Gibbs energy of solvation calculated with MD simulations in benzene at $\SI{298}{\kelvin}$ and $\SI{1}{\bar}$.}
    \label{tab:contributions_benzene}
\end{table}

\section{\label{sec:conclusion}Conclusion}
We present a Density Functional Theory framework based on the PC-SAFT equation of state for the prediction of solvation free energies in non-polar solvents.
In our approach, the solute is described in full atomistic detail, based on a molecular force field, while the solvent is described in a more coarse-grained manner, based on the molecular model of PC-SAFT.
Compared to other methods, this hybrid approach leads to a favourable balance between computational efficiency, complexity, and accuracy.
The approach is fully predictive, not requiring any adjustable parameters or empirical corrections.
The only input for the SFE calculations are the PC-SAFT parameters of the solvent, molecular force field parameters and coordinates of the individual Lennard-Jones interaction sites of the solute.
The coordinates of the solute interaction sites can be taken from a snapshot generated by molecular dynamics simulation or directly from the topology of the molecular force field.
Following the idea of Percus' test-particle theory, we first applied the DFT framework to the calculation of the free energy of self-solvation. 
By definition, the self-solvation free energy equals the residual chemical potential within the canonical ensemble. A comparison between the self-solvation free energy as calculated by the DFT framework and the residual chemical potential calculated based on the bulk model (PC-SAFT equation of state) thus provides an unambiguous consistency test.
The consistency test was applied to \textit{n}-alkanes at various temperatures and densities. Excellent agreement is obtained.
The extension to mixtures is straightforward and we calculate residual chemical potentials of mixtures within an accuracy of $\SI{1}{\kilo\joule\per\mole}$. Furthermore, we predict solvation free energies of molecules from different chemical groups in the solvents cyclohexane, hexane and benzene.
For systems without strong solute-solvent Coulomb interactions, the calculated solvation free energies accurately compare to the solvation free energy obtained by MD simulations and experiments.
Besides validating the accuracy of our DFT framework, these favourable results confirm the applicability of the Berthelot-Lorentz combining rules for calculating cross interactions between coarse-grained PC-SAFT parameters and atomistic force-field parameters, which we consider an important insight.
Extensions to systems with strong multi-polar, associating, or electrostatic interactions are the subject of future work.

\begin{acknowledgments}
This work was funded by Deutsche Forschungsgemeinschaft (DFG, German Research Foundation) – Project Number 327154368 – SFB 1313. Further, this work was funded by Deutsche Forschungsgemeinschaft (DFG, German Research Foundation) under Germanys Excellence Strategy - EXC 2075-390740016. We appreciate the support in visualising the density fields (figs. \ref{fig:density_hexane} and \ref{fig:density_profiles}) by Stefan Scheller and Tom Ertl.
\end{acknowledgments}





\section{\label{app:implementation}Implementation Details}
DFT calculations are performed on a 3-dimensional Cartesian frame with $N_x\cdot N_y\cdot N_z=256^3$ grid points using a cubic system with side length 40~\r{A}. The convolution integrals appearing in the Helmholtz energy functionals are solved by the Convolution Theorem of the Fourier Transform
\begin{equation}
    f\otimes g =\int f(\rb^\prime)g(\rb-\rb^\prime) \mathrm{d}\rb^\prime=\mathcal{F}^{-1}\left[\mathcal{F}\left[f\right]\cdot \mathcal{F}\left[g\right]\right]
\end{equation}
The 3-dimensional Fourier Transforms are performed by the Intel MKL library. The initial guess $\rho^0_i(\rb)$ for the density profile is the ideal gas solution to the Euler-Lagrange equation, however, we limit the maximum density to eliminate nonphysically high density in the vicinity of the solute.
\begin{equation}
    \rho^0_i(\rb)=\rho^\mathrm{bulk}_i\cdot \min\left(\exp(-\beta V_i^\mathrm{ext}(\rb),1\right)
\end{equation}
where $\rho_i^\mathrm{bulk}$ is the number density of component $i$ of the surrounding bulk phase. The equilibrium density profile is calculated by solving the Euler-Lagrange equations of eq. \ref{eq:ELE} using a damped Picard iteration scheme. The calculation consumes $\SI{14}{\giga\byte}$ of memory and the run time for one calculation is about 6-8 minutes on a AMD Ryzen 3900 CPU. The Intel MKL library inherently uses OpenMP for the 3-dimensional Fourier Transform and no further parallelisation steps are taken. The run time can be reduced to 1 minute and the memory consumption to $\SI{4}{\giga\byte}$ for systems containing $N_x\cdot N_y\cdot N_z=128^3$ grid cells without significant changes in the resulting solvation free energies.

\bibliography{aipsamp}

\providecommand{\noopsort}[1]{}\providecommand{\singleletter}[1]{#1}%
\begin{thebibliography}{71}%
\makeatletter
\providecommand \@ifxundefined [1]{%
 \@ifx{#1\undefined}
}%
\providecommand \@ifnum [1]{%
 \ifnum #1\expandafter \@firstoftwo
 \else \expandafter \@secondoftwo
 \fi
}%
\providecommand \@ifx [1]{%
 \ifx #1\expandafter \@firstoftwo
 \else \expandafter \@secondoftwo
 \fi
}%
\providecommand \natexlab [1]{#1}%
\providecommand \enquote  [1]{``#1''}%
\providecommand \bibnamefont  [1]{#1}%
\providecommand \bibfnamefont [1]{#1}%
\providecommand \citenamefont [1]{#1}%
\providecommand \href@noop [0]{\@secondoftwo}%
\providecommand \href [0]{\begingroup \@sanitize@url \@href}%
\providecommand \@href[1]{\@@startlink{#1}\@@href}%
\providecommand \@@href[1]{\endgroup#1\@@endlink}%
\providecommand \@sanitize@url [0]{\catcode `\\12\catcode `\$12\catcode
  `\&12\catcode `\#12\catcode `\^12\catcode `\_12\catcode `\%12\relax}%
\providecommand \@@startlink[1]{}%
\providecommand \@@endlink[0]{}%
\providecommand \url  [0]{\begingroup\@sanitize@url \@url }%
\providecommand \@url [1]{\endgroup\@href {#1}{\urlprefix }}%
\providecommand \urlprefix  [0]{URL }%
\providecommand \Eprint [0]{\href }%
\providecommand \doibase [0]{http://dx.doi.org/}%
\providecommand \selectlanguage [0]{\@gobble}%
\providecommand \bibinfo  [0]{\@secondoftwo}%
\providecommand \bibfield  [0]{\@secondoftwo}%
\providecommand \translation [1]{[#1]}%
\providecommand \BibitemOpen [0]{}%
\providecommand \bibitemStop [0]{}%
\providecommand \bibitemNoStop [0]{.\EOS\space}%
\providecommand \EOS [0]{\spacefactor3000\relax}%
\providecommand \BibitemShut  [1]{\csname bibitem#1\endcsname}%
\let\auto@bib@innerbib\@empty
\bibitem [{\citenamefont {Hirata}(2003)}]{hirata2003molecular}%
  \BibitemOpen
  \bibfield  {author} {\bibinfo {author} {\bibfnamefont {F.}~\bibnamefont
  {Hirata}},\ }\href@noop {} {\emph {\bibinfo {title} {Molecular theory of
  solvation}}},\ Vol.~\bibinfo {volume} {24}\ (\bibinfo  {publisher} {Springer
  Science \& Business Media},\ \bibinfo {year} {2003})\BibitemShut {NoStop}%
\bibitem [{\citenamefont {Feig}\ \emph {et~al.}(2004)\citenamefont {Feig},
  \citenamefont {Onufriev}, \citenamefont {Lee}, \citenamefont {Im},
  \citenamefont {Case},\ and\ \citenamefont
  {Brooks~III}}]{feig2004performance}%
  \BibitemOpen
  \bibfield  {author} {\bibinfo {author} {\bibfnamefont {M.}~\bibnamefont
  {Feig}}, \bibinfo {author} {\bibfnamefont {A.}~\bibnamefont {Onufriev}},
  \bibinfo {author} {\bibfnamefont {M.~S.}\ \bibnamefont {Lee}}, \bibinfo
  {author} {\bibfnamefont {W.}~\bibnamefont {Im}}, \bibinfo {author}
  {\bibfnamefont {D.~A.}\ \bibnamefont {Case}}, \ and\ \bibinfo {author}
  {\bibfnamefont {C.~L.}\ \bibnamefont {Brooks~III}},\ }\bibfield  {title}
  {\enquote {\bibinfo {title} {Performance comparison of generalized born and
  poisson methods in the calculation of electrostatic solvation energies for
  protein structures},}\ }\href@noop {} {\bibfield  {journal} {\bibinfo
  {journal} {Journal of computational chemistry}\ }\textbf {\bibinfo {volume}
  {25}},\ \bibinfo {pages} {265--284} (\bibinfo {year} {2004})}\BibitemShut
  {NoStop}%
\bibitem [{\citenamefont {Garrido}\ \emph {et~al.}(2012)\citenamefont
  {Garrido}, \citenamefont {Economou}, \citenamefont {Queimada}, \citenamefont
  {Jorge},\ and\ \citenamefont {Macedo}}]{garrido2012prediction}%
  \BibitemOpen
  \bibfield  {author} {\bibinfo {author} {\bibfnamefont {N.~M.}\ \bibnamefont
  {Garrido}}, \bibinfo {author} {\bibfnamefont {I.~G.}\ \bibnamefont
  {Economou}}, \bibinfo {author} {\bibfnamefont {A.~J.}\ \bibnamefont
  {Queimada}}, \bibinfo {author} {\bibfnamefont {M.}~\bibnamefont {Jorge}}, \
  and\ \bibinfo {author} {\bibfnamefont {E.~A.}\ \bibnamefont {Macedo}},\
  }\bibfield  {title} {\enquote {\bibinfo {title} {Prediction of the
  n-hexane/water and 1-octanol/water partition coefficients for environmentally
  relevant compounds using molecular simulation},}\ }\href@noop {} {\bibfield
  {journal} {\bibinfo  {journal} {AIChE journal}\ }\textbf {\bibinfo {volume}
  {58}},\ \bibinfo {pages} {1929--1938} (\bibinfo {year} {2012})}\BibitemShut
  {NoStop}%
\bibitem [{\citenamefont {Jorgensen}\ and\ \citenamefont
  {Ravimohan}(1985)}]{jorgensen1985monte}%
  \BibitemOpen
  \bibfield  {author} {\bibinfo {author} {\bibfnamefont {W.~L.}\ \bibnamefont
  {Jorgensen}}\ and\ \bibinfo {author} {\bibfnamefont {C.}~\bibnamefont
  {Ravimohan}},\ }\bibfield  {title} {\enquote {\bibinfo {title} {Monte carlo
  simulation of differences in free energies of hydration},}\ }\href@noop {}
  {\bibfield  {journal} {\bibinfo  {journal} {The Journal of chemical physics}\
  }\textbf {\bibinfo {volume} {83}},\ \bibinfo {pages} {3050--3054} (\bibinfo
  {year} {1985})}\BibitemShut {NoStop}%
\bibitem [{\citenamefont {Riquelme}\ \emph {et~al.}(2018)\citenamefont
  {Riquelme}, \citenamefont {Lara}, \citenamefont {Mobley}, \citenamefont
  {Verstraelen}, \citenamefont {Matamala},\ and\ \citenamefont
  {Vöhringer-Martinez}}]{riquelme2018hydration}%
  \BibitemOpen
  \bibfield  {author} {\bibinfo {author} {\bibfnamefont {M.}~\bibnamefont
  {Riquelme}}, \bibinfo {author} {\bibfnamefont {A.}~\bibnamefont {Lara}},
  \bibinfo {author} {\bibfnamefont {D.~L.}\ \bibnamefont {Mobley}}, \bibinfo
  {author} {\bibfnamefont {T.}~\bibnamefont {Verstraelen}}, \bibinfo {author}
  {\bibfnamefont {A.~R.}\ \bibnamefont {Matamala}}, \ and\ \bibinfo {author}
  {\bibfnamefont {E.}~\bibnamefont {Vöhringer-Martinez}},\ }\bibfield  {title}
  {\enquote {\bibinfo {title} {Hydration free energies in the freesolv database
  calculated with polarized iterative hirshfeld charges},}\ }\href@noop {}
  {\bibfield  {journal} {\bibinfo  {journal} {Journal of chemical information
  and modeling}\ }\textbf {\bibinfo {volume} {58}},\ \bibinfo {pages}
  {1779--1797} (\bibinfo {year} {2018})}\BibitemShut {NoStop}%
\bibitem [{\citenamefont {Roux}\ and\ \citenamefont
  {Simonson}(1999)}]{roux1999implicit}%
  \BibitemOpen
  \bibfield  {author} {\bibinfo {author} {\bibfnamefont {B.}~\bibnamefont
  {Roux}}\ and\ \bibinfo {author} {\bibfnamefont {T.}~\bibnamefont
  {Simonson}},\ }\bibfield  {title} {\enquote {\bibinfo {title} {Implicit
  solvent models},}\ }\href@noop {} {\bibfield  {journal} {\bibinfo  {journal}
  {Biophysical chemistry}\ }\textbf {\bibinfo {volume} {78}},\ \bibinfo {pages}
  {1--20} (\bibinfo {year} {1999})}\BibitemShut {NoStop}%
\bibitem [{\citenamefont {Bashford}\ and\ \citenamefont
  {Case}(2000)}]{bashford2000generalized}%
  \BibitemOpen
  \bibfield  {author} {\bibinfo {author} {\bibfnamefont {D.}~\bibnamefont
  {Bashford}}\ and\ \bibinfo {author} {\bibfnamefont {D.~A.}\ \bibnamefont
  {Case}},\ }\bibfield  {title} {\enquote {\bibinfo {title} {Generalized born
  models of macromolecular solvation effects},}\ }\href@noop {} {\bibfield
  {journal} {\bibinfo  {journal} {Annual review of physical chemistry}\
  }\textbf {\bibinfo {volume} {51}},\ \bibinfo {pages} {129--152} (\bibinfo
  {year} {2000})}\BibitemShut {NoStop}%
\bibitem [{\citenamefont {Mongan}\ \emph {et~al.}(2007)\citenamefont {Mongan},
  \citenamefont {Simmerling}, \citenamefont {McCammon}, \citenamefont {Case},\
  and\ \citenamefont {Onufriev}}]{mongan2007generalized}%
  \BibitemOpen
  \bibfield  {author} {\bibinfo {author} {\bibfnamefont {J.}~\bibnamefont
  {Mongan}}, \bibinfo {author} {\bibfnamefont {C.}~\bibnamefont {Simmerling}},
  \bibinfo {author} {\bibfnamefont {J.~A.}\ \bibnamefont {McCammon}}, \bibinfo
  {author} {\bibfnamefont {D.~A.}\ \bibnamefont {Case}}, \ and\ \bibinfo
  {author} {\bibfnamefont {A.}~\bibnamefont {Onufriev}},\ }\bibfield  {title}
  {\enquote {\bibinfo {title} {Generalized born model with a simple, robust
  molecular volume correction},}\ }\href@noop {} {\bibfield  {journal}
  {\bibinfo  {journal} {Journal of chemical theory and computation}\ }\textbf
  {\bibinfo {volume} {3}},\ \bibinfo {pages} {156--169} (\bibinfo {year}
  {2007})}\BibitemShut {NoStop}%
\bibitem [{\citenamefont {Marenich}\ \emph {et~al.}(2007)\citenamefont
  {Marenich}, \citenamefont {Olson}, \citenamefont {Kelly}, \citenamefont
  {Cramer},\ and\ \citenamefont {Truhlar}}]{marenich2007self}%
  \BibitemOpen
  \bibfield  {author} {\bibinfo {author} {\bibfnamefont {A.~V.}\ \bibnamefont
  {Marenich}}, \bibinfo {author} {\bibfnamefont {R.~M.}\ \bibnamefont {Olson}},
  \bibinfo {author} {\bibfnamefont {C.~P.}\ \bibnamefont {Kelly}}, \bibinfo
  {author} {\bibfnamefont {C.~J.}\ \bibnamefont {Cramer}}, \ and\ \bibinfo
  {author} {\bibfnamefont {D.~G.}\ \bibnamefont {Truhlar}},\ }\bibfield
  {title} {\enquote {\bibinfo {title} {Self-consistent reaction field model for
  aqueous and nonaqueous solutions based on accurate polarized partial
  charges},}\ }\href@noop {} {\bibfield  {journal} {\bibinfo  {journal}
  {Journal of Chemical Theory and Computation}\ }\textbf {\bibinfo {volume}
  {3}},\ \bibinfo {pages} {2011--2033} (\bibinfo {year} {2007})}\BibitemShut
  {NoStop}%
\bibitem [{\citenamefont {Gendre}, \citenamefont {Ramirez},\ and\ \citenamefont
  {Borgis}(2009)}]{gendre2009classical}%
  \BibitemOpen
  \bibfield  {author} {\bibinfo {author} {\bibfnamefont {L.}~\bibnamefont
  {Gendre}}, \bibinfo {author} {\bibfnamefont {R.}~\bibnamefont {Ramirez}}, \
  and\ \bibinfo {author} {\bibfnamefont {D.}~\bibnamefont {Borgis}},\
  }\bibfield  {title} {\enquote {\bibinfo {title} {Classical density functional
  theory of solvation in molecular solvents: Angular grid implementation},}\
  }\href@noop {} {\bibfield  {journal} {\bibinfo  {journal} {Chemical Physics
  Letters}\ }\textbf {\bibinfo {volume} {474}},\ \bibinfo {pages} {366--370}
  (\bibinfo {year} {2009})}\BibitemShut {NoStop}%
\bibitem [{\citenamefont {Zhao}\ \emph {et~al.}(2011)\citenamefont {Zhao},
  \citenamefont {Ramirez}, \citenamefont {Vuilleumier},\ and\ \citenamefont
  {Borgis}}]{zhao2011molecular}%
  \BibitemOpen
  \bibfield  {author} {\bibinfo {author} {\bibfnamefont {S.}~\bibnamefont
  {Zhao}}, \bibinfo {author} {\bibfnamefont {R.}~\bibnamefont {Ramirez}},
  \bibinfo {author} {\bibfnamefont {R.}~\bibnamefont {Vuilleumier}}, \ and\
  \bibinfo {author} {\bibfnamefont {D.}~\bibnamefont {Borgis}},\ }\bibfield
  {title} {\enquote {\bibinfo {title} {Molecular density functional theory of
  solvation: From polar solvents to water},}\ }\href@noop {} {\bibfield
  {journal} {\bibinfo  {journal} {The Journal of chemical physics}\ }\textbf
  {\bibinfo {volume} {134}},\ \bibinfo {pages} {194102} (\bibinfo {year}
  {2011})}\BibitemShut {NoStop}%
\bibitem [{\citenamefont {Borgis}, \citenamefont {Gendre},\ and\ \citenamefont
  {Ramirez}(2012)}]{borgis2012molecular}%
  \BibitemOpen
  \bibfield  {author} {\bibinfo {author} {\bibfnamefont {D.}~\bibnamefont
  {Borgis}}, \bibinfo {author} {\bibfnamefont {L.}~\bibnamefont {Gendre}}, \
  and\ \bibinfo {author} {\bibfnamefont {R.}~\bibnamefont {Ramirez}},\
  }\bibfield  {title} {\enquote {\bibinfo {title} {Molecular density functional
  theory: Application to solvation and electron-transfer thermodynamics in
  polar solvents},}\ }\href@noop {} {\bibfield  {journal} {\bibinfo  {journal}
  {The Journal of Physical Chemistry B}\ }\textbf {\bibinfo {volume} {116}},\
  \bibinfo {pages} {2504--2512} (\bibinfo {year} {2012})}\BibitemShut {NoStop}%
\bibitem [{\citenamefont {Jeanmairet}\ \emph {et~al.}(2013)\citenamefont
  {Jeanmairet}, \citenamefont {Levesque}, \citenamefont {Vuilleumier},\ and\
  \citenamefont {Borgis}}]{jeanmairet2013molecular}%
  \BibitemOpen
  \bibfield  {author} {\bibinfo {author} {\bibfnamefont {G.}~\bibnamefont
  {Jeanmairet}}, \bibinfo {author} {\bibfnamefont {M.}~\bibnamefont
  {Levesque}}, \bibinfo {author} {\bibfnamefont {R.}~\bibnamefont
  {Vuilleumier}}, \ and\ \bibinfo {author} {\bibfnamefont {D.}~\bibnamefont
  {Borgis}},\ }\bibfield  {title} {\enquote {\bibinfo {title} {Molecular
  density functional theory of water},}\ }\href@noop {} {\bibfield  {journal}
  {\bibinfo  {journal} {The Journal of Physical Chemistry Letters}\ }\textbf
  {\bibinfo {volume} {4}},\ \bibinfo {pages} {619--624} (\bibinfo {year}
  {2013})}\BibitemShut {NoStop}%
\bibitem [{\citenamefont {Jeanmairet}\ \emph {et~al.}(2015)\citenamefont
  {Jeanmairet}, \citenamefont {Levesque}, \citenamefont {Sergiievskyi},\ and\
  \citenamefont {Borgis}}]{jeanmairet2015molecular}%
  \BibitemOpen
  \bibfield  {author} {\bibinfo {author} {\bibfnamefont {G.}~\bibnamefont
  {Jeanmairet}}, \bibinfo {author} {\bibfnamefont {M.}~\bibnamefont
  {Levesque}}, \bibinfo {author} {\bibfnamefont {V.}~\bibnamefont
  {Sergiievskyi}}, \ and\ \bibinfo {author} {\bibfnamefont {D.}~\bibnamefont
  {Borgis}},\ }\bibfield  {title} {\enquote {\bibinfo {title} {Molecular
  density functional theory for water with liquid-gas coexistence and correct
  pressure},}\ }\href@noop {} {\bibfield  {journal} {\bibinfo  {journal} {The
  Journal of Chemical Physics}\ }\textbf {\bibinfo {volume} {142}},\ \bibinfo
  {pages} {154112} (\bibinfo {year} {2015})}\BibitemShut {NoStop}%
\bibitem [{\citenamefont {Zhao}, \citenamefont {Jin},\ and\ \citenamefont
  {Wu}(2011)}]{zhao2011new}%
  \BibitemOpen
  \bibfield  {author} {\bibinfo {author} {\bibfnamefont {S.}~\bibnamefont
  {Zhao}}, \bibinfo {author} {\bibfnamefont {Z.}~\bibnamefont {Jin}}, \ and\
  \bibinfo {author} {\bibfnamefont {J.}~\bibnamefont {Wu}},\ }\bibfield
  {title} {\enquote {\bibinfo {title} {New theoretical method for rapid
  prediction of solvation free energy in water},}\ }\href@noop {} {\bibfield
  {journal} {\bibinfo  {journal} {The Journal of Physical Chemistry B}\
  }\textbf {\bibinfo {volume} {115}},\ \bibinfo {pages} {6971--6975} (\bibinfo
  {year} {2011})}\BibitemShut {NoStop}%
\bibitem [{\citenamefont {Luukkonen}\ \emph {et~al.}(2020)\citenamefont
  {Luukkonen}, \citenamefont {Levesque}, \citenamefont {Belloni},\ and\
  \citenamefont {Borgis}}]{luukkonen2020hydration}%
  \BibitemOpen
  \bibfield  {author} {\bibinfo {author} {\bibfnamefont {S.}~\bibnamefont
  {Luukkonen}}, \bibinfo {author} {\bibfnamefont {M.}~\bibnamefont {Levesque}},
  \bibinfo {author} {\bibfnamefont {L.}~\bibnamefont {Belloni}}, \ and\
  \bibinfo {author} {\bibfnamefont {D.}~\bibnamefont {Borgis}},\ }\bibfield
  {title} {\enquote {\bibinfo {title} {Hydration free energies and solvation
  structures with molecular density functional theory in the hypernetted chain
  approximation},}\ }\href@noop {} {\bibfield  {journal} {\bibinfo  {journal}
  {The Journal of Chemical Physics}\ }\textbf {\bibinfo {volume} {152}},\
  \bibinfo {pages} {064110} (\bibinfo {year} {2020})}\BibitemShut {NoStop}%
\bibitem [{\citenamefont {Mobley}\ and\ \citenamefont
  {Guthrie}(2014)}]{mobley2014freesolv}%
  \BibitemOpen
  \bibfield  {author} {\bibinfo {author} {\bibfnamefont {D.~L.}\ \bibnamefont
  {Mobley}}\ and\ \bibinfo {author} {\bibfnamefont {J.~P.}\ \bibnamefont
  {Guthrie}},\ }\bibfield  {title} {\enquote {\bibinfo {title} {Freesolv: a
  database of experimental and calculated hydration free energies, with input
  files},}\ }\href@noop {} {\bibfield  {journal} {\bibinfo  {journal} {Journal
  of computer-aided molecular design}\ }\textbf {\bibinfo {volume} {28}},\
  \bibinfo {pages} {711--720} (\bibinfo {year} {2014})}\BibitemShut {NoStop}%
\bibitem [{\citenamefont {Jeanmairet}, \citenamefont {Levesque},\ and\
  \citenamefont {Borgis}(2020)}]{jeanmairet2020tackling}%
  \BibitemOpen
  \bibfield  {author} {\bibinfo {author} {\bibfnamefont {G.}~\bibnamefont
  {Jeanmairet}}, \bibinfo {author} {\bibfnamefont {M.}~\bibnamefont
  {Levesque}}, \ and\ \bibinfo {author} {\bibfnamefont {D.}~\bibnamefont
  {Borgis}},\ }\bibfield  {title} {\enquote {\bibinfo {title} {Tackling solvent
  effects by coupling electronic and molecular density functional theory},}\
  }\href@noop {} {\bibfield  {journal} {\bibinfo  {journal} {Journal of
  Chemical Theory and Computation}\ } (\bibinfo {year} {2020})}\BibitemShut
  {NoStop}%
\bibitem [{\citenamefont {Blum}\ and\ \citenamefont
  {Torruella}(1972)}]{blum1972invariant1}%
  \BibitemOpen
  \bibfield  {author} {\bibinfo {author} {\bibfnamefont {L.}~\bibnamefont
  {Blum}}\ and\ \bibinfo {author} {\bibfnamefont {A.}~\bibnamefont
  {Torruella}},\ }\bibfield  {title} {\enquote {\bibinfo {title} {Invariant
  expansion for two-body correlations: Thermodynamic functions, scattering, and
  the ornstein—zernike equation},}\ }\href@noop {} {\bibfield  {journal}
  {\bibinfo  {journal} {The Journal of Chemical Physics}\ }\textbf {\bibinfo
  {volume} {56}},\ \bibinfo {pages} {303--310} (\bibinfo {year}
  {1972})}\BibitemShut {NoStop}%
\bibitem [{\citenamefont {Blum}(1972)}]{blum1972invariant2}%
  \BibitemOpen
  \bibfield  {author} {\bibinfo {author} {\bibfnamefont {L.}~\bibnamefont
  {Blum}},\ }\bibfield  {title} {\enquote {\bibinfo {title} {Invariant
  expansion. ii. the ornstein-zernike equation for nonspherical molecules and
  an extended solution to the mean spherical model},}\ }\href@noop {}
  {\bibfield  {journal} {\bibinfo  {journal} {The Journal of Chemical Physics}\
  }\textbf {\bibinfo {volume} {57}},\ \bibinfo {pages} {1862--1869} (\bibinfo
  {year} {1972})}\BibitemShut {NoStop}%
\bibitem [{\citenamefont {Chandler}, \citenamefont {McCoy},\ and\ \citenamefont
  {Singer}(1986{\natexlab{a}})}]{chandler1986density1}%
  \BibitemOpen
  \bibfield  {author} {\bibinfo {author} {\bibfnamefont {D.}~\bibnamefont
  {Chandler}}, \bibinfo {author} {\bibfnamefont {J.~D.}\ \bibnamefont {McCoy}},
  \ and\ \bibinfo {author} {\bibfnamefont {S.~J.}\ \bibnamefont {Singer}},\
  }\bibfield  {title} {\enquote {\bibinfo {title} {Density functional theory of
  nonuniform polyatomic systems. i. general formulation},}\ }\href@noop {}
  {\bibfield  {journal} {\bibinfo  {journal} {The Journal of chemical physics}\
  }\textbf {\bibinfo {volume} {85}},\ \bibinfo {pages} {5971--5976} (\bibinfo
  {year} {1986}{\natexlab{a}})}\BibitemShut {NoStop}%
\bibitem [{\citenamefont {Chandler}, \citenamefont {McCoy},\ and\ \citenamefont
  {Singer}(1986{\natexlab{b}})}]{chandler1986density2}%
  \BibitemOpen
  \bibfield  {author} {\bibinfo {author} {\bibfnamefont {D.}~\bibnamefont
  {Chandler}}, \bibinfo {author} {\bibfnamefont {J.~D.}\ \bibnamefont {McCoy}},
  \ and\ \bibinfo {author} {\bibfnamefont {S.~J.}\ \bibnamefont {Singer}},\
  }\bibfield  {title} {\enquote {\bibinfo {title} {Density functional theory of
  nonuniform polyatomic systems. ii. rational closures for integral
  equations},}\ }\href@noop {} {\bibfield  {journal} {\bibinfo  {journal} {The
  Journal of chemical physics}\ }\textbf {\bibinfo {volume} {85}},\ \bibinfo
  {pages} {5977--5982} (\bibinfo {year} {1986}{\natexlab{b}})}\BibitemShut
  {NoStop}%
\bibitem [{\citenamefont {Liu}, \citenamefont {Zhao},\ and\ \citenamefont
  {Wu}(2013)}]{liu2013site}%
  \BibitemOpen
  \bibfield  {author} {\bibinfo {author} {\bibfnamefont {Y.}~\bibnamefont
  {Liu}}, \bibinfo {author} {\bibfnamefont {S.}~\bibnamefont {Zhao}}, \ and\
  \bibinfo {author} {\bibfnamefont {J.}~\bibnamefont {Wu}},\ }\bibfield
  {title} {\enquote {\bibinfo {title} {A site density functional theory for
  water: Application to solvation of amino acid side chains},}\ }\href@noop {}
  {\bibfield  {journal} {\bibinfo  {journal} {Journal of chemical theory and
  computation}\ }\textbf {\bibinfo {volume} {9}},\ \bibinfo {pages}
  {1896--1908} (\bibinfo {year} {2013})}\BibitemShut {NoStop}%
\bibitem [{\citenamefont {Liu}, \citenamefont {Fu},\ and\ \citenamefont
  {Wu}(2013)}]{liu2013high}%
  \BibitemOpen
  \bibfield  {author} {\bibinfo {author} {\bibfnamefont {Y.}~\bibnamefont
  {Liu}}, \bibinfo {author} {\bibfnamefont {J.}~\bibnamefont {Fu}}, \ and\
  \bibinfo {author} {\bibfnamefont {J.}~\bibnamefont {Wu}},\ }\bibfield
  {title} {\enquote {\bibinfo {title} {High-throughput prediction of the
  hydration free energies of small molecules from a classical density
  functional theory},}\ }\href@noop {} {\bibfield  {journal} {\bibinfo
  {journal} {The Journal of Physical Chemistry Letters}\ }\textbf {\bibinfo
  {volume} {4}},\ \bibinfo {pages} {3687--3691} (\bibinfo {year}
  {2013})}\BibitemShut {NoStop}%
\bibitem [{\citenamefont {Yu}\ and\ \citenamefont
  {Wu}(2002)}]{yu2002structures}%
  \BibitemOpen
  \bibfield  {author} {\bibinfo {author} {\bibfnamefont {Y.-X.}\ \bibnamefont
  {Yu}}\ and\ \bibinfo {author} {\bibfnamefont {J.}~\bibnamefont {Wu}},\
  }\bibfield  {title} {\enquote {\bibinfo {title} {Structures of hard-sphere
  fluids from a modified fundamental-measure theory},}\ }\href@noop {}
  {\bibfield  {journal} {\bibinfo  {journal} {The Journal of chemical physics}\
  }\textbf {\bibinfo {volume} {117}},\ \bibinfo {pages} {10156--10164}
  (\bibinfo {year} {2002})}\BibitemShut {NoStop}%
\bibitem [{\citenamefont {Jackson}, \citenamefont {Chapman},\ and\
  \citenamefont {Gubbins}(1988)}]{jackson1988phase1}%
  \BibitemOpen
  \bibfield  {author} {\bibinfo {author} {\bibfnamefont {G.}~\bibnamefont
  {Jackson}}, \bibinfo {author} {\bibfnamefont {W.~G.}\ \bibnamefont
  {Chapman}}, \ and\ \bibinfo {author} {\bibfnamefont {K.~E.}\ \bibnamefont
  {Gubbins}},\ }\bibfield  {title} {\enquote {\bibinfo {title} {Phase
  equilibria of associating fluids: Spherical molecules with multiple bonding
  sites},}\ }\href@noop {} {\bibfield  {journal} {\bibinfo  {journal}
  {Molecular Physics}\ }\textbf {\bibinfo {volume} {65}},\ \bibinfo {pages}
  {1--31} (\bibinfo {year} {1988})}\BibitemShut {NoStop}%
\bibitem [{\citenamefont {Chapman}, \citenamefont {Jackson},\ and\
  \citenamefont {Gubbins}(1988)}]{chapman1988phase2}%
  \BibitemOpen
  \bibfield  {author} {\bibinfo {author} {\bibfnamefont {W.~G.}\ \bibnamefont
  {Chapman}}, \bibinfo {author} {\bibfnamefont {G.}~\bibnamefont {Jackson}}, \
  and\ \bibinfo {author} {\bibfnamefont {K.~E.}\ \bibnamefont {Gubbins}},\
  }\bibfield  {title} {\enquote {\bibinfo {title} {Phase equilibria of
  associating fluids: chain molecules with multiple bonding sites},}\
  }\href@noop {} {\bibfield  {journal} {\bibinfo  {journal} {Molecular
  Physics}\ }\textbf {\bibinfo {volume} {65}},\ \bibinfo {pages} {1057--1079}
  (\bibinfo {year} {1988})}\BibitemShut {NoStop}%
\bibitem [{\citenamefont {Chapman}\ \emph {et~al.}(1989)\citenamefont
  {Chapman}, \citenamefont {Gubbins}, \citenamefont {Jackson},\ and\
  \citenamefont {Radosz}}]{chapman1989saft}%
  \BibitemOpen
  \bibfield  {author} {\bibinfo {author} {\bibfnamefont {W.~G.}\ \bibnamefont
  {Chapman}}, \bibinfo {author} {\bibfnamefont {K.~E.}\ \bibnamefont
  {Gubbins}}, \bibinfo {author} {\bibfnamefont {G.}~\bibnamefont {Jackson}}, \
  and\ \bibinfo {author} {\bibfnamefont {M.}~\bibnamefont {Radosz}},\
  }\bibfield  {title} {\enquote {\bibinfo {title} {Saft: equation-of-state
  solution model for associating fluids},}\ }\href@noop {} {\bibfield
  {journal} {\bibinfo  {journal} {Fluid Phase Equilibria}\ }\textbf {\bibinfo
  {volume} {52}},\ \bibinfo {pages} {31--38} (\bibinfo {year}
  {1989})}\BibitemShut {NoStop}%
\bibitem [{\citenamefont {Chapman}\ \emph {et~al.}(1990)\citenamefont
  {Chapman}, \citenamefont {Gubbins}, \citenamefont {Jackson},\ and\
  \citenamefont {Radosz}}]{chapman1990new}%
  \BibitemOpen
  \bibfield  {author} {\bibinfo {author} {\bibfnamefont {W.~G.}\ \bibnamefont
  {Chapman}}, \bibinfo {author} {\bibfnamefont {K.~E.}\ \bibnamefont
  {Gubbins}}, \bibinfo {author} {\bibfnamefont {G.}~\bibnamefont {Jackson}}, \
  and\ \bibinfo {author} {\bibfnamefont {M.}~\bibnamefont {Radosz}},\
  }\bibfield  {title} {\enquote {\bibinfo {title} {New reference equation of
  state for associating liquids},}\ }\href@noop {} {\bibfield  {journal}
  {\bibinfo  {journal} {Industrial \& engineering chemistry research}\ }\textbf
  {\bibinfo {volume} {29}},\ \bibinfo {pages} {1709--1721} (\bibinfo {year}
  {1990})}\BibitemShut {NoStop}%
\bibitem [{\citenamefont {Wertheim}(1984{\natexlab{a}})}]{wertheim1984fluids1}%
  \BibitemOpen
  \bibfield  {author} {\bibinfo {author} {\bibfnamefont {M.}~\bibnamefont
  {Wertheim}},\ }\bibfield  {title} {\enquote {\bibinfo {title} {Fluids with
  highly directional attractive forces. i. statistical thermodynamics},}\
  }\href@noop {} {\bibfield  {journal} {\bibinfo  {journal} {Journal of
  statistical physics}\ }\textbf {\bibinfo {volume} {35}},\ \bibinfo {pages}
  {19--34} (\bibinfo {year} {1984}{\natexlab{a}})}\BibitemShut {NoStop}%
\bibitem [{\citenamefont {Wertheim}(1984{\natexlab{b}})}]{wertheim1984fluids2}%
  \BibitemOpen
  \bibfield  {author} {\bibinfo {author} {\bibfnamefont {M.}~\bibnamefont
  {Wertheim}},\ }\bibfield  {title} {\enquote {\bibinfo {title} {Fluids with
  highly directional attractive forces. ii. thermodynamic perturbation theory
  and integral equations},}\ }\href@noop {} {\bibfield  {journal} {\bibinfo
  {journal} {Journal of statistical physics}\ }\textbf {\bibinfo {volume}
  {35}},\ \bibinfo {pages} {35--47} (\bibinfo {year}
  {1984}{\natexlab{b}})}\BibitemShut {NoStop}%
\bibitem [{\citenamefont {Wertheim}(1986{\natexlab{a}})}]{wertheim1986fluids1}%
  \BibitemOpen
  \bibfield  {author} {\bibinfo {author} {\bibfnamefont {M.}~\bibnamefont
  {Wertheim}},\ }\bibfield  {title} {\enquote {\bibinfo {title} {Fluids with
  highly directional attractive forces. iii. multiple attraction sites},}\
  }\href@noop {} {\bibfield  {journal} {\bibinfo  {journal} {Journal of
  statistical physics}\ }\textbf {\bibinfo {volume} {42}},\ \bibinfo {pages}
  {459--476} (\bibinfo {year} {1986}{\natexlab{a}})}\BibitemShut {NoStop}%
\bibitem [{\citenamefont {Wertheim}(1986{\natexlab{b}})}]{wertheim1986fluids2}%
  \BibitemOpen
  \bibfield  {author} {\bibinfo {author} {\bibfnamefont {M.}~\bibnamefont
  {Wertheim}},\ }\bibfield  {title} {\enquote {\bibinfo {title} {Fluids with
  highly directional attractive forces. iv. equilibrium polymerization},}\
  }\href@noop {} {\bibfield  {journal} {\bibinfo  {journal} {Journal of
  statistical physics}\ }\textbf {\bibinfo {volume} {42}},\ \bibinfo {pages}
  {477--492} (\bibinfo {year} {1986}{\natexlab{b}})}\BibitemShut {NoStop}%
\bibitem [{\citenamefont {Gross}\ and\ \citenamefont
  {Sadowski}(2001)}]{gross2001perturbed}%
  \BibitemOpen
  \bibfield  {author} {\bibinfo {author} {\bibfnamefont {J.}~\bibnamefont
  {Gross}}\ and\ \bibinfo {author} {\bibfnamefont {G.}~\bibnamefont
  {Sadowski}},\ }\bibfield  {title} {\enquote {\bibinfo {title}
  {Perturbed-chain saft: An equation of state based on a perturbation theory
  for chain molecules},}\ }\href@noop {} {\bibfield  {journal} {\bibinfo
  {journal} {Industrial \& engineering chemistry research}\ }\textbf {\bibinfo
  {volume} {40}},\ \bibinfo {pages} {1244--1260} (\bibinfo {year}
  {2001})}\BibitemShut {NoStop}%
\bibitem [{\citenamefont {Sauer}\ and\ \citenamefont
  {Gross}(2017)}]{sauer2017classical}%
  \BibitemOpen
  \bibfield  {author} {\bibinfo {author} {\bibfnamefont {E.}~\bibnamefont
  {Sauer}}\ and\ \bibinfo {author} {\bibfnamefont {J.}~\bibnamefont {Gross}},\
  }\bibfield  {title} {\enquote {\bibinfo {title} {Classical density functional
  theory for liquid--fluid interfaces and confined systems: A functional for
  the perturbed-chain polar statistical associating fluid theory equation of
  state},}\ }\href@noop {} {\bibfield  {journal} {\bibinfo  {journal}
  {Industrial \& Engineering Chemistry Research}\ }\textbf {\bibinfo {volume}
  {56}},\ \bibinfo {pages} {4119--4135} (\bibinfo {year} {2017})}\BibitemShut
  {NoStop}%
\bibitem [{\citenamefont {Sauer}\ \emph {et~al.}(2018)\citenamefont {Sauer},
  \citenamefont {Terzis}, \citenamefont {Theiss}, \citenamefont {Weigand},\
  and\ \citenamefont {Gross}}]{sauer2018prediction}%
  \BibitemOpen
  \bibfield  {author} {\bibinfo {author} {\bibfnamefont {E.}~\bibnamefont
  {Sauer}}, \bibinfo {author} {\bibfnamefont {A.}~\bibnamefont {Terzis}},
  \bibinfo {author} {\bibfnamefont {M.}~\bibnamefont {Theiss}}, \bibinfo
  {author} {\bibfnamefont {B.}~\bibnamefont {Weigand}}, \ and\ \bibinfo
  {author} {\bibfnamefont {J.}~\bibnamefont {Gross}},\ }\bibfield  {title}
  {\enquote {\bibinfo {title} {Prediction of contact angles and density
  profiles of sessile droplets using classical density functional theory based
  on the pcp-saft equation of state},}\ }\href@noop {} {\bibfield  {journal}
  {\bibinfo  {journal} {Langmuir}\ }\textbf {\bibinfo {volume} {34}},\ \bibinfo
  {pages} {12519--12531} (\bibinfo {year} {2018})}\BibitemShut {NoStop}%
\bibitem [{\citenamefont {Sauer}\ and\ \citenamefont
  {Gross}(2019)}]{sauer2019prediction}%
  \BibitemOpen
  \bibfield  {author} {\bibinfo {author} {\bibfnamefont {E.}~\bibnamefont
  {Sauer}}\ and\ \bibinfo {author} {\bibfnamefont {J.}~\bibnamefont {Gross}},\
  }\bibfield  {title} {\enquote {\bibinfo {title} {Prediction of adsorption
  isotherms and selectivities: Comparison between classical density functional
  theory based on the perturbed-chain statistical associating fluid theory
  equation of state and ideal adsorbed solution theory},}\ }\href@noop {}
  {\bibfield  {journal} {\bibinfo  {journal} {Langmuir}\ }\textbf {\bibinfo
  {volume} {35}},\ \bibinfo {pages} {11690--11701} (\bibinfo {year}
  {2019})}\BibitemShut {NoStop}%
\bibitem [{\citenamefont {Rehner}\ and\ \citenamefont
  {Gross}(2018)}]{rehner2018surface}%
  \BibitemOpen
  \bibfield  {author} {\bibinfo {author} {\bibfnamefont {P.}~\bibnamefont
  {Rehner}}\ and\ \bibinfo {author} {\bibfnamefont {J.}~\bibnamefont {Gross}},\
  }\bibfield  {title} {\enquote {\bibinfo {title} {Surface tension of droplets
  and tolman lengths of real substances and mixtures from density functional
  theory},}\ }\href@noop {} {\bibfield  {journal} {\bibinfo  {journal} {The
  Journal of chemical physics}\ }\textbf {\bibinfo {volume} {148}},\ \bibinfo
  {pages} {164703} (\bibinfo {year} {2018})}\BibitemShut {NoStop}%
\bibitem [{\citenamefont {Ben-Naim}(2013)}]{ben2013solvation}%
  \BibitemOpen
  \bibfield  {author} {\bibinfo {author} {\bibfnamefont {A.~Y.}\ \bibnamefont
  {Ben-Naim}},\ }\href@noop {} {\emph {\bibinfo {title} {Solvation
  thermodynamics}}}\ (\bibinfo  {publisher} {Springer Science \& Business
  Media},\ \bibinfo {year} {2013})\BibitemShut {NoStop}%
\bibitem [{\citenamefont {Roth}\ \emph {et~al.}(2002)\citenamefont {Roth},
  \citenamefont {Evans}, \citenamefont {Lang},\ and\ \citenamefont
  {Kahl}}]{roth2002fundamental}%
  \BibitemOpen
  \bibfield  {author} {\bibinfo {author} {\bibfnamefont {R.}~\bibnamefont
  {Roth}}, \bibinfo {author} {\bibfnamefont {R.}~\bibnamefont {Evans}},
  \bibinfo {author} {\bibfnamefont {A.}~\bibnamefont {Lang}}, \ and\ \bibinfo
  {author} {\bibfnamefont {G.}~\bibnamefont {Kahl}},\ }\bibfield  {title}
  {\enquote {\bibinfo {title} {Fundamental measure theory for hard-sphere
  mixtures revisited: the white bear version},}\ }\href@noop {} {\bibfield
  {journal} {\bibinfo  {journal} {Journal of Physics: Condensed Matter}\
  }\textbf {\bibinfo {volume} {14}},\ \bibinfo {pages} {12063} (\bibinfo {year}
  {2002})}\BibitemShut {NoStop}%
\bibitem [{\citenamefont {Tripathi}\ and\ \citenamefont
  {Chapman}(2005{\natexlab{a}})}]{tripathi2005microstructure1}%
  \BibitemOpen
  \bibfield  {author} {\bibinfo {author} {\bibfnamefont {S.}~\bibnamefont
  {Tripathi}}\ and\ \bibinfo {author} {\bibfnamefont {W.~G.}\ \bibnamefont
  {Chapman}},\ }\bibfield  {title} {\enquote {\bibinfo {title} {Microstructure
  of inhomogeneous polyatomic mixtures from a density functional formalism for
  atomic mixtures},}\ }\href@noop {} {\bibfield  {journal} {\bibinfo  {journal}
  {The Journal of chemical physics}\ }\textbf {\bibinfo {volume} {122}},\
  \bibinfo {pages} {094506} (\bibinfo {year} {2005}{\natexlab{a}})}\BibitemShut
  {NoStop}%
\bibitem [{\citenamefont {Tripathi}\ and\ \citenamefont
  {Chapman}(2005{\natexlab{b}})}]{tripathi2005microstructure2}%
  \BibitemOpen
  \bibfield  {author} {\bibinfo {author} {\bibfnamefont {S.}~\bibnamefont
  {Tripathi}}\ and\ \bibinfo {author} {\bibfnamefont {W.~G.}\ \bibnamefont
  {Chapman}},\ }\bibfield  {title} {\enquote {\bibinfo {title} {Microstructure
  and thermodynamics of inhomogeneous polymer blends and solutions},}\
  }\href@noop {} {\bibfield  {journal} {\bibinfo  {journal} {Physical review
  letters}\ }\textbf {\bibinfo {volume} {94}},\ \bibinfo {pages} {087801}
  (\bibinfo {year} {2005}{\natexlab{b}})}\BibitemShut {NoStop}%
\bibitem [{\citenamefont {Zmpitas}\ and\ \citenamefont
  {Gross}(2016)}]{zmpitas2016detailed}%
  \BibitemOpen
  \bibfield  {author} {\bibinfo {author} {\bibfnamefont {W.}~\bibnamefont
  {Zmpitas}}\ and\ \bibinfo {author} {\bibfnamefont {J.}~\bibnamefont
  {Gross}},\ }\bibfield  {title} {\enquote {\bibinfo {title} {Detailed
  pedagogical review and analysis of wertheim's thermodynamic perturbation
  theory},}\ }\href@noop {} {\bibfield  {journal} {\bibinfo  {journal} {Fluid
  Phase Equilibria}\ }\textbf {\bibinfo {volume} {428}},\ \bibinfo {pages}
  {121--152} (\bibinfo {year} {2016})}\BibitemShut {NoStop}%
\bibitem [{\citenamefont {Jain}, \citenamefont {Dominik},\ and\ \citenamefont
  {Chapman}(2007)}]{jain2007modified}%
  \BibitemOpen
  \bibfield  {author} {\bibinfo {author} {\bibfnamefont {S.}~\bibnamefont
  {Jain}}, \bibinfo {author} {\bibfnamefont {A.}~\bibnamefont {Dominik}}, \
  and\ \bibinfo {author} {\bibfnamefont {W.~G.}\ \bibnamefont {Chapman}},\
  }\bibfield  {title} {\enquote {\bibinfo {title} {Modified interfacial
  statistical associating fluid theory: A perturbation density functional
  theory for inhomogeneous complex fluids},}\ }\href@noop {} {\bibfield
  {journal} {\bibinfo  {journal} {The Journal of chemical physics}\ }\textbf
  {\bibinfo {volume} {127}},\ \bibinfo {pages} {244904} (\bibinfo {year}
  {2007})}\BibitemShut {NoStop}%
\bibitem [{\citenamefont {Mairhofer}, \citenamefont {Xiao},\ and\ \citenamefont
  {Gross}(2018)}]{mairhofer2018classical}%
  \BibitemOpen
  \bibfield  {author} {\bibinfo {author} {\bibfnamefont {J.}~\bibnamefont
  {Mairhofer}}, \bibinfo {author} {\bibfnamefont {B.}~\bibnamefont {Xiao}}, \
  and\ \bibinfo {author} {\bibfnamefont {J.}~\bibnamefont {Gross}},\ }\bibfield
   {title} {\enquote {\bibinfo {title} {A classical density functional theory
  for vapor-liquid interfaces consistent with the heterosegmented
  group-contribution perturbed-chain polar statistical associating fluid
  theory},}\ }\href@noop {} {\bibfield  {journal} {\bibinfo  {journal} {Fluid
  Phase Equilibria}\ }\textbf {\bibinfo {volume} {472}},\ \bibinfo {pages}
  {117--127} (\bibinfo {year} {2018})}\BibitemShut {NoStop}%
\bibitem [{\citenamefont {Stierle}\ \emph {et~al.}(2020)\citenamefont
  {Stierle}, \citenamefont {Sauer}, \citenamefont {Eller}, \citenamefont
  {Theiss}, \citenamefont {Rehner}, \citenamefont {Ackermann},\ and\
  \citenamefont {Gross}}]{stierle2020guide}%
  \BibitemOpen
  \bibfield  {author} {\bibinfo {author} {\bibfnamefont {R.}~\bibnamefont
  {Stierle}}, \bibinfo {author} {\bibfnamefont {E.}~\bibnamefont {Sauer}},
  \bibinfo {author} {\bibfnamefont {J.}~\bibnamefont {Eller}}, \bibinfo
  {author} {\bibfnamefont {M.}~\bibnamefont {Theiss}}, \bibinfo {author}
  {\bibfnamefont {P.}~\bibnamefont {Rehner}}, \bibinfo {author} {\bibfnamefont
  {P.}~\bibnamefont {Ackermann}}, \ and\ \bibinfo {author} {\bibfnamefont
  {J.}~\bibnamefont {Gross}},\ }\bibfield  {title} {\enquote {\bibinfo {title}
  {Guide to efficient solution of pc-saft classical density functional theory
  in various coordinate systems using fast fourier and similar transforms},}\
  }\href@noop {} {\bibfield  {journal} {\bibinfo  {journal} {Fluid Phase
  Equilibria}\ }\textbf {\bibinfo {volume} {504}},\ \bibinfo {pages} {112306}
  (\bibinfo {year} {2020})}\BibitemShut {NoStop}%
\bibitem [{\citenamefont {Rodinger}\ and\ \citenamefont
  {Pom{\`e}s}(2005)}]{rodinger2005enhancing}%
  \BibitemOpen
  \bibfield  {author} {\bibinfo {author} {\bibfnamefont {T.}~\bibnamefont
  {Rodinger}}\ and\ \bibinfo {author} {\bibfnamefont {R.}~\bibnamefont
  {Pom{\`e}s}},\ }\bibfield  {title} {\enquote {\bibinfo {title} {Enhancing the
  accuracy, the efficiency and the scope of free energy simulations},}\
  }\href@noop {} {\bibfield  {journal} {\bibinfo  {journal} {Current opinion in
  structural biology}\ }\textbf {\bibinfo {volume} {15}},\ \bibinfo {pages}
  {164--170} (\bibinfo {year} {2005})}\BibitemShut {NoStop}%
\bibitem [{\citenamefont {Shirts}, \citenamefont {Mobley},\ and\ \citenamefont
  {Chodera}(2007)}]{shirts2007alchemical}%
  \BibitemOpen
  \bibfield  {author} {\bibinfo {author} {\bibfnamefont {M.~R.}\ \bibnamefont
  {Shirts}}, \bibinfo {author} {\bibfnamefont {D.~L.}\ \bibnamefont {Mobley}},
  \ and\ \bibinfo {author} {\bibfnamefont {J.~D.}\ \bibnamefont {Chodera}},\
  }\bibfield  {title} {\enquote {\bibinfo {title} {Alchemical free energy
  calculations: ready for prime time?}}\ }\href@noop {} {\bibfield  {journal}
  {\bibinfo  {journal} {Annual reports in computational chemistry}\ }\textbf
  {\bibinfo {volume} {3}},\ \bibinfo {pages} {41--59} (\bibinfo {year}
  {2007})}\BibitemShut {NoStop}%
\bibitem [{\citenamefont {Beutler}\ \emph {et~al.}(1994)\citenamefont
  {Beutler}, \citenamefont {Mark}, \citenamefont {van Schaik}, \citenamefont
  {Gerber},\ and\ \citenamefont {Van~Gunsteren}}]{beutler1994avoiding}%
  \BibitemOpen
  \bibfield  {author} {\bibinfo {author} {\bibfnamefont {T.~C.}\ \bibnamefont
  {Beutler}}, \bibinfo {author} {\bibfnamefont {A.~E.}\ \bibnamefont {Mark}},
  \bibinfo {author} {\bibfnamefont {R.~C.}\ \bibnamefont {van Schaik}},
  \bibinfo {author} {\bibfnamefont {P.~R.}\ \bibnamefont {Gerber}}, \ and\
  \bibinfo {author} {\bibfnamefont {W.~F.}\ \bibnamefont {Van~Gunsteren}},\
  }\bibfield  {title} {\enquote {\bibinfo {title} {Avoiding singularities and
  numerical instabilities in free energy calculations based on molecular
  simulations},}\ }\href@noop {} {\bibfield  {journal} {\bibinfo  {journal}
  {Chemical physics letters}\ }\textbf {\bibinfo {volume} {222}},\ \bibinfo
  {pages} {529--539} (\bibinfo {year} {1994})}\BibitemShut {NoStop}%
\bibitem [{\citenamefont {Zacharias}, \citenamefont {Straatsma},\ and\
  \citenamefont {McCammon}(1994)}]{zacharias1994separation}%
  \BibitemOpen
  \bibfield  {author} {\bibinfo {author} {\bibfnamefont {M.}~\bibnamefont
  {Zacharias}}, \bibinfo {author} {\bibfnamefont {T.}~\bibnamefont
  {Straatsma}}, \ and\ \bibinfo {author} {\bibfnamefont {J.}~\bibnamefont
  {McCammon}},\ }\bibfield  {title} {\enquote {\bibinfo {title}
  {Separation-shifted scaling, a new scaling method for lennard-jones
  interactions in thermodynamic integration},}\ }\href@noop {} {\bibfield
  {journal} {\bibinfo  {journal} {The Journal of chemical physics}\ }\textbf
  {\bibinfo {volume} {100}},\ \bibinfo {pages} {9025--9031} (\bibinfo {year}
  {1994})}\BibitemShut {NoStop}%
\bibitem [{\citenamefont {Kirkwood}(1935)}]{kirkwood1935statistical}%
  \BibitemOpen
  \bibfield  {author} {\bibinfo {author} {\bibfnamefont {J.~G.}\ \bibnamefont
  {Kirkwood}},\ }\bibfield  {title} {\enquote {\bibinfo {title} {Statistical
  mechanics of fluid mixtures},}\ }\href@noop {} {\bibfield  {journal}
  {\bibinfo  {journal} {The Journal of chemical physics}\ }\textbf {\bibinfo
  {volume} {3}},\ \bibinfo {pages} {300--313} (\bibinfo {year}
  {1935})}\BibitemShut {NoStop}%
\bibitem [{\citenamefont {Zwanzig}(1954)}]{zwanzig1954high}%
  \BibitemOpen
  \bibfield  {author} {\bibinfo {author} {\bibfnamefont {R.~W.}\ \bibnamefont
  {Zwanzig}},\ }\bibfield  {title} {\enquote {\bibinfo {title}
  {High-temperature equation of state by a perturbation method. i. nonpolar
  gases},}\ }\href@noop {} {\bibfield  {journal} {\bibinfo  {journal} {The
  Journal of Chemical Physics}\ }\textbf {\bibinfo {volume} {22}},\ \bibinfo
  {pages} {1420--1426} (\bibinfo {year} {1954})}\BibitemShut {NoStop}%
\bibitem [{\citenamefont {Bennett}(1976)}]{bennett1976efficient}%
  \BibitemOpen
  \bibfield  {author} {\bibinfo {author} {\bibfnamefont {C.~H.}\ \bibnamefont
  {Bennett}},\ }\bibfield  {title} {\enquote {\bibinfo {title} {Efficient
  estimation of free energy differences from monte carlo data},}\ }\href@noop
  {} {\bibfield  {journal} {\bibinfo  {journal} {Journal of Computational
  Physics}\ }\textbf {\bibinfo {volume} {22}},\ \bibinfo {pages} {245--268}
  (\bibinfo {year} {1976})}\BibitemShut {NoStop}%
\bibitem [{\citenamefont {Shirts}\ and\ \citenamefont
  {Chodera}(2008)}]{shirts2008statistically}%
  \BibitemOpen
  \bibfield  {author} {\bibinfo {author} {\bibfnamefont {M.~R.}\ \bibnamefont
  {Shirts}}\ and\ \bibinfo {author} {\bibfnamefont {J.~D.}\ \bibnamefont
  {Chodera}},\ }\bibfield  {title} {\enquote {\bibinfo {title} {Statistically
  optimal analysis of samples from multiple equilibrium states},}\ }\href@noop
  {} {\bibfield  {journal} {\bibinfo  {journal} {The Journal of chemical
  physics}\ }\textbf {\bibinfo {volume} {129}},\ \bibinfo {pages} {124105}
  (\bibinfo {year} {2008})}\BibitemShut {NoStop}%
\bibitem [{\citenamefont {Shirts}\ and\ \citenamefont
  {Pande}(2005)}]{shirts2005comparison}%
  \BibitemOpen
  \bibfield  {author} {\bibinfo {author} {\bibfnamefont {M.~R.}\ \bibnamefont
  {Shirts}}\ and\ \bibinfo {author} {\bibfnamefont {V.~S.}\ \bibnamefont
  {Pande}},\ }\bibfield  {title} {\enquote {\bibinfo {title} {Comparison of
  efficiency and bias of free energies computed by exponential averaging, the
  bennett acceptance ratio, and thermodynamic integration},}\ }\href@noop {}
  {\bibfield  {journal} {\bibinfo  {journal} {The Journal of chemical physics}\
  }\textbf {\bibinfo {volume} {122}},\ \bibinfo {pages} {144107} (\bibinfo
  {year} {2005})}\BibitemShut {NoStop}%
\bibitem [{\citenamefont {Ytreberg}, \citenamefont {Swendsen},\ and\
  \citenamefont {Zuckerman}(2006)}]{ytreberg2006comparison}%
  \BibitemOpen
  \bibfield  {author} {\bibinfo {author} {\bibfnamefont {F.~M.}\ \bibnamefont
  {Ytreberg}}, \bibinfo {author} {\bibfnamefont {R.~H.}\ \bibnamefont
  {Swendsen}}, \ and\ \bibinfo {author} {\bibfnamefont {D.~M.}\ \bibnamefont
  {Zuckerman}},\ }\bibfield  {title} {\enquote {\bibinfo {title} {Comparison of
  free energy methods for molecular systems},}\ }\href@noop {} {\bibfield
  {journal} {\bibinfo  {journal} {The Journal of chemical physics}\ }\textbf
  {\bibinfo {volume} {125}},\ \bibinfo {pages} {184114} (\bibinfo {year}
  {2006})}\BibitemShut {NoStop}%
\bibitem [{\citenamefont {Paliwal}\ and\ \citenamefont
  {Shirts}(2011)}]{paliwal2011benchmark}%
  \BibitemOpen
  \bibfield  {author} {\bibinfo {author} {\bibfnamefont {H.}~\bibnamefont
  {Paliwal}}\ and\ \bibinfo {author} {\bibfnamefont {M.~R.}\ \bibnamefont
  {Shirts}},\ }\bibfield  {title} {\enquote {\bibinfo {title} {A benchmark test
  set for alchemical free energy transformations and its use to quantify error
  in common free energy methods},}\ }\href@noop {} {\bibfield  {journal}
  {\bibinfo  {journal} {Journal of chemical theory and computation}\ }\textbf
  {\bibinfo {volume} {7}},\ \bibinfo {pages} {4115--4134} (\bibinfo {year}
  {2011})}\BibitemShut {NoStop}%
\bibitem [{\citenamefont {Klimovich}, \citenamefont {Shirts},\ and\
  \citenamefont {Mobley}(2015)}]{Klimovich:2015er}%
  \BibitemOpen
  \bibfield  {author} {\bibinfo {author} {\bibfnamefont {P.~V.}\ \bibnamefont
  {Klimovich}}, \bibinfo {author} {\bibfnamefont {M.~R.}\ \bibnamefont
  {Shirts}}, \ and\ \bibinfo {author} {\bibfnamefont {D.~L.}\ \bibnamefont
  {Mobley}},\ }\bibfield  {title} {\enquote {\bibinfo {title} {Guidelines for
  the analysis of free energy calculations},}\ }\href {\doibase
  10.1007/s10822-015-9840-9} {\bibfield  {journal} {\bibinfo  {journal} {J
  Comput Aided Mol Des}\ }\textbf {\bibinfo {volume} {29}},\ \bibinfo {pages}
  {397--411} (\bibinfo {year} {2015})}\BibitemShut {NoStop}%
\bibitem [{\citenamefont {Berendsen}, \citenamefont {van~der Spoel},\ and\
  \citenamefont {van Drunen}(1995)}]{berendsen1995gromacs}%
  \BibitemOpen
  \bibfield  {author} {\bibinfo {author} {\bibfnamefont {H.~J.}\ \bibnamefont
  {Berendsen}}, \bibinfo {author} {\bibfnamefont {D.}~\bibnamefont {van~der
  Spoel}}, \ and\ \bibinfo {author} {\bibfnamefont {R.}~\bibnamefont {van
  Drunen}},\ }\bibfield  {title} {\enquote {\bibinfo {title} {Gromacs: a
  message-passing parallel molecular dynamics implementation},}\ }\href@noop {}
  {\bibfield  {journal} {\bibinfo  {journal} {Computer physics communications}\
  }\textbf {\bibinfo {volume} {91}},\ \bibinfo {pages} {43--56} (\bibinfo
  {year} {1995})}\BibitemShut {NoStop}%
\bibitem [{\citenamefont {Lindahl}, \citenamefont {Hess},\ and\ \citenamefont
  {Van Der~Spoel}(2001)}]{lindahl2001gromacs}%
  \BibitemOpen
  \bibfield  {author} {\bibinfo {author} {\bibfnamefont {E.}~\bibnamefont
  {Lindahl}}, \bibinfo {author} {\bibfnamefont {B.}~\bibnamefont {Hess}}, \
  and\ \bibinfo {author} {\bibfnamefont {D.}~\bibnamefont {Van Der~Spoel}},\
  }\bibfield  {title} {\enquote {\bibinfo {title} {Gromacs 3.0: a package for
  molecular simulation and trajectory analysis},}\ }\href@noop {} {\bibfield
  {journal} {\bibinfo  {journal} {Molecular modeling annual}\ }\textbf
  {\bibinfo {volume} {7}},\ \bibinfo {pages} {306--317} (\bibinfo {year}
  {2001})}\BibitemShut {NoStop}%
\bibitem [{\citenamefont {Van Der~Spoel}\ \emph {et~al.}(2005)\citenamefont
  {Van Der~Spoel}, \citenamefont {Lindahl}, \citenamefont {Hess}, \citenamefont
  {Groenhof}, \citenamefont {Mark},\ and\ \citenamefont
  {Berendsen}}]{van2005gromacs}%
  \BibitemOpen
  \bibfield  {author} {\bibinfo {author} {\bibfnamefont {D.}~\bibnamefont {Van
  Der~Spoel}}, \bibinfo {author} {\bibfnamefont {E.}~\bibnamefont {Lindahl}},
  \bibinfo {author} {\bibfnamefont {B.}~\bibnamefont {Hess}}, \bibinfo {author}
  {\bibfnamefont {G.}~\bibnamefont {Groenhof}}, \bibinfo {author}
  {\bibfnamefont {A.~E.}\ \bibnamefont {Mark}}, \ and\ \bibinfo {author}
  {\bibfnamefont {H.~J.}\ \bibnamefont {Berendsen}},\ }\bibfield  {title}
  {\enquote {\bibinfo {title} {Gromacs: fast, flexible, and free},}\
  }\href@noop {} {\bibfield  {journal} {\bibinfo  {journal} {Journal of
  computational chemistry}\ }\textbf {\bibinfo {volume} {26}},\ \bibinfo
  {pages} {1701--1718} (\bibinfo {year} {2005})}\BibitemShut {NoStop}%
\bibitem [{\citenamefont {Wang}\ \emph {et~al.}(2004)\citenamefont {Wang},
  \citenamefont {Wolf}, \citenamefont {Caldwell}, \citenamefont {Kollman},\
  and\ \citenamefont {Case}}]{wang2004development}%
  \BibitemOpen
  \bibfield  {author} {\bibinfo {author} {\bibfnamefont {J.}~\bibnamefont
  {Wang}}, \bibinfo {author} {\bibfnamefont {R.~M.}\ \bibnamefont {Wolf}},
  \bibinfo {author} {\bibfnamefont {J.~W.}\ \bibnamefont {Caldwell}}, \bibinfo
  {author} {\bibfnamefont {P.~A.}\ \bibnamefont {Kollman}}, \ and\ \bibinfo
  {author} {\bibfnamefont {D.~A.}\ \bibnamefont {Case}},\ }\bibfield  {title}
  {\enquote {\bibinfo {title} {Development and testing of a general amber force
  field},}\ }\href@noop {} {\bibfield  {journal} {\bibinfo  {journal} {Journal
  of computational chemistry}\ }\textbf {\bibinfo {volume} {25}},\ \bibinfo
  {pages} {1157--1174} (\bibinfo {year} {2004})}\BibitemShut {NoStop}%
\bibitem [{\citenamefont {Ryckaert}, \citenamefont {Ciccotti},\ and\
  \citenamefont {Berendsen}(1977)}]{ryckaert1977numerical}%
  \BibitemOpen
  \bibfield  {author} {\bibinfo {author} {\bibfnamefont {J.-P.}\ \bibnamefont
  {Ryckaert}}, \bibinfo {author} {\bibfnamefont {G.}~\bibnamefont {Ciccotti}},
  \ and\ \bibinfo {author} {\bibfnamefont {H.~J.}\ \bibnamefont {Berendsen}},\
  }\bibfield  {title} {\enquote {\bibinfo {title} {Numerical integration of the
  cartesian equations of motion of a system with constraints: molecular
  dynamics of n-alkanes},}\ }\href@noop {} {\bibfield  {journal} {\bibinfo
  {journal} {Journal of computational physics}\ }\textbf {\bibinfo {volume}
  {23}},\ \bibinfo {pages} {327--341} (\bibinfo {year} {1977})}\BibitemShut
  {NoStop}%
\bibitem [{\citenamefont {Hockney}(1970)}]{hockney1970potential}%
  \BibitemOpen
  \bibfield  {author} {\bibinfo {author} {\bibfnamefont {R.~W.}\ \bibnamefont
  {Hockney}},\ }\bibfield  {title} {\enquote {\bibinfo {title} {The potential
  calculation and some applications},}\ }\href@noop {} {\bibfield  {journal}
  {\bibinfo  {journal} {Methods Comput. Phys.}\ }\textbf {\bibinfo {volume}
  {9}},\ \bibinfo {pages} {136} (\bibinfo {year} {1970})}\BibitemShut {NoStop}%
\bibitem [{\citenamefont {Parrinello}\ and\ \citenamefont
  {Rahman}(1980)}]{parrinello1980crystal}%
  \BibitemOpen
  \bibfield  {author} {\bibinfo {author} {\bibfnamefont {M.}~\bibnamefont
  {Parrinello}}\ and\ \bibinfo {author} {\bibfnamefont {A.}~\bibnamefont
  {Rahman}},\ }\bibfield  {title} {\enquote {\bibinfo {title} {Crystal
  structure and pair potentials: A molecular-dynamics study},}\ }\href@noop {}
  {\bibfield  {journal} {\bibinfo  {journal} {Physical review letters}\
  }\textbf {\bibinfo {volume} {45}},\ \bibinfo {pages} {1196} (\bibinfo {year}
  {1980})}\BibitemShut {NoStop}%
\bibitem [{\citenamefont {Percus}(1962)}]{percus1962approximation}%
  \BibitemOpen
  \bibfield  {author} {\bibinfo {author} {\bibfnamefont {J.}~\bibnamefont
  {Percus}},\ }\bibfield  {title} {\enquote {\bibinfo {title} {Approximation
  methods in classical statistical mechanics},}\ }\href@noop {} {\bibfield
  {journal} {\bibinfo  {journal} {Physical Review Letters}\ }\textbf {\bibinfo
  {volume} {8}},\ \bibinfo {pages} {462} (\bibinfo {year} {1962})}\BibitemShut
  {NoStop}%
\bibitem [{\citenamefont {Percus}(1964)}]{percus1964equilibrium}%
  \BibitemOpen
  \bibfield  {author} {\bibinfo {author} {\bibfnamefont {J.}~\bibnamefont
  {Percus}},\ }\bibfield  {title} {\enquote {\bibinfo {title} {The equilibrium
  theory of classical fluids},}\ }\href@noop {} {\bibfield  {journal} {\bibinfo
   {journal} {by HL Frisch and JL Lebowitz, Benjamin, New York}\ } (\bibinfo
  {year} {1964})}\BibitemShut {NoStop}%
\bibitem [{\citenamefont {Politzer}, \citenamefont {Murray},\ and\
  \citenamefont {Clark}(2013)}]{politzer2013halogen}%
  \BibitemOpen
  \bibfield  {author} {\bibinfo {author} {\bibfnamefont {P.}~\bibnamefont
  {Politzer}}, \bibinfo {author} {\bibfnamefont {J.~S.}\ \bibnamefont
  {Murray}}, \ and\ \bibinfo {author} {\bibfnamefont {T.}~\bibnamefont
  {Clark}},\ }\bibfield  {title} {\enquote {\bibinfo {title} {Halogen bonding
  and other $\sigma$-hole interactions: a perspective},}\ }\href@noop {}
  {\bibfield  {journal} {\bibinfo  {journal} {Physical Chemistry Chemical
  Physics}\ }\textbf {\bibinfo {volume} {15}},\ \bibinfo {pages} {11178--11189}
  (\bibinfo {year} {2013})}\BibitemShut {NoStop}%
\bibitem [{\citenamefont {Guti{\'e}rrez}\ \emph {et~al.}(2016)\citenamefont
  {Guti{\'e}rrez}, \citenamefont {Lin}, \citenamefont {Vanommeslaeghe},
  \citenamefont {Lemkul}, \citenamefont {Armacost}, \citenamefont
  {Brooks~III},\ and\ \citenamefont
  {MacKerell~Jr}}]{gutierrez2016parametrization}%
  \BibitemOpen
  \bibfield  {author} {\bibinfo {author} {\bibfnamefont {I.~S.}\ \bibnamefont
  {Guti{\'e}rrez}}, \bibinfo {author} {\bibfnamefont {F.-Y.}\ \bibnamefont
  {Lin}}, \bibinfo {author} {\bibfnamefont {K.}~\bibnamefont {Vanommeslaeghe}},
  \bibinfo {author} {\bibfnamefont {J.~A.}\ \bibnamefont {Lemkul}}, \bibinfo
  {author} {\bibfnamefont {K.~A.}\ \bibnamefont {Armacost}}, \bibinfo {author}
  {\bibfnamefont {C.~L.}\ \bibnamefont {Brooks~III}}, \ and\ \bibinfo {author}
  {\bibfnamefont {A.~D.}\ \bibnamefont {MacKerell~Jr}},\ }\bibfield  {title}
  {\enquote {\bibinfo {title} {Parametrization of halogen bonds in the charmm
  general force field: Improved treatment of ligand--protein interactions},}\
  }\href@noop {} {\bibfield  {journal} {\bibinfo  {journal} {Bioorganic \&
  medicinal chemistry}\ }\textbf {\bibinfo {volume} {24}},\ \bibinfo {pages}
  {4812--4825} (\bibinfo {year} {2016})}\BibitemShut {NoStop}%
\bibitem [{\citenamefont {Frodl}\ and\ \citenamefont
  {Dietrich}(1992)}]{frodl1992bulk}%
  \BibitemOpen
  \bibfield  {author} {\bibinfo {author} {\bibfnamefont {P.}~\bibnamefont
  {Frodl}}\ and\ \bibinfo {author} {\bibfnamefont {S.}~\bibnamefont
  {Dietrich}},\ }\bibfield  {title} {\enquote {\bibinfo {title} {Bulk and
  interfacial properties of polar and molecular fluids},}\ }\href@noop {}
  {\bibfield  {journal} {\bibinfo  {journal} {Physical Review A}\ }\textbf
  {\bibinfo {volume} {45}},\ \bibinfo {pages} {7330} (\bibinfo {year}
  {1992})}\BibitemShut {NoStop}%
\bibitem [{\citenamefont {Reindl}, \citenamefont {Bier},\ and\ \citenamefont
  {Dietrich}(2017)}]{reindl2017electrolyte}%
  \BibitemOpen
  \bibfield  {author} {\bibinfo {author} {\bibfnamefont {A.}~\bibnamefont
  {Reindl}}, \bibinfo {author} {\bibfnamefont {M.}~\bibnamefont {Bier}}, \ and\
  \bibinfo {author} {\bibfnamefont {S.}~\bibnamefont {Dietrich}},\ }\bibfield
  {title} {\enquote {\bibinfo {title} {Electrolyte solutions at curved
  electrodes. ii. microscopic approach},}\ }\href@noop {} {\bibfield  {journal}
  {\bibinfo  {journal} {The Journal of chemical physics}\ }\textbf {\bibinfo
  {volume} {146}},\ \bibinfo {pages} {154704} (\bibinfo {year}
  {2017})}\BibitemShut {NoStop}%
\end{thebibliography}%

\widetext
\clearpage

\setcounter{section}{0}
\setcounter{equation}{0}
\setcounter{figure}{0}
\setcounter{table}{0}
\setcounter{page}{1}
\makeatletter

\renewcommand\thepage{S\arabic{page}} 
\renewcommand\thefigure{S\arabic{figure}} 
\renewcommand\thetable{S\arabic{table}} 
\renewcommand\theequation{S\arabic{equation}} 

\begin{center} \Large{Supporting Information}\end{center}
\vspace{4ex}
\begin{center} \Large{Predicting Solvation Free Energies in Non-Polar Solvents using Classical Density Functional Theory based on the PC-SAFT equation of state}\end{center}
\begin{center}
\vspace{2ex}
Johannes Eller, Tanja Matzerath, Thijs van Westen and Joachim Gross*
\end{center}

\noindent
Institute of Thermodynamics and Thermal Process Engineering, University of Stuttgart, Pfaffenwaldring 9, 70569 Stuttgart, Germany\\[2.0ex]

\noindent
*To whom correspondence should be addressed. \\
E-mail: joachim.gross@itt.uni-stuttgart.de\\[2.0ex]

\noindent
(Date: \today)

\newpage
Here, we derive explicit relations for the chemical and pseudo-chemical potentials from their respective partition functions in the canonical and isobaric-isothermal ensemble.
We assume that the electronic and nuclear degrees of freedom are separable and independent of the intramolecular configuration.
In our notation, $\mathbf{r}$ denotes the center of mass coordinate of the molecule and $\mathbf{R}$ denotes the intramolecular configuration relative to the center of mass coordinate.
Further, $\mathbf{r}^N$ is a short-hand notation for $\mathbf{r}^N=\left\lbrace\rb_i,i=1,\dots,N\right\rbrace$ and $\mathbf{X}=\left\lbrace \mathbf{r},\mathbf{R} \right\rbrace$.
For brevity, we only consider pure-component solvents, although the application to mixtures is straightforward.

\section{\label{SI:mu(N,V,T)}Chemical potential in the canonical ensemble}
The solute chemical potential $\mu_s(N_s,N,V,T)$ in the canonical ensemble at constant number of molecules $N_s$ of solute species $s$, solvent molecules $N$, volume $V$ and temperature $T$ is defined as the difference in Helmholtz energy upon adding one solute molecule to the system
\begin{equation}
    \mu_s(N_s,N,V,T)=F(N_s+1,N,V,T)-F(N_s,N,V,T)
    \label{eq:mu_s(N,V,T)}
\end{equation}
The Helmholtz energy, in turn, is related to the canonical partition function $Q(N_s,N,V,T)$ with
\begin{equation}
    F(N_s,N,V,T)=-k_BT\ln{Q(N_s,N,V,T)}
    \label{eq:F=Q}
\end{equation}
where $k_B$ is the Boltzmann constant.
The canonical partition function $Q(N_s,N,V,T)$ is given by
\begin{align}
    Q(N_s,N,V,T)&=\frac{q_s^{N_s}~q^N}{\Lambda_s^{3N_s}\Lambda^{3N}N_s\,!~N\,!}\int \dr^{N_s}\int \diff \mathbf{R}^{N_s} \int \dr^N \int \diff \mathbf{R}^{N}~\exp{\left(-\beta U\left(\mathbf{r}^N,\rb^{N_s},\mathbf{R}^N,\mathbf{R}^{N_s}\right)\right)}\nonumber\\
    &=\frac{q_s^{N_s}~q^N}{\Lambda_s^{3N_s}\Lambda^{3N}N_s\,!~N\,!}\int \diff \mathbf{X}^{N_s} \int \diff \mathbf{X}^{N}~ \exp{\left(-\beta U\left(\mathbf{X}^{N_s},\mathbf{X}^{N}\right)\right)}
    \label{eq:Q(N,V,T)}
\end{align}
where $\beta=(k_B T)^{-1}$ is the inverse temperature, $q_s$ and $q$ are the intramolecular partition functions capturing electronic and nuclear degrees of freedom, and $\Lambda_s$ and $\Lambda$ are the de Broglie wavelengths for both solute and solvent, respectively. The factor $N_s\,!~N\,!$ accounts for the number of molecule permutations of molecular configurations of non-distinguishable molecules. The configuration integral of the Boltzmann factor of the total potential energy $U$, including both, (bonded and non-bonded) intra- and intermolecular contributions, runs over the center of mass coordinates $\mathbf{r}^{N_s},~\mathbf{r}^N$ and the intramolecular configurations $\mathbf{R}^{N_s},~\mathbf{R}^N$ of the solvent and solute molecules.
Inserting eqs. \eqref{eq:F=Q}, \eqref{eq:Q(N,V,T)} in eq. \eqref{eq:mu_s(N,V,T)} yields
\begin{align}
    \mu_s(N_s,N,V,T)&=-k_BT\ln{\frac{Q(N_s+1,N,V,T)}{Q(N_s,N,V,T)}}\nonumber\\
    &=-k_BT\ln{\frac{q_s}{\Lambda_s^3 (N_s+1)}\frac{\int \diff \mathbf{X}^{N_s+1} \int \diff \mathbf{X}^N~ \exp{\left(-\beta U(\mathbf{X}^{N_s+1},\mathbf{X}^N)\right)}}{\int \diff \mathbf{X}^{N_s} \int \diff \mathbf{X}^N~\exp{\left(-\beta U(\mathbf{X}^{N_s},\mathbf{X}^N)\right)}}}
    \label{eq:mu_NVT}
\end{align}
The total potential energy $U(\mathbf{X}^{N_s+1},\mathbf{X}^N)$ can be separated into the potential energy of only the solvent molecules of the system $U(\mathbf{X}^{N_s},\mathbf{X}^N)$ and the energy contribution $\Delta U(\mathbf{X}^{N_s},\mathbf{X}^N;\mathbf{X}_{N_s+1})$ arising from the insertion of the $N_s+1$-th solute molecule.
\begin{align}
    U\left(\mathbf{X}^{N_s+1},\mathbf{X}^N\right)&=U\left(\mathbf{X}^{N_s},\mathbf{X}^N\right)+\Delta U_s\left(\mathbf{X}^{N_s},\mathbf{X}^N;\mathbf{X}_{N_s+1}\right)\nonumber\\
    &=U\left(\mathbf{X}^{N_s},\mathbf{X}^N\right)+U_{B}\left(\mathbf{X}^{N_s},\mathbf{X}^{N};\mathbf{X}_{N_s+1}\right)+U_{s,\mathrm{intra}}\left(\mathbf{R}_{N_s+1}\right)
\label{eq:U_B_org}
\end{align}
In the second equality, we decompose $\Delta U_s$ of the considered solute molecule into the intermolecular energy with all other solvent molecules and an intramolecular energy contribution $U_{s,\mathrm{intra}}(\mathbf{R}_{N_s+1})$.
The first part is referred to as the binding energy $U_B(\mathbf{X}^{N_s},\mathbf{X}^{N};\mathbf{X}_{N_s+1})$.
Additionally, we introduce relative coordinates $\rb^{\prime N_s}$ and $\rb^{\prime N}$, as
\begin{equation}
\begin{aligned}
    \rb^{\prime N_s}&=\rb^{N_s}-\mathbf{r}_{N_s+1}\\
    \rb^{\prime N}&=\rb^{N\phantom{_s}}-\mathbf{r}_{N_s+1}
\end{aligned}
\label{eq:relative_coord}
\end{equation}
with the short-hand notation $\mathbf{X}^{\prime}=\left\lbrace\mathbf{r}^\prime,\mathbf{R}\right\rbrace$.
The total energy is then
\begin{equation}
    U\left(\mathbf{X}^{\prime N_s+1},\mathbf{X}^{\prime N}\right)=U\left(\mathbf{X}^{\prime N_s},\mathbf{X}^{\prime N}\right) + U_{B}\left(\mathbf{X}^{\prime N_s},\mathbf{X}^{\prime N};\mathbf{R}_{N_s+1}\right)+U_{s,\mathrm{intra}}\left(\mathbf{R}_{N_s+1}\right)
\label{eq:U_B}
\end{equation}
The integration over the centre of mass position of the solute $\dr_{N_s+1}$ in eq. \eqref{eq:mu_NVT} can now be performed independently, simply yielding the volume $V$ of the system. The chemical potential is obtained as
\begin{multline}
    \mu_s(N_s,N,V,T)=-k_BT\ln{\frac{q_s~q_s^\mathrm{intra}\; V}{\Lambda_s^3 (N_s+1)} }\\
    -k_BT\ln{ \int \diff\mathbf{X}^{\prime N_s} \int \diff\mathbf{X}^{\prime N} \int \diff \mathbf{R}_{N_s+1}}~\exp{\left(-\beta U_{B}\left(\mathbf{X}^{\prime N_s},\mathbf{X}^{\prime N};\mathbf{R}_{N_s+1}\right)\right)} \\
    \times \frac{ \exp{\left(-\beta U_{s,\mathrm{intra}}\left(\mathbf{R}_{N_s+1}\right)\right)} } {q_s^\mathrm{intra}}~ \frac{\exp{\left(-\beta U\left(\mathbf{X}^{\prime N_s},\mathbf{X}^{\prime N}\right) \right)}}{\int \diff\mathbf{X}^{N_s} \int \diff\mathbf{X}^N \exp{\left(-\beta U\left(\mathbf{X}^{N_s},\mathbf{X}^{N}\right) \right)}}
    \label{eq:mu_s(NVT)_2}
\end{multline}
where we multiplied and divided by the intramolecular partition sum of the solute molecule in the (hypothetical) low-density limit, defined as
\begin{equation}
    q_s^\mathrm{intra} = \int \diff \mathbf{R}_{N_s+1}~\exp{\left(-\beta U_{s,\mathrm{intra}}\left(\mathbf{R}_{N_s+1}\right)\right)}
\label{eq:q_s_intra}
\end{equation}
Of course, we could alternatively consider any actual or hypothetical state in eq.~\eqref{eq:q_s_intra}.
The first term on the right hand side of eq.~\eqref{eq:mu_s(NVT)_2} can be identified as the ideal gas chemical potential $\mu_s^\mathrm{ig}(N_s,N,V,T)$ of the solute in the canonical ensemble, which can also be expressed in terms of the solute density $\rho_s=(N_s+1)/V$ in the system. The second term on the right hand side is referred to as the residual chemical potential $\mu_s^\mathrm{res}(N_s,N,V,T)$.
To further simplify the notation, we introduce the canonical ensemble average $\left\langle ~\right\rangle_{NVT}$, as
\begin{equation}
    \left\langle A\right\rangle_{NVT}=\int\diff\mathbf{X}^N~A~\frac{\exp{\left(-\beta U(\mathbf{X}^N)\right)}}{\int\diff\mathbf{X}^N~\exp{\left(-\beta U(\mathbf{X}^N)\right)}}
    \label{eq:NVT_average}
\end{equation}
We can then write the residual chemical potential in terms of a canonical average, with
\begin{equation}
    \mu_s^\mathrm{res}(N_s,N,V,T)=-k_BT\ln{\left\langle \int \diff \mathbf{R}_{N_s+1} \exp{\left(-\beta U_B(\mathbf{X}^{\prime N_s},\mathbf{X}^{\prime N};\mathbf{R}_{N_s+1})\right)} \frac{\exp{\left(-\beta U_{s,\mathrm{intra}}(\mathbf{R}_{N_s+1})\right)}}{q_s^\mathrm{intra}} \right\rangle_{N_sNVT}}
    \label{eq:mu_s(NVT)_3}
\end{equation}
To further condense the notation, we introduce the ensemble average over the intramolecular configurations $\left\langle ~\right\rangle_{\mathbf{R}_{N_s+1}}$ of the solute molecule according to
\begin{equation}
    \left\langle A\right\rangle_{\mathbf{R}_{N_s+1}}=\int \diff \mathbf{R}_{N_s+1}~A~\frac{\exp{\left(-\beta U_{s,\mathrm{intra}}(\mathbf{R}_{N_s+1})\right)}}{q_s^\mathrm{intra}}
    \label{eq:angle_average}
\end{equation}
Combining eqs. \eqref{eq:mu_s(NVT)_2}, \eqref{eq:mu_s(NVT)_3} and \eqref{eq:angle_average}, we arrive at the following compact result for the chemical potential of the solute
\begin{equation}
    \mu_s(N_s,N,V,T)=k_BT\ln{\frac{\rho_s \Lambda_s^3}{q_s~q_s^\mathrm{intra}}}-k_BT\ln{\left\langle\left\langle\exp{\left(-\beta U_B(\mathbf{X}^{\prime N_s},\mathbf{X}^{\prime N};\mathbf{R}_{N_s+1})\right)}\right\rangle_{\mathbf{R}_{N_s+1}}\right\rangle_{N_s NVT}}
    \label{eq:muNVT}
\end{equation}
The first and second term on the right-hand-side define the ideal and residual contribution to the chemical potential within the canonical ensemble, respectively. We obtain
\begin{align}
    \mu_s^\mathrm{ig}(N_s,N,V,T) &=k_BT\ln{\frac{\rho_s \Lambda_s^3}{q_s~q_s^\mathrm{intra}}}
    \label{eq:muigNVT}
    \\
    \mu_s^\mathrm{res}(N_s,N,V,T) &=-k_BT\ln{\left\langle\left\langle\exp{\left(-\beta U_B(\mathbf{X}^{\prime N_s},\mathbf{X}^{\prime N};\mathbf{R}_{N_s+1})\right)}\right\rangle_{\mathbf{R}_{N_s+1}}\right\rangle_{N_s NVT}}
    \label{eq:muresNVT}
\end{align}
The ideal-gas chemical potential contains a contribution $q_s^\mathrm{intra}$ due to the ensemble average over the solute configurations $\left\langle~\right\rangle_{\mathbf{R}_{N_s+1}}$.

\section{\label{SI:mu*(N,V,T)}Pseudo-chemical potential in the canonical ensemble}
The pseudo-chemical potential $\mu_s^\ast(N_s,N,V,T)$ corresponds to the change in Helmholtz energy due to adding the $(N_s+1)$-th solute molecule at a fixed center of mass location $\mathbf{r}_{N_s+1}=\mathbf{r}_0$ to the system. The pseudo-chemical potential is defined as
\begin{align}
    \mu_s^\ast(N_s,N,V,T)&=F(N_s+1,N,V,T;\rb_0)-F(N_s,N,V,T)\nonumber\\
    &=-k_BT\ln{\frac{Q(N_s+1,N,V,T;\rb_0)}{Q(N_s,N,V,T)}}
\end{align}
The canonical partition function of the system reads
\begin{align}
    Q(N_s+1,N,V,T;\rb_0)&=\frac{q_s^{N_s+1}~q^N}{\Lambda_s^{3N_s}\Lambda^{3N}N_s\,!~N\,!}\int \diff\mathbf{X}^{N_s} \int \diff\mathbf{X}^N \int \diff \mathbf{R}_{N_s+1}~\exp{\left(-\beta U\left(\mathbf{X}^N,\mathbf{X}^{N_s};\rb_0,\mathbf{R}_{N_s+1}\right)\right)}
\end{align}
The following differences to the usual canonical partition function arise due to the fixed position of the $(N_s+1)$-th molecule. The fixed molecule is distinguishable from the remaining $N_s$ molecules, therefore, the number of possible molecule permutations is only $N_s\,!~N\,!$. The integration of the momenta and the molecule configurations is only performed over $N_s$ molecules and thus yields only $\Lambda_s^{3N_s}$ de Broglie wavelengths.
The pseudo-chemical potential can then be expressed as
\begin{align}
    \mu_s^\ast(N_s,N,V,T)&=-k_BT\ln{q_s}-k_BT\ln{\int \diff\mathbf{X}^{N_s} \int \diff\mathbf{X}^N \int \diff \mathbf{R}_{N_s+1}}\nonumber\\
    &\qquad\times\frac{\exp{\left(-\beta U\left(\mathbf{X}^N,\mathbf{X}^{N_s};\rb_0,\mathbf{R}_{N_s+1}\right)\right)}}{\int \diff\mathbf{X}^{N_s} \int \diff\mathbf{X}^N \exp{\left(-\beta U\left(\mathbf{X}^N,\mathbf{X}^{N_s}\right)\right)}}
\end{align}
Following the steps taken in section \ref{SI:mu(N,V,T)}, separating the total potential energy and introducing relative coordinates, gives the pseudo-chemical potential according to
\begin{multline}
    \mu_s^\ast(N_s,N,V,T)=-k_BT\ln{\left(q_s~q_s^\mathrm{intra}\right)}\\
    -k_BT\ln{\int\diff\mathbf{X}^{\prime N_s}\int \diff \mathbf{X}^{\prime N}\int\diff \mathbf{R}_{N_s+1}}~ \exp{\left(-\beta U_B\left(\mathbf{X}^{\prime N},\mathbf{X}^{\prime N_s};\mathbf{R}_{N_s+1}\right)\right)}\\
    \times \frac{ \exp{\left(-\beta U_{s,\mathrm{intra}}\left(\mathbf{R}_{N_s+1}\right)\right)} }{q_s^\mathrm{intra}}~ \frac{\exp{\left(-\beta U\left(\mathbf{X}^{\prime N},\mathbf{X}^{\prime N_s}\right)\right)}}{\int \diff\mathbf{X}^{N_s} \int \diff\mathbf{X}^N\exp{\left(-\beta U\left(\mathbf{X}^N,\mathbf{X}^{N_s}\right)\right)}}
\end{multline}
The second term on the right hand side can be expressed as a canonical average in terms of the binding energy $U_B$, leading to
\begin{multline}
     \mu_s^\ast(N_s,N,V,T)=-k_BT\ln{\left(q_s~q_s^\mathrm{intra}\right)}\\
     -k_BT\ln{\left\langle\int\diff \mathbf{R}_{N_s+1}~\exp{\left(-\beta U_B\left(\mathbf{X}^N,\mathbf{X}^{N_s}; \mathbf{R}_{N_s+1}\right) \right)}~\frac{ \exp{\left(-\beta U_{s,\mathrm{intra}}\left(\mathbf{R}_{N_s+1}\right) \right)} }{q_s^\mathrm{intra}} \right\rangle_{N_s NVT}}
\end{multline}
and the integration over all configurations of the solute molecule resembles an ensemble average, as defined in eq. \eqref{eq:angle_average}, so that the pseudo-chemical potential in the canonical ensemble can be expressed as
\begin{equation}
     \mu_s^\ast(N_s,N,V,T)=-k_BT\ln{\left(q_s~q_s^\mathrm{intra}\right)} 
     -k_BT\ln{\left\langle\left\langle\exp{\left(-\beta U_B\left(\mathbf{X}^N,\mathbf{X}^{N_s};\mathbf{R}_{N_s+1}\right)\right)}\right\rangle_{\mathbf{R}_{N_s+1}}\right\rangle_{N_s NVT}}
\end{equation}
The two terms on the right-hand side define the ideal and residual contributions to the pseudo-chemical potential, as
\begin{align}
    \mu_s^{\ast,\mathrm{ig}}(N_s,N,V,T) &= -k_BT\ln{\left(q_s~q_s^\mathrm{intra}\right)}
    \label{eq:mu_*ig(N,V,T)}
    \\
    \mu_s^{\ast,\mathrm{res}}(N_s,N,V,T) &= 
    -k_BT\ln{\left\langle\left\langle\exp{\left(-\beta U_B\left(\mathbf{X}^N,\mathbf{X}^{N_s};\mathbf{R}_{N_s+1}\right)\right)}\right\rangle_{\mathbf{R}_{N_s+1}}\right\rangle_{N_s NVT}}
    \label{eq:mures*NVT}
\end{align}

\noindent
The ideal contribution thus equals the chemical potential of the solute, at fixed position $\mathbf{r}_{N_s+1}=\mathbf{r}_0$ and without interactions to the remaining system.

\section{\label{SI:mu(N,p,T}Chemical potential in the isobaric-isothermal ensemble}
The solute chemical potential $\mu_s(N_s,N,p,T)$ in the isobaric-isothermal ensemble is determined by the difference in the Gibbs energy
\begin{align}
    \mu_s(N_s,N,p,T)&=G(N_s+1,N,p,T)-G(N_s,N,p,T)\nonumber\\
    &=-k_BT\ln{\frac{\Delta(N_s+1,N,p,T)}{\Delta(N_s,N,p,T)}}
\end{align}
Here, $\Delta(N_s,N,p,T)$ is the isobaric-isothermal partition function given by 
\begin{align}
    \Delta(N_s,N,p,T)&=\frac{1}{V_0}\int_0^\infty \diff{V}~\exp{\left(-\beta pV\right)}~Q(N_s,N,V,T)\nonumber\\
    &=\frac{q_s^{N_s}~q^N}{\Lambda_s^{3N_s}\Lambda^{3N}N_s\,!~N\,!}\frac{1}{V_0}\int_0^\infty \diff{V}~\exp{\left(-\beta pV\right)} \int \diff \mathbf{X}^{N_s} \int \diff \mathbf{X}^N~\exp{\left(-\beta U\left(\mathbf{X}^N,\mathbf{X}^{N_s}\right)\right)}
\end{align}
with the reference volume $V_0$ needed to make the partition function dimensionless.
Using the relative coordinates of eq. \eqref{eq:relative_coord} and decomposing the total potential energy into the potential energy of the solvent molecules $U$, the binding energy $U_B$ and the intramolecular potential energy of the added solute molecule $U_{s,\mathrm{intra}}$ as in eq. \eqref{eq:U_B}, the chemical potential can now be expressed as

\begin{multline}
    \mu_s(N_s,N,p,T)=-k_BT\ln {\frac{q_s~q_s^\mathrm{intra}}{\Lambda_s^3 (N_s+1)}} \\
    -k_BT\ln \Biggl( {\int_0^\infty \diff{V}~V\exp{\left(-\beta pV\right)}\int\diff\mathbf{X}^{\prime N_s} \int\diff\mathbf{X}^{\prime N} \int\diff{\mathbf{R}}_{N_s+1}} 
    \\
    \times\exp{\left(-\beta U_B\left(\mathbf{X}^{\prime N},\mathbf{X}^{\prime N_s};\mathbf{R}_{N_s+1}\right)\right)}~\frac{ \exp{\left(-\beta U_{s,\mathrm{intra}}\left(\mathbf{R}_{N_s+1}\right)\right)} }{q_s^\mathrm{intra}}
    \\
    \times \frac{\exp{\left(-\beta U\left(\mathbf{X}^{\prime N_s},\mathbf{X}^{\prime N}\right)\right)}}{\int_0^\infty \diff{V} \exp{\left(-\beta pV\right)}\int\diff\mathbf{X}^{N_s} \int\diff\mathbf{X}^N \exp{\left(-\beta U\left(\mathbf{X}^{N_s},\mathbf{X}^N\right)\right)}} \Biggr)
\end{multline}
\\
\noindent
In the second term on the right-hand side we can identify an $NpT$ ensemble average $\left\langle~\right\rangle_{NpT}$

\begin{equation}
    \left\langle A\right\rangle_{NpT}=\frac{\int_0^\infty \diff{V} \exp{\left(-\beta pV\right)}~
    \int\diff\mathbf{X}^N~A~  \exp{\left(-\beta U(\mathbf{X}^N)\right)}}{\int_0^\infty \diff{V} \exp{\left(-\beta pV\right)}~
    \int\diff\mathbf{X}^N~ 
    \exp{\left(-\beta U(\mathbf{X}^N)\right)}}
\end{equation}

\noindent
leading to

\begin{multline}
    \mu_s(N_s,N,p,T)=
    -k_BT\ln{\frac{q_s~q_s^\mathrm{intra}}{\Lambda_s^3 (N_s+1)}}\\
    -k_BT\ln{\left\langle V \int\diff \mathbf{R}_{N_s+1}~\exp{\left(-\beta U_B\left(\mathbf{X}^{N_s},\mathbf{X}^N;\mathbf{R}_{N_s+1}\right)\right)}~\frac{ \exp{\left(-\beta U_{s,\mathrm{intra}}\left(\mathbf{R}_{N_s+1}\right)\right)} }{q_s^\mathrm{intra}}
    \right\rangle_{N_sNpT}}
\end{multline}

\noindent
Rewriting this in terms of the ensemble average over the intramolecular configurations of the solute molecule $\mathbf{R}_{N_s+1}$, as defined in eq.~\eqref{eq:angle_average}, we obtain

\begin{equation}
    \begin{split}
    \mu_s(N_s,N,p,T)=&-k_BT\ln{\frac{q_s~q_s^\mathrm{intra}}{\Lambda_s^3 (N_s+1)}} \\
    &-k_BT\ln{ \left\langle V \left\langle \exp{\left(-\beta U_B\left(\mathbf{X}^{N_s},\mathbf{X}^N;\mathbf{R}_{N_s+1}\right)\right)}\right\rangle_{\mathbf{R}_{N_s+1}}\right\rangle_{N_sNpT}}
    \end{split}
\end{equation}

\noindent
Rewriting the volume $V$ to the compressibility factor $Z(p) \equiv pV/(N+N_s)k_BT$, we obtain

\begin{equation}
    \begin{split}
    \mu_s(N_s,N,p,T)=&-k_BT\ln{\frac{q_s~q_s^\mathrm{intra} (N+N_s)k_B T}{\Lambda_s^3 (N_s +1) p}} \\
    &-k_BT\ln{ \left\langle Z(p) \left\langle \exp{\left(-\beta U_B\left(\mathbf{X}^{N_s},\mathbf{X}^N;\mathbf{R}_{N_s+1}\right)\right)}\right\rangle_{\mathbf{R}_{N_s+1}}\right\rangle_{N_sNpT}}
    \end{split}
\end{equation}

\noindent
Due to rewriting the volume $V$, the second term in this equation now reduces to zero in the case of a hypothetical ideal gas at the specified conditions $N_s$, $N$, $p$ and $T$ (since for the ideal gas $U_B=0$ and $Z^{\mathrm{ig}}(p) \equiv 1$). The two terms on the right-hand side of this equation thus define the ideal and residual contributions to the solute's chemical potential within the isobaric-isothermal ensemble, and we can write

\begin{align}
    \mu_s^{\mathrm{ig}}(N_s,N,p,T) &=
    -k_BT\ln{\frac{q_s~q_s^\mathrm{intra} (N+N_s)k_B T}{\Lambda_s^3 (N_s +1) p}}
    \\
    \mu_s^{\mathrm{res}}(N_s,N,p,T) &=
    -k_BT\ln{ \left\langle Z(p) \left\langle \exp{\left(-\beta U_B\left(\mathbf{X}^{N_s},\mathbf{X}^N;\mathbf{R}_{N_s+1}\right)\right)}\right\rangle_{\mathbf{R}_{N_s+1}}\right\rangle_{N_sNpT}}
\end{align}

\noindent
The ideal and residual contributions to the chemical potential are thus different than those in the canonical ensemble, eqs.~\eqref{eq:muigNVT}-\eqref{eq:muresNVT}.

\section{\label{SI:mu*(N,p,T}Pseudo-chemical potential in the isobaric-isothermal ensemble}
The pseudo-chemical potential $\mu_s^\ast(N_s,N,p,T)$ of the solute in the isobaric-isothermal ensemble corresponds to the change in Gibbs energy by adding an additional solute molecule at a fixed location $\mathbf{r}_{N_s+1}=\rb_0$
\begin{align}
    \mu_s^\ast(N_s,N,p,T)&=G(N_s+1,N,p,T;\rb_0)-G(N_s,N,p,T)\nonumber\\
    &=-k_BT\ln{\frac{\Delta(N_s+1,N,p,T;\rb_0)}{\Delta(N_s,N,p,T)}}
\end{align}
The isobaric-isothermal partition function of the system with the fixed molecule, using the decomposition of total potential energy of the system according to eqs. \eqref{eq:U_B_org}, is given by
\begin{multline}
    \Delta(N_s+1,N,p,T;\rb_0)=\frac{q_s^{N_s+1}~q^N}{\Lambda_s^{3N_s}\Lambda^{3N}N_s\,!~N\,!}\frac{1}{V_0}\int_0^\infty \diff{V}~\exp{\left(-\beta pV\right)} \int\diff\mathbf{X}^{\prime N_s} \int\diff\mathbf{X}^{\prime N} \int\diff \mathbf{R}_{N_s+1}\\
    \times~\exp{\left(-\beta U_B\left(\mathbf{X}^{\prime N_s},\mathbf{X}^{\prime N};\mathbf{R}_{N_s+1}\right)\right)}~\exp{\left(-\beta U_{s,\mathrm{intra}}\left(\mathbf{R}_{N_s+1}\right)\right)}~\exp{\left(-\beta U\left(\mathbf{X}^{\prime N_s},\mathbf{X}^{\prime N}\right)\right)}
\end{multline}
The pseudo-chemical potential then reads
\begin{multline}
    \mu_s^\ast(N_s,N,p,T)=-k_BT\ln{\left(q_s~q_s^\mathrm{intra}\right)}\\
    -k_BT\ln{\int_0^\infty \diff{V}~\exp{\left(-\beta pV\right)} \int\diff\mathbf{X}^{\prime N_s} \int\diff\mathbf{X}^{\prime N} \int\diff \mathbf{R}_{N_s+1}}~\exp{\left(-\beta U_B\left(\mathbf{X}^{\prime N_s},\mathbf{X}^{\prime N};\mathbf{R}_{N_s+1}\right)\right)}\\
    \times~\frac{ \exp{\left(-\beta U_{s,\mathrm{intra}}\left(\mathbf{R}_{N_s+1}\right)\right)} }{q_s^\mathrm{intra}}~\frac{\exp{\left(-\beta U\left(\mathbf{X}^{\prime N_s},\mathbf{X}^{\prime N}\right)\right)}}{\int_0^\infty \diff{V} \exp{\left(-\beta pV\right)}\int\diff\mathbf{X}^{N_s} \int\diff\mathbf{X}^N~\exp{\left(-\beta U(\mathbf{X}^N,\mathbf{X}^{N_s})\right)}}
\end{multline}
Introducing the $NpT$ ensemble average $\left\langle~\right\rangle_{NpT}$ and the average over the solute configuration $\left\langle\right\rangle_{\mathbf{R}_{N_s+1}}$, the final expression of the pseudo-chemical potential reads
\begin{equation}
    \mu_s^\ast(N_s,N,p,T)=-k_BT\ln{\left(q_s~q_s^\mathrm{intra}\right)} - k_BT\ln{\left\langle\left\langle\exp{\left(-\beta U_B\right(\mathbf{X}^{N_s},\mathbf{X}^N;\mathbf{R}_{N_s+1}\left)\right)}\right\rangle_{\mathbf{R}_{N_s+1}}\right\rangle_{N_sNpT}}
\end{equation}
The ideal and residual contributions to the pseudo-chemical potential are thus
\begin{align}
    \mu_s^{\ast,\mathrm{ig}}(N_s,N,p,T) &= -k_BT\ln{\left(q_s~q_s^\mathrm{intra}\right)}
    \\
    \mu_s^{\ast,\mathrm{res}}(N_s,N,p,T) &=
    k_BT\ln{\left\langle\left\langle\exp{\left(-\beta U_B\right(\mathbf{X}^{N_s},\mathbf{X}^N;\mathbf{R}_{N_s+1}\left)\right)}\right\rangle_{\mathbf{R}_{N_s+1}}\right\rangle_{N_sNpT}}
\end{align}
which, contrary to the ideal and residual contributions to the full chemical potential, are found to be independent of the considered ensemble, see eqs. \eqref{eq:mu_*ig(N,V,T)}-\eqref{eq:mures*NVT}.

\section{PC-SAFT parameters}

\begin{table}[h]
    \centering
    \begin{tabular}{lccc}
        \toprule
              \textbf{Solvent} & $m$ & $\sigma/$\r{A} & $\epsilon/k_B/\SI{}{\kelvin}$ \\
        \midrule
        \midrule
            methane & 1.0    & 3.7039 & 150.03 \\
        \midrule
            ethane & 1.6069 & 3.5206 & 191.42 \\
        \midrule
            propane & 2.0020 & 3.6184 & 208.11 \\
        \midrule
            butane & 2.3316 & 3.7086 & 222.88 \\
        \midrule
            pentane & 2.6896 & 3.7729 & 231.20 \\
        \midrule
            hexane & 3.0576 & 3.7983 & 236.77 \\
        \midrule
            heptane & 3.4831 & 3.8049 & 238.40 \\
        \midrule 
            octane & 3.8176 & 3.8373 & 242.78 \\
        \midrule
            nonane & 4.2079 & 3.8448 & 244.51 \\
        \midrule 
            decane & 4.6627 & 3.8384 & 243.87 \\
        \midrule 
            cyclohexane & 2.5303 & 3.8499 & 278.11 \\
        \midrule 
            benzene & 2.4653 & 3.6478 & 287.35 \\
        \bottomrule
    \end{tabular}
    \caption{PC-SAFT parameter for \textit{n}-alkanes, cyclohexane and benzene.}
    \label{tab:PC-SAFT parameter}
\end{table}

\end{document}